\documentclass[12pt,]{article}
\usepackage{amsmath}
\usepackage{amsfonts} \usepackage{amssymb}
\newtheorem{theorem}{Theorem}[section]
\newtheorem{definition}[theorem]{Definition}
\newtheorem{lemma}[theorem]{Lemma}

\newtheorem{proposition}[theorem]{Proposition}
\newtheorem{corollary}[theorem]{Corollary}

\numberwithin{equation}{section}

\begin{document}

\title{
Conformal structures of static vacuum data}

\author {Helmut Friedrich$^{1,2}$\\
$^1$ Max-Planck-Institut f\"ur Gravitationsphysik\\
Am M\"uhlenberg 1, 14476 Golm, Germany \\
$^2$ Department of Mathematics and Statistics\\
University of Otago, P. O. Box 56, Dunedin 9010,\\ New Zealand}

\maketitle                
 
\begin{abstract}

In the Cauchy problem for asymptotically flat vacuum  data the solution-jets along the cylinder at space-like infinity develop in general  logarithmic singularities at  the critical sets at which the cylinder  touches future/past null infinity. The tendency of these  singularities to spread along the null generators of null infinity obstructs the development of a smooth conformal structure at null infinity. 
For the solution-jets arising from time reflection symmetric data   to extend smoothly to the critical sets it  is  {\it necessary}  that the Cotton tensor of the initial three-metric $h$ satisfies a certain conformally invariant condition $(*)$ at space-like infinity, it is 
{\it sufficient}  that $h$ be asymptotically static at space-like infinity.
The purpose of this article is to characterize the gap between these conditions. We show that with the class of metrics which satisfy condition $(*)$  on the Cotton tensor and a  
certain non-degeneracy requirement is associated  a one-form $\kappa$ with conformally invariant differential $d\kappa$. We provide two criteria.
If $h$ is real analytic, $\kappa$ is closed, and  one of it integrals satisfies a certain equation then $h$ is conformal to static data near  space-like infinity.
If $h$ is smooth, $\kappa$ is asymptotically closed, and  one of it integrals satisfies a certain equation asymptotically 
then $h$ is asymptotically conformal to static data at space-like infinity.

\end{abstract}

\newpage

{\footnotesize

\section{Introduction}

The purpose of this  article is to  characterize  in the class of general time reflection symmetric, asymptotically flat vacuum data the subset  of  three-metrics which are {\it conformal to static data 
in some neighbourhood of} or {\it asymptotically conformal to static data at}   space-like infinity. We begin by indicating  the motivations which determine  the particular type of characterization which will be useful for us. 

The correctness of Penrose's idea  that gravitational fields behave asymptotically such as to admit a conformal boundary at infinity of a certain smoothness (\cite{penrose:1963},  \cite{penrose:scri}) has been confirmed  under fairly general assumptions for solutions of the Einstein vacuum field equations  with non-vanishing cosmological constant. 
In the de Sitter-type case (cosmological constant $\lambda > 0$ 
if the signature $(-, +, +,+)$ is used), it has been shown that vacuum  solutions can be obtained by prescribing data on an oriented, compact three-manifold which is to acquire  the meaning of the time-like conformal boundary at future (say) time-like and null infinity. There is complete freedom to prescribe a three-dimensional conformal structure, represented by some Riemannian three-metric $h_{ab}$,  and an $h$-divergence free symmetric  tensor field $w_{ab}$. Neither smallness assumptions on the fields $h_{ab}$ and  $w_{ab}$ nor restrictions on the Yamabe class of the conformal structure of $h_{ab}$ are required (\cite{friedrich:1deSitter}). Moreover, de Sitter space has been shown to be non-linearly stable  in the strong sense  that  standard Cauchy data sufficiently close to de Sitter data develop into solutions whose conformal structure extends smoothly  to future/past time-like and null infinity   
 (\cite{friedrich:n-geod}). Finally, these results have been shown to generalize to the Einstein equations coupled to conformally sufficiently well behaved matter fields (\cite{friedrich:global}). 
 
In the case of anti-de Sitter-type solutions the idea of the conformal boundary  turned out to be  instrumental in proving the existence of solutions to the vacuum field equations with negative cosmological constant which admit a smooth, time-like  conformal boundary at null and space-like infinity  (\cite{friedrich:AdS}).  The corresponding {\it initial boundary value problem} for Einstein's field equations  admits  the standard freedom of prescribing Cauchy data on a space-like initial slice  (with the appropriate asymptotic behaviour at space-like infinity) as well as the freedom to prescribe a Lorentzian conformal structure  on the boundary.
As usual in such problems, the data have to satisfy consistency conditions where the two hypersurfaces meet. 
That this particular  initial boundary value problem is quite natural for the Einstein equations is corroborated by the fact that it admits a  formulation which is manifestly covariant, a feature which should not be taken for granted (cf. \cite{friedrich:geom-unique}).  

In the case $\lambda < 0$  the non-linear stability question is 
 intrinsically  different from that in the cases where $\lambda \ge 0$. The conformal boundary at space-like and null infinity is time-like so that the spaces are not globally hyperbolic in the ususal sense and stability statements will have to include specifications of boundary conditions and choices of boundary data. Moreover, even  in the case of the conformally flat anti-de Sitter (covering) space there does not exist a smooth and finite conformal representation of past/future time-like infinity.

The case of vanishing cosmological constant and asymptotically flat initial data differs 
from the cases $\lambda \neq 0$ in principle.  Smooth vacuum data for the Cauchy problem in some suitable class of weighted Sobolev spaces cannot be prescribed completely freely any longer if the null cone  structure of a solution is to  extend smoothly to null infinity.   For this to be the case there must be imposed conditions on the Cauchy data   near space-like infinity  (\cite{friedrich:i-null}, \cite{friedrich:spin-2}).  The situation is not precisely understood yet. We consider the present work as a further step towards identifying the necessary restrictions and the remaining freedom.  

It should be emphasized that a detailed  understanding of the domains where space-like and null infinity meet is not of pure academic interest but has practical consequences.  As shown in \cite{friedrich:kannar1} and further elaborated on in \cite{j.a.v.kroon:2007} it  would  provide useful information about the  relation between the nature of the Cauchy data and the structure of fields induced on null infinity and it  would clarify the role of various concepts such as the Newman-Penrose constants and the Metzner-Bondi-Sachs group
which have been discussed at length in the literature. Moreover, various questions will reduce to straightforward, though possibly lengthy, calculations. 
On a different level,  it offers possibilities  to calculate numerically on finite grids the future and past evolution of asymptotically flat space-times globally  from standard Cauchy data, including their radiation field and other asymptotic quantities of physical interest. 
 In any case, if one wishes to make use of the geometric advantages of smooth conformal boundaries, sufficient insight into the geometrical/physical meaning of the required restrictions on the Cauchy data  is indispensible to understand whether the insistence on smooth conformal asymptotics unduly restricts our possibilities to model physical systems. 

The stability results on the hyperboloidal initial value problem (\cite{friedrich:n-geod}, \cite{friedrich:global}) show that the  field equations decide on the asymptotic  smoothness at null infinity in any neighbourhood of space-like infinity. 
An insight into the underlying mechanisms  requires a careful analysis of  the nature of the evolution process and its precise dependence on the data in the region where space-like infinity meets null infinity. 
 
 Cutler and Wald made an early  attempt to circumvent this difficulty by preparing asymptotically flat Cauchy data for which the behaviour of the corresponding solution space-time
was sufficiently well controlled  near space-like infinity to decide whether  it
 admits  hyperboloidal slices to which the results of  (\cite{friedrich:n-geod}, \cite{friedrich:global}) apply. 
 This led to a first, though very special,  result on the existence of radiative solutions to the Einstein-Maxwell equations which admit a complete, smooth conformal boundary at null infinity (\cite{cutler:wald}). The remarkable work on the existence of asymptotically flat solutions to the vacuum constraints which coincide inside a prescribed ball with some given initial data and outside a larger ball with some exactly static or stationary data, initiated  by Corvino  (\cite{corvino:2000})  and generalized by Corvino and Schoen  (\cite{corvino:schoen})  and by Chru\'sciel and Delay (\cite{chrusciel:delay:2003}),  led to a similar construction 
of a fairly  large class of solutions to the vacuum field equations which admit a complete, smooth conformal structure at null infinity (\cite{corvino: 2007}, \cite{chrusciel:delay:2002}). However, all these solutions seem to be  special by  arising from initial data which are static or stationary in a full neighbourhood of space-like infinity. There remains the question whether such strong conditions are necessary to ensure a smooth structure at null infinity.

In the standard conformal compactification of Minkowski space space-like infinity is represented by a point $i^0$ to which the conformal structure extends smoothly (\cite{penrose:scri}). Any Cauchy hypersurface of Minkowski space approaches that point.
If one stipulates a similar picture  for asymptotically flat solutions to the vacuum field equations with non-vanishing mass one finds that Cauchy data for the conformal field equations, which are underlying the work quoted above, become strongly singular when the Cauchy surface approaches the point $i^0$. This fact, which reflects the slow fall-off behaviour of the gravitational field near space-like infinity, is the stumbling block which prevents us from obtaining in the case $\lambda = 0$ a non-linear stability result which would be as strong and unrestricted as the one obtained in the de Sitter case.

It is a  surprising feature of  the vacuum field equation that they admit  a different picture in which the Cauchy problem for asymptotically flat initial data becomes finite and regular  \cite{friedrich:i-null}. In this setting, which relies on a particular type of conformal gauge which away from the initial hypersurface is provided by the conformal space-time structure itself, the Cauchy data (assumed here,  for convenience, to have only one end and simple topology) for the conformal field equations are given on a three-manifold $\tilde{S}$ diffeomeorphic to 
an open ball in $\mathbb{R}^3$ whose boundary $I^0 = \partial \tilde{S} \sim S^2$ represents  space-like infinity with respect to the initial slice. The vacuum solution arising from these data lives on a manifold $\tilde{M}$  which is a bounded neighbourhood of 
$\{0\} \times \tilde{S}$ in $\mathbb{R} \times \tilde{S}$, where we identified 
$\tilde{S}$ with $\{0\} \times \tilde{S}$.

In a particular conformal gauge on the initial slice $\tilde{S}$,  the Cauchy data and the conformal field equations extend smoothly to all of $\bar{S} = \tilde{S} \cup I^0$. This  allows us to attach  to $\tilde{M}$  in the given gauge the set $I = ]-1, 1[ \times I^0$
as a boundary such that $\{0\} \times I^0$ identifies with $I^0$.  This cylinder represents
space-like infinity with respect to the solution space-time. It is  important to note  that this construction is in no way arbitrary and that the boundary $I$ is smoothly generated in the given gauge by the (uniquely extended) conformal field equations from the set $I^0$. What was a Cauchy problem initially is represented now by an initial boundary value problem. 
This problem is, however,  of a very special kind. The conformal  field equations extend to $I$ so that only the differential operators tangential to $I$ have non-vanishing coefficients at points of $I$. Consequently, there is no freedom to prescribe boundary data.

In the given gauge the space-time metric extends  in a smooth but degenerate way to the set $I$. This creates no problems. The main objective leading to the picture outlined above was the desire to obtain a setting which would allow us to discuss  in detail the {\it field equations} near space-like and null infinity and it turns out that the equations  are regular and hyperbolic on $I$ and extend with this property also to 
 future/past null infinity ${\cal J}^{\pm}$ if the conformal structure extends to these sets with sufficient smoothness.

What has been referred to above loosely as the  `region where space-like infinity meets null infinity' is made precise in the present  setting by referring to an (arbitrarily small) neighbourhood of the   {\it critical sets} $I^{\pm} = \{\pm 1\} \times I^0$.  At these sets, which define the future/past boundary of the cylinder $I$, the sets future/past null infinity ${\cal J}^{\pm}$ come arbitrarily close to $I$ 
and the sets $I^{\pm}$ can in fact be thought of as representing the set of past/future endpoints of the null generators on  future/past null infinity.  The field equations are not any longer  hyperbolic  at the critical sets and the decision which initial data evolve into solutions that admit a smooth  conformal structure at null infinity takes place precisely at these sets.

While a cursory glance at the construction of \cite{friedrich:i-null}  may emphasize  unavoidable similarities between  this and earlier  attempts to relate the structures at space-like and null infinity, the representation of the field equations and in particular the explicit description of the critical sets obtained here mark essential differences. 
The setting discussed in 
\cite{ashtekar:hansen:1978} and related later work hardly lends itself for a detailed study of the field equations. The sets $I$ and $I^{\pm}$ are compressed there  into one point $i^0$,  the rich structure unfolding on $I$ and $I^{\pm}$ cannot be seen, and questions of smoothness are resolved by fiat. In \cite{beig:schmidt:1982} is proposed an analogue of the Bondi expansion for the analysis of space-like infinity. This allows for  some discussion of the field equations. In particular, the hierachy of linear equations discussed there bears some similarity to the transport equations on the cylinder $I$ referred to below. It is hard to see, however, how the setting could be used to  analyse the precise relations between space-like and null infinity and questions of  smoothness. The critical sets `live beyond' this setting. 

The setting of \cite{friedrich:i-null} requires some `decent' behaviour of the Cauchy data near space-like infinity but the general  class of data for which it makes sense and for which some smoothness is preserved on the cylinder $I$ has not been worked out yet. To allow for a convenient  first analysis there have been considered in  \cite{friedrich:i-null} time reflection symmetric data which admit a smooth conformal compactification at space-like infinity. These data include the static data. In \cite{dain:friedrich} have been discussed {\it non-time reflection symmetric} data whose underlying metrics admit a smooth conformal compactification at space-like infinity. But even this requirement  can be largely weakened to include, in particular, 
stationary Cauchy data, which  do not admit  a smooth compactification (whose asymptotics  is still somewhat special though (\cite{dain:stationary})). It has been shown in  \cite{friedrich:cargese} that for static  data 
and more recently in  \cite{acena:kroon} that for stationary data the 
complete setting outlined above, including the sets $I^{\pm}$ where space-like infinity $I$ touches the set future/past null infinity ${\cal J}^{\pm}$ as well as the fields induced there, is 
smooth.

For sufficiently  general data the loss of hyperbolicity at the critical sets leads to a loss of smoothness at $I^{\pm}$. The fact that the cylinder $I$ is a total characteristic has the consequence 
that on $I$ the unknown $u$ of the conformal field equations (which comprises a number of tensor fields)  and its derivatives transverse to $I$, i.e. the solution-jets $J^p_I(u)$, $p = 0, 1, 2, \ldots$, on $I$,
are governed by hyperbolic transport equations interior to $I$ (cf. 
\cite{friedrich:i-null}, \cite{friedrich:cargese} for more details). The derivatives of $u$ of order  $p \ge 1$ are subject to linear equations 
with forcing terms $F^p$ which depend on the solution-jets  
$J^q_I(u)$, $q \le p - 1$, of lower order. The linear equations have been analysed in  \cite{friedrich:i-null} and in cases which admit sufficient control on the  forcing terms the solutions can be discussed in detail.
It turned out that in general the solution-jets $J^p_I(u)$ are smooth on $I$ but can develop logarithmic singularities at the critical sets $I^{\pm}$. Because the forcing terms $F^p$ become quite complicated with increasing order $p$ a full analysis of the transport equations is still outstanding, simply because of the heavy algebra involved.

There are good reasons to assume that the smoothness of the  solution-jets $J^p_I(u)$ at $I^{\pm}$ controls the smoothness of the conformal structure at null infinity (\cite {friedrich:spin-2}). Therefore it is important to identify those properties of the initial data which give rise to obstructions to the smoothness of the solution-jets. The most complete results have been obtained so far in the case of time reflection symmetric data which admit a smooth conformal compactification.  To explain the situation we use a conformal gauge on the initial slice which is different from the one indicated above. 
The conformal data are then derived from a smooth (in our convention negative definite) Riemannian metric $h$ of positive Yamabe class on a three manifold $S \sim S^3$ on which there exist a point $i \in S$ and  function  $\Omega \in C^2(S) \cap C^{\infty}(\tilde{S})$ so that 
\begin{equation}
\label{fact-cond}
\Omega(i) = 0, \,\,\, D_a\Omega(i) = 0, \,\,\, D_aD_b\Omega(i) = - 2\,h_{ab}, \,\,\,
\Omega > 0 \,\,\,\mbox{on}\,\,\,\tilde{S} = S \setminus \{i\},
\end{equation}
where $D$ denotes the Levi-Civita connection defined by $h$,
and where  $\tilde{h} = \Omega^{-2}\,h$ satisfies the vacuum constraints 
\[
R[\tilde{h}] = 0 \quad \mbox{on} \quad \tilde{S}. 
\]
Two types of obstructions to the smoothness on the critical sets have been identified in the case of such data.

\vspace{.3cm}

\noindent
$(i)$ {\it There arise obstructions related to the conformal structure of $h$}. 

\vspace{.3cm}

\noindent
It has been shown in
 \cite{friedrich:i-null} that a {\it necessary} condition on $h$ for the solution-jets $J^p_I(u)$ to extend smoothly to the critical sets is that the dualized Cotton tensor of $h$, given by 
\[ 
B_{ab} = \frac{1}{2}\,(D_{c}R_{da}  - \frac{1}{4}\,D_{c}R\,h_{da})\,\epsilon_b\,^{cd},
\]
where $R_{ab}$, $R$, and $\epsilon_{abc}$  denote the Ricci tensor, the Ricci scalar, and  the totally antisymmetric tensor density which defines the volume element of $h$, satisfies  
\[
(*) \quad \quad \quad D_{\{a_1} \cdots D_{a_p}\,B_{ab\}}(i) = 0 \quad \mbox{for} \quad
p = 0, 1, 2, 3, \ldots .
\]
Here the curved brackets indicate taking the {\it symmetric trace free part
of the tensor with respect to the indices in the curved brackets}. At the lowest order $p$ at which 
$(*)$ would be violated the solution-jet $J^p_I(u)$ would develop logarithmic singularities at the critical sets which then generate logarithmic singularities also at higher orders.

Condition $(*)$ has been observed as a regularity requirement    for the first time  in 
 \cite{friedrich:stat-rad}, which studies  the case of vanishing ADM mass. There it has been observed in particular (cf. also \cite{beig:1991}) that

\vspace{.1cm}

\noindent
$-$ $(*)$ is conformally invariant (though a single  member of the 
sequence  is in general not \\
\hspace*{.45cm}conformally invariant) and thus imposes a restriction on the conformal structure of $h$,

\vspace{.1cm}
\noindent
$-$ data which behave near $i$ like conformally compactified asymptotically flat static data\\
\hspace*{.45cm}(discussed in detail below) present a non-trivial class of data which satisfy $(*)$.

\vspace{.3cm}

\noindent
$(ii)$ {\it There arise obstructions related to the conformal scaling  of $h$}.

\vspace{.3cm}

\noindent
Suppose that there is a neighbourhood $U$ of $i$ in $S$ on which the function $\Omega$
and the metric  $h$  are such that the metric
$\tilde{h} = \Omega^{-2}\,h$ on  $\tilde{U} = U \setminus \{i\}$, which has 
an asymptotically flat end at $i$,  satisfies the static vacuum field equations on $\tilde{U} $ (see for details below). 
As pointed out above, the solution-jets $J^p_I(u)$, where $u$ are the Cauchy data for the conformal field equations derived from $h$ and $\Omega$, then extend smoothly to $I^{\pm}$.

Assuming $U$ suitably small, we can consider  other conformal factors $\Omega' \in C^2(U) \cap C^{\infty}(\tilde{U})$ 
which satisfy conditions (\ref{fact-cond}) and which are such that 
$R[\Omega'^{-2}\,\tilde{h}] = 0$ on $\tilde{U}$. Such functions are obtained by solving the Lichnerowicz equation with suitable boundary data. One can then write 
$\Omega' = \Omega\,\theta$ with a positive function 
$\theta \in C^2(U) \cap C^{\infty}(\tilde{U})$ of the form $\theta = (1 + \Omega^{1/2}\,W)^{-2}$ where $W$ is 
a solution to $(\Delta_h - \frac{1}{8}\,R[h])W = 0$ near $i$. Assuming $\theta$ to satisfy the conditions $\theta(i) = 1$, $D_a\theta(i) = 0$, $D_aD_b\theta(i) = 0$, which ensure in particular that the static vacuum data $\tilde{h}$ and the vacuum data $\theta^{-2}\,\tilde{h}$ on $\tilde{U}$ have the same mass (and which avoid certain subtleties), J. Valiente Kroon states in particular  (and in  somewhat different terms) the following result  in  \cite{j.a.v.kroon:2010}, \cite{j.a.v.kroon:2011}:

\vspace{.1cm}

\noindent
The solution-jets $J^p_I(u')$, where $u'$ are the Cauchy data for the conformal field equations derived from $h$ and $\Omega'$, are free from logarithmic singularities and extend at all orders $p$ smoothly to the critical sets  if and only if $\theta - 1$ vanishes at $i$ at all orders.

\vspace{.2cm}

Additional information is required to understand the precise scope of these results. For convenience it has been assumed   in  \cite{friedrich:i-null}  and consequently in \cite{j.a.v.kroon:2011} that the metric $h$ is real analytic in some neighbourhood of the point $i$. The analysis of  \cite{dain:friedrich} shows, however, that this is not necessary. The properties of the fundamental solution of the conformal Laplace operator with pole at $i$ which are needed in the two articles hold also if $h$ is only $C^{\infty}$. 
 Moreover, only the derivatives of the data $h$ and $\Omega$ at $i$ up to some related finite order $p' = p'(p)$ are needed  to determine for a given order $p$ the solution-jet $J^p_I(u)$ on $I$ and thus on $I^{\pm}$. This implies that the results referred to above only depend on the Taylor coefficients in an expansion of the data $h$ and $\Omega$ at $i$. As a consequence, the results are less restrictive than may  appear at first sight.

To explain that only the asymptotic behaviour of $h$ needs to be restricted in case $(i)$ considered above,
let $h$ and $\Omega$ denote again a metric and a conformal factor on $U$ such that (\ref{fact-cond}) holds and 
$\tilde{h} = \Omega^{-2}\,h$
satisfies the static field equations on $\tilde{U}$. Consider metrics which are of the form
\begin{equation}
\label{gen-as-stat-metrics} 
h' = \vartheta^2\left\{(1 - \phi)\,h+ \phi\,\hat{h}\right\},
\end{equation}
where $\vartheta$ is a smooth positive conformal factor and
$\hat{h}$ some smooth Riemannian metric on $S$, the function $\phi \in C^{\infty}(S)$  
vanishes at all orders at $i$, is positive and $\le 1$ on $S \setminus \{i\}$, and equal to $1$ on a neighbourhood of $S \setminus U$. 
The metric $(1 - \phi)\,h + \phi\,\hat{h}$  coincides then at all orders with the metric $h$ at $i$ and it follows by the observations above that with $h$ also $h'$ satisfies $(*)$. Besides data which are conformal to  static  data in some punctured neighbourhood of $i$ there is thus quite a general  class of  non-analytic $C^{\infty}$ data which satisfy condition $(*)$ and which are {\it not conformal to static data in any punctured neighbourhood of $i$}.

To explain the role of the asymptotic behaviour of $h$ in case $(ii)$, let $\tilde{h}'$ be another  metric on $\tilde{U}$ with an asymptotically flat end at 
$i$. Assume that it satisfies the vacuum constraint $R[\tilde{h}'] = 0$ but not necessarily the static field equations in any neighbourhood of $i$. If this metric is such that $h' = \Omega^2\,\tilde{h}'$ extends smoothly (resp. with a certain smoothness) to all of $U$ and
$h' - h$ vanishes at all orders (up to some order) at $i$, then the solution-jets $J^p_I(u')$ determined by the pair $(h', \Omega)$ extend smoothly 
(up to some order) to the critical sets. If $\Omega'$ were a different conformal factor with $R[\Omega'^{-2}\,h'] = 0$ on $\tilde{U}$, we could subject the pair $(h', \Omega')$ to the analysis of
\cite{j.a.v.kroon:2011} and would find again that (up to some order) the solution-jets determined by these data extend smoothly to the critical sets if and only if $\Omega' = \Omega$ (up to some order). This seems to  indicate that the discussion of \cite{j.a.v.kroon:2011} does not require {\it staticity on some neighbourhood of $i$} but only {\it asymptotic staticity at $i$}. There remains, however, the question whether metrics $h'$ which bear the relations to static metrics indicated above do exist.
 If the requirements are relaxed, as indicated by the statements in brackets,  
results of \cite{chrusciel:delay:2003} apply which ensure the existence of asymptotically flat, not necessarily static, vacuum data which coincide near space-like infinity up to an {\it arbitrarily prescribed} order with some static vacuum data. 
While  the restriction to finite differentiability may in fact be sufficient  for practical purposes because the violations of smoothness of the solution-jets become less severe for increasing order $p$
 (\cite{friedrich:i-null}, \cite{friedrich:cargese}), we concentrate here  on the $C^{\infty}$ case to get a complete understanding of the situation. 
Therefore it is of interest that the techniques and arguments provided  by \cite{chrusciel:delay:2003} 
allow one to show the existence of metrics $\tilde{h}'$ for which 
$h' - h$ vanishes at all orders at $i$ (\cite{chrusciel:2012p}).

The discussion above suggests that {\it staticity}, possibly in an asymptotic sense,  plays a distinguished role when it comes to deciding whether  a time reflection symmetric vacuum solution develops smooth conformal asymptotics at null infinity. The connection is fairly clear for the class of data considered in case $(ii)$, the relation is not so tight in case $(i)$.
In  the examples  (\ref{gen-as-stat-metrics}) staticity is used, though only  in an asymptotic sense. It is not clear, however,  whether this is necessary.  {\it So far it cannot be excluded that there  exist data for which the solution-jets extend smoothly to the critical set while the data
are not asymptotically static at $i$.}

\vspace{.1cm}

\noindent
Several steps are required to find the answer: \\
$-$ Assuming $(*)$, one needs to control the algebraic structure of the forcing terms $F^p$ 
well\\
\hspace*{.4cm}enough to decide whether the solution jets $J^p_I(u)$ develop 
logarithmic singularities at\\
\hspace*{.4cm}$I^{\pm}$, \\
$-$ if they do, the precise conditions on the initial data  $u|_{\bar{S}}$  need to be derived which ensure \\
\hspace*{.4cm}that the singular terms drop out, \\
$-$ these conditions must be translated into meaningful, possibly geometric, statements on\\
\hspace*{.4cm}the initial data $h$ and $\Omega$. 

\vspace{.1cm}

\noindent
These steps are already quite difficult in the case $(ii)$ where a regular static reference solution is available and everything reduces to controlling the one function $\theta$. They are more difficult  in the case $(i)$, where no comparison solution is available
and even {\it changes in the conformal structure} of the metric $h$ may be required.
It is not clear what to expect in the third step. Recognizing condition  $(*)$  in the context of  \cite{friedrich:i-null} was considerably simplified because it was  known already from  \cite{friedrich:stat-rad}. 
It is not obvious at all, however,  whether and which conditions beyond $(*)$ have necessarily to 
be imposed on the conformal structure of the data  to guarantee regularity at null infinity if the scaling of the metric is chosen appropriately.
{\it But certainly we will have to understand how the set of conformally static data is embedded in the set of data which satisfy $(*)$ and by which conditions the conformally static data are singled out. }

As a preparation for this investigation has been derived in  \cite{friedrich:statconv} a 
characterization of the static data which is complete and well adapted to our analysis. 
Secondly,
it has been shown in \cite{friedrich:confstatic},   \cite{friedrich:exactstat} that  the map which relates static data to their conformal classes is injective  if a few exceptional cases, that admit conformal Killing fields, are omitted.

In the present article we derive under a non-degeneracy assumption, which excludes, in particular, the exceptional cases,  a criterion that allows one to decide whether a metric which  satisfies  $(*)$ is conformal to static vacuum data up to some order,  in some asymptotic sense, or on some neighbourhood of $i$.

Though we mentioned  above that analyticity is not required in the present context, it turns out technically convenient to assume it in a first analysis because it allows one to put certain questions into a geometric setting and to recognize arguments  which are difficult to see in the context of messy recursive procedures. All our basics results can be derived recursively in the $C^{\infty}$ setting, however, and the arguments are given after obtaining the analytic results.

A second aspect which permeates the following  analysis is the fact that we are only interested in the structure of the data on an unspecified small 
neighbourhood of the point $i$ which represents space-like infinity.
Without pointing it out  in each case, this neighbourhood will always be assumed to (and, in fact, can) be chosen such that the requirements and statements we make are correct.

We comment on the structure of the article.
In section \ref{c-problem} we state our problem more precisely, recall a few results on static data, give formal definitions of notions of asymptotic staticity, and comment on the $C^{\infty}$ case. 

In section \ref{d-critrel} we derive the {\it basic equation} (of second order) which needs to be solvable for a conformal factor $\omega$  in order for a metric $h$ to be conformal to a  static datum. This equation poses three problems: it  contains a term which  is highly singular
at the point $i$, it is highly overdetermined, and it does  not seem to  fix the differential of $\omega$ at $i$ though the results of  \cite{friedrich:confstatic} and   \cite{friedrich:exactstat} tell us that this differential cannot be chosen arbitrarily.

In section  \ref{h-co-gauge} we discuss properties of a conformal gauge which is used in 
section \ref{f-hadamard} to control the properties of a fundamental solution. Assuming $h$ to be real analytic near $i$, we study in section \ref{g-holomorphic}  the extension of our setting into the complex domain. With the same assumption it is shown  in section  \ref{analyticity of f} that condition $(*)$ is precisely the condition which ensures the regularity of our basic equation. After this it is shown that the same result holds in  the $C^{\infty}$ case. 

In the subsequent sections we only consider metrics which satisfy $(*)$.
In section \ref{j-subconds}  consequences of the overdeterminedness of the basic equation are analysed. 
It turns out that this is done most conveniently in terms of a certain conformal density 
$t_{ab}$ which exists and is smooth for metrics satisfying $(*)$. 
It shares various properties with the (dualized) Cotton tensor. We use it to rewrite our basic equation and to derive a certain {\it consistency  condition}.
This  equation, which can be read as a PDE of first order for $\omega$ that  is  implicit and again overdetermined, allows us to derive our main results, which are stated in Theorems \ref{conf-stat-cond} and \ref{as-conf-stat-cond}.
With a certain non-degeneracy assumption the consistency condition  determines the value of 
$d\omega$ at $i$ uniquely. This  solves the third problem mentioned above.
The  condition furthermore  suggests  the definition a 1-form $\kappa$ whose differential $d \kappa$ is  conformally invariant.  The requirements that this 1-form be closed and one of its integrals satisfy a certain equation provide a criterion under which the metric is conformal to a static one near $i$  if it is assumed to be analytic near $i$. 
Theorem \ref{as-conf-stat-cond} considers the $C^{\infty}$ case and asymptotic staticity.
 It should be said that the non-degeneracy assumption 
is made  only for convenience. Cases where it is violated simply need a more detailed analysis.
In the final part of section   \ref{j-subconds} we analyse some aspects of the criterion in more detail.

In section \ref{k-datevol} we finally indicate  how our criterion is to be used in the analysis of the solution-jets on the cylinder at space-like infinity. With the results mentioned above, 
showing that our criteria are implied by the regularity requirement on the solution-jets at the critical sets comes close to showing that the {\it necessary and sufficient} condition for the solution-jets to extend smoothly to the critical sets is that the initial metric $h$ behaves  at space-like infinity asymptotically like static data\footnote{The complete proof requires the understanding of certain degenerate situations. 
The analysis of \cite{j.a.v.kroon:2011} needs to be extended to deal with the fact that the relation between static data and their conformal classes is not one-to-one  (\cite{friedrich:confstatic}, \cite {friedrich:exactstat}) and the present analysis should be extended to cover also the cases which have been excluded by the non-degeneracy requirement  on $t_{ab}$  in Theorem \ref{conf-stat-cond}.}. While the need for such a condition was certainly  not forseen when the concept of the conformal boundary was introduced, it appears to be a most natural requirement if there must be imposed specific restrictions on time reflection symmetric  Cauchy data at space-like infinity at all.

In the articles \cite{friedrich:statconv}, \cite{friedrich:confstatic}, \cite{friedrich:exactstat} have only been considered conformal extensions of asymptotically flat static vacuum data. They are defined explicitly and uniquely in terms of structures associated with the static data. Because the data considered in this article  are more general we add an appendix in which it is shown that the conformal extensions are unique up to conformal diffeomorphisms and that conformal maps of asymptotically flat spaces induce under a suitable assumption conformal maps of the conformally extended spaces. This removes, in particular,  an assumption made in the articles \cite{friedrich:confstatic}, \cite{friedrich:exactstat}. After this work was completed it was pointed out to us that some results of the appendix have been already discussed in 
\cite{chrusciel:1989}.
 
There is some inevitable overlap of the present analysis with that of  \cite{friedrich:i-null} and some of the arguments are given in more detail here because we expect to  need them  in subsequent work. We do not point out the relationship  in each case.

\vspace{.2cm}

\noindent
{\bf Acknowledgements}: I would like to thank Sergio Dain and Robin Graham for discussions and the members of the relativity group of Dunedin for discussions and hospitality.

\section{The problem}
\label{c-problem}

We are  interested in certain asymptotic properties of 
smooth, asymptotically flat (for convenience one end only), time reflection symmetric initial data sets for Einstein's vacuum field equations which admit smooth conformal compactifications and satisfy an additional  decay condition at  infinity. In technical terms, which allow us to specify this condition,  it means that we consider smooth, compact,  three-dimensional  ({\it negative definite})  Riemannian
spaces 
$(S, h)$ with a distinguished point $i \in S$
such that 

\vspace{.1cm}

\noindent
(i) $(S, h)$ has  positive Yamabe number, i.e.
\begin{equation}
\label{yamabe-number}
Y(S, h) 
= -  \inf_{\{\vartheta \in C^{\infty}(S), \vartheta >
0\}}\, \frac{\int_S 
(D_a\vartheta\,D^a\vartheta + \frac{1}{8}\,R\,\vartheta^2)
\,d\mu}
{(\int_S\,\vartheta^6\,d\mu)^{1/3}} > 0,
\end{equation}

\vspace{.1cm}

\noindent
(ii)  $h$ is not locally conformally flat near $i$,

\vspace{.1cm}

\noindent
(iii)  $h$ satisfies the condition 
\begin{equation}
\label{regcond}
D_{\{a_1} \cdots D_{a_p}\,B_{ab\}}(i) = 0 \quad \mbox{for} \quad
p = 0, 1, 2, 3, \ldots , p_*.
\end{equation}

\vspace{.2cm}

\noindent
Here $D$ denotes the covariant derivative operator,
$d\mu$ the volume element defined by $h$,  
\[
R_{ab}, \quad R, \quad L_{ab} = R_{ab} - \frac{1}{4}\,R\,h_{ab}, \quad 
B_{acd} = D_{[c}\,L_{d]a},
\quad
B_{ab} = 1/2\,\,B_{acd}\,\epsilon_b\,^{cd},
\]
the Ricci tensor,  the Ricci scalar,  the Schouten tensor, the Cotton tensor of $h$ respectively. By $T_{\{a_1  \ldots a_q\} a_{q + 1}  \ldots a_{q + p}}$ we denote the tensor obtained from a given tensor $T_{a_1  \ldots a_q a_{q + 1}  \ldots a_{q + p}}$ by taking the part of it which is symmetric and $h$-trace free with respect to the indices enclosed by the curly brackets. Finally,
$p_*$ is either a fixed positive  integer or $p_* = \infty$ in which case 
the requirement on $B_{ab}$  is to be satisfied for all $p \in \mathbb{N}$  and 
(\ref{regcond}) coincides with condition $(*)$ of the introduction.
 
\vspace{.2cm}

Condition (i)  ensures the existence of a unique function 
$\Omega \in C^2(S) \cap C^{\infty}(\tilde{S})$ which satisfies
\begin{equation}
\label{Omega-and-derivs-at-i}
\Omega = \,0, \quad D_a\Omega = \,0, \quad D_aD_b\Omega = - 2\,h_{ab}
\quad \mbox{at} \quad i, \quad \quad 
\Omega > 0 \quad \mbox{on} \quad \tilde{S} \equiv S \setminus \{i\},
\end{equation}
and the equation
\begin{equation}
\label{Omega-equ}
L_h\,(\Omega^{- 1/2}) = 4\,\pi\,\delta_i \quad \mbox{on} \quad S.
\end{equation} 
Here $L_h = \Delta_h - \frac{1}{8}\,R[h]$ denotes the conformal
Laplacian where $\Delta_h = D_aD^a$  and $\delta_i$ the Dirac density of weight one at $i$ (and thus in $i$-centered normal coordinates the usual Dirac measure $\delta_0$). These properties imply that   $\tilde{h} \equiv  \Omega^{-2}\,h$ has vanishing Ricci scalar and  the space 
$(\tilde{S}, \tilde{h})$ represents in fact an asymptotically flat initial data set for Einstein's vacuum field equations with vanishing second fundamental form for which  the point $i$ represents space-like infinity. 

Condition (ii) and the positive mass theorem  imply that these initial data sets have ADM mass 
\begin{equation}
\label{ADM-mass-pos}
m > 0.
\end{equation}
Our main reason for requiring $B_{ab}$ not to vanish identically {\it on some neighbourhood of $i$} is, however,  that otherwise the data would be conformal to the static Schwarzschild data on that  neighbourhood  and the following analysis would become trivial.

\vspace{.2cm}

The evolution in time of  initial data as described above has been analysed  
in \cite{friedrich:i-null} in a neighbourhood of the `critical sets' at which  space-like infinity `touches' null infinity. In this context, condition (iii) has been obtained as a necessary requirement  on the data for the solution space-time  to extend in a certain sense smoothly to the critical sets.
Understood as a {\it sequence of conditions for $0 \le p \le p_*$}, (\ref{regcond}) 
is conformally invariant (\cite{friedrich:stat-rad}) and therefore relevant   for the initial data  
$(\tilde{S}, \tilde{h})$.

\vspace{.2cm}

To analyse the strength of condition (\ref{regcond}), which is our main interest,  it suffices to consider the metric $h$ and its associated structures on some open neighbourhood $U$ of $i$. This neighbourhood can and will be assumed to be sufficiently small for the following statements to be correct. With the restriction to such a neighbourhood we will loose sight of the global condition $(i)$ which will therefore be ignored in the following.

We recall some facts which have been discussed in detail, though in somewhat different notation, in \cite{dain:friedrich}. 
Assuming $h$ to be smooth, a convenient local representation of the function $\Omega$ near $i$ is obtained as follows. 
Given the ADM-mass $m > 0$ of the initial data set we  write $\mu = m^2/4$. 
Then there exist $C^{\infty}$  functions $\rho$, $w$  on $U$ which satisfy
\begin{equation}
\label{sigmaval}
\rho = 0,\,\,\,\,\,\,D_a\rho = 0,\,\,\,\,\,\,
D_a D_b \rho = - 2\,\mu\,h_{ab}
\quad \mbox{at} \quad i,
\quad \quad  \rho > 0 \quad \mbox{on} \quad \tilde{U} \equiv U \setminus \{i\},
\end{equation} 
and
\begin{equation}
\label{srtsigmaequ}
L_h\,(\rho^{-1/2} + w) = \frac{4\,\pi}{\sqrt{\mu}}\,\delta_i  \quad \mbox{on} \quad U.
\end{equation} 
the factor $\mu^{-1/2}$ is inserted on the right hand side  to simplify some of the the expressions below.
{\it The $C^{\infty}$ function $L_h(w)$ vanishes at all orders at the point $i$} and the function $w$ is determined up to replacements $w \rightarrow w + w'$ with smooth solutions $w'$ to $L_h\,(w') = 0$. 
Because this equation admits solutions with $w'(i) \neq 0$ we can assume $w$ to be chosen such that $w(i) = 0$.
Assuming (\ref{sigmaval}), we can write equation (\ref{srtsigmaequ}) equivalently
\begin{equation}
\label{an2fequ}
0 = \Sigma[ h, \rho ] \equiv  2\,\rho\,s - D_a\rho\,D^a\rho + \frac{1}{6}\,R\,\rho^2 
- \check{\rho}
 \quad \mbox{on} \quad U
\quad \mbox{with} \quad
s = \frac{1}{3}\,\Delta_h\,\rho,
\end{equation}
where $\check{\rho} =  \frac{4}{3}\,\rho^{5/2}\,L_h(w)$ is smooth and vanishes at all orders at $i$.
 
{\it  If $h$ is (real) analytic the function $\rho$ is analytic and 
 $w = 0$, $\check{\rho} = 0$ in}  (\ref{srtsigmaequ}) and  (\ref{an2fequ}) (\cite{garabedian}). 
Taking derivatives of the right hand side of  equation (\ref{an2fequ}), requiring them to vanish at $i$ and using  (\ref{sigmaval}), one finds that $ D_aD_bD_c\rho(i) = 0$. Given this, a
 direct calculation shows that the Taylor expansion of $\rho$ at $i$ is determined uniquely by
 (\ref{sigmaval}),  (\ref{an2fequ}) because $\check{\rho}$ vanishes at all orders at $i$. 
 In the analytic case the function $\rho$  is  thus determined uniquely by $(U, h, \mu)$ and can in fact be considered as a functional of $h$ multiplied by $\mu$. In the following it will be important that 
{\it the expression of the Taylor  coefficients  of $\rho$ at $i$ in terms  of  quantities derived from $h$ and $\mu$ are independent of $h$ being smooth or analytic}.

\vspace{.1cm}

The function $\rho$ and the associated tensor field
\begin{equation}
\label{n1fequ}
 \Sigma_{ab}[h, \rho]  \equiv D_a D_b \rho -
s\,h_{ab} + \rho\,(1 - \rho)\,s_{ab},
\end{equation}
with $s$ given as in  (\ref{an2fequ}) and the trace free part $s_{ab} = R_{ab} - \frac{1}{3}\,R\,h_{ab}$ of the Ricci tensor,
will play central roles in the following discussion.
We set
\begin{equation}
\label{B1intco}
\Lambda_a[h, \rho]  \equiv D_a\,s + (1 - \rho)\,s_{ab}\,D^b\,\rho
+ \frac{1}{6}\,R\,D_a\rho + \frac{1}{12}\,\rho\,(1 - \rho)\,D_aR,
\end{equation}
\begin{equation}
\label{Xi-def}
 \Xi_{bca}[h, \rho]  \equiv  (1 - \rho)\,B_{bca}  
- 2\,D_{[c}\rho\,s_{a]b} 
- D^d\,\rho\,s_{d[c}\,h_{a]b}.
\end{equation}
\begin{equation}
\label{dual-Xi-def}
\Xi_{ab}[h, \rho] \equiv \frac{1}{2}\,\Xi_{acd}\,\epsilon_b\,^{cd} =  (1 - \rho)\,B_{ab}  
- D_{c}\rho\,s_{d(a}\,\epsilon_{b)}\,^{cd}.  
\end{equation}
A direct calculation which uses the Bianchi identity
\begin{equation}
\label{s-r-Bianchi-id}
D^as_{ab} = \frac{1}{6}\,D_bR,
\end{equation}
and the decomposition of the curvature tensor
\begin{equation}
\label{n=3-Riem-dcomp}
R_{abcd} = 2\,(h_{a[c}\,L_{d]b} + h_{b[d}\,L_{c]a}),
\end{equation}
 valid in three dimensions, then gives (suppressing the argument $[h, \rho]$) the identities 

\begin{equation}
\label{Lambda-def}
\Lambda_a = \frac{1}{2}\,D^e\,\Sigma_{ea} 
\end{equation}

\begin{equation}
\label{Bcott}
D_{[c}\, \Sigma_{a]b} + 
\frac{1}{2}\,D^e\,\Sigma_{e[c}\,h_{a]b} = \rho \,\,\Xi_{bca},
\end{equation}

\begin{equation}
\label{Cid}
D_a\,\Sigma = \,2\,\rho\,\Lambda_a - 2\,D^b\rho\,\Sigma_{ba} + \frac{1}{6}\,\rho^3\,D_aR
- D_a \check{\rho}.
\end{equation}

\vspace{.2cm}

\noindent
{\it Static vacuum data}

\vspace{.2cm}

\noindent
Let us assume  that $h$ satisfies the equations
\begin{equation} 
\label{conf-static-vacuum-equ}
R[h] = 0 
\quad \mbox{and} \quad 
\Sigma_{ab}[h, \rho] = 0
\quad \mbox{on} \quad U.
\end{equation}
The identities above then imply 
that  $\Lambda_a = 0$, $\Xi_{abc} = 0$, $D_a(\Sigma + \hat{\rho}) = 0$ on $U$ whence 
$\Sigma = - \check{\rho}$ on $U$ because 
$\Sigma + \check{\rho} = 0$ at $i$ by (\ref{sigmaval}). 

\vspace{.1cm}

If  we assume in addition that $h$ is analytic then $\check{\rho} = 0$ whence $\Sigma = 0$.
Setting  then 
\begin{equation}
\label{conf-vac-transition}
v = \frac{1 -   \sqrt{\rho}}{1 +   \sqrt{\rho}}
\quad \mbox{and}\quad 
\tilde{h} = \Omega_*^{-2}\,h \quad \mbox{with}\quad 
\Omega_* = \frac{\rho}{\mu\,(1 + \sqrt{\rho})^2}
\quad \mbox{on} \quad \tilde{U} = U \setminus \{i\},
\end{equation}
and using the transformation law of the Ricci tensor under conformal rescalings, one finds that $v$ and $\tilde{h}$ satisfy the equations 
\begin{equation} 
\label{static-vacuum-equ}
R_{ab}[\tilde{h}] - \frac{1}{v}\,\tilde{D}_a\tilde{D}_b v = 0, \quad \Delta_{\tilde{h}}\,v = 0,
\end{equation}
which imply  that the Lorentz metric $\tilde{g} = v^2\,dt^2 + \tilde{h}$ defines  a Lorentzian  asymptotically flat  solution to Einstein's vacuum field equations on $M = \mathbb{R} \times \tilde{U}$ which is static. 

It is well known (\cite{beig:simon}) that asymptotically flat solutions to the  {\it static vacuum field equations} 
(\ref{static-vacuum-equ})  admit conformal extensions at space-like infinity with a rescaled metric 
$h$ which is real analytic near $i$ in suitable coordinates (so that (\ref{srtsigmaequ}) holds with $w = 0$) and satisfies the {\it conformal static field equations}  (\ref{conf-static-vacuum-equ}).

It follows then that   
\begin{equation}
\label{stat-vac-dual-Xi-def}
0 = \Xi_{ab}[h, \mu] =  (1 - \rho)\,B_{ab}  
- D_{c}\rho\,s_{d(a}\,\epsilon_{b)}\,^{cd},
\end{equation}
which implies 
\begin{equation}
\label{Drho-Drho-contr-stat-vac-dual-Xi-def}
0 = B_{ab}\,D^a\rho\,D^b\rho \quad \mbox{near} \quad i,
\end{equation}
and thus  condition (\ref{regcond}) with $p_* = \infty$
 (\cite{friedrich:stat-rad}, see also section \ref{g-holomorphic}). Because of the conformal invariance of  (\ref{regcond}), data which are conformal to static vacuum data near space-like infinity  thus represent a class of data which satisfy condition  (\ref{regcond}) in a non-trivial way. This implies the existence of a large class of time-reflection symmetric, asymptotically flat vacuum data which are $C^{\infty}$ and satisfy 
 (\ref{regcond}) in a non-trivial way.  
The examples pointed out in the introduction have in common, however,  that they behave  asymptotically like metrics which are conformal to static vacuum metrics.

\vspace{.1cm}

If above we had just required that $h$ were $C^{\infty}$,  that $R[h] = 0$ near $i$ (a conformal gauge that  can locally always be achieved),  and that $\Sigma_{ab}[h, \rho]$ only vanished at all orders at $i$, we would have concluded 
that $\tilde{h}$ and $v$ satisfy the static vacuum equations asymptotically in the sense 
that the fields on the left hand sides of equations (\ref{static-vacuum-equ}) have, in terms of coordinates 
exhibiting the asymptotic flatness of the metric $\tilde{h}$, arbitrarily fast fall-off  behaviour at space-like infinity.
Similarly we would have concluded that the fields on the right hand sides of equations
(\ref{stat-vac-dual-Xi-def}) and (\ref{Drho-Drho-contr-stat-vac-dual-Xi-def}) vanish at all orders at $i$ from which we would again have obtained (\ref{regcond}) with $p_* = \infty$.

\vspace{.1cm}

 {\it In this article we would like to answer the following question. Suppose $h$ is a smooth metric 
which satisfies  condition $(*)$, does there exist a criterion which allows us to decide whether
$h$ is in fact conformal to a metric which behaves at space-like infinity
asymptotically  like a static datum ?} 

\vspace{.1cm}

Though the meaning of the question should be intuitively clear the following definition allows us to explain some subtle distinctions.

\begin{definition}
\label{as-stat}
Let $(\tilde{S}, \tilde{h})$ denote a smooth,  time reflection symmetric, asymptotically flat  vacuum data set (one end only) and $\tilde{S}'$ a (in the following suitably chosen) neighbourhood of space-like infinity in $\tilde{S}$. We say
that $(\tilde{S}, \tilde{h})$  is 

\vspace{.1cm}

\noindent
{\rm (i)  asymptotically static of order $j$} for some given positive  integer 
$j$ if there exists a \\ 
\hspace*{.5cm}static, asymptotically flat  vacuum data set $(\tilde{N}, \tilde{k}, v)$ and a diffeomorphism \\ 
\hspace*{.5cm}$\psi: \tilde{N} \rightarrow \tilde{S}'$ so that 
$ \psi^*\tilde{h} -\tilde{k} = O(|z|^{- j})$ as $|z| \rightarrow \infty$, 
where $z^a$ denote coordinates  \\ 
\hspace*{.5cm}on $\tilde{N}$  in which
 $\tilde{k}_{ab}  = - (1 + \frac{2\,m}{|z|})\,\delta_{ab} + O(|z|^{- (1 + \epsilon)})$ as 
 $|z| \rightarrow \infty$  with some 
$\epsilon > 0$,

\vspace{.1cm}

\noindent
{\rm (ii) weakly asymptotically static}  if it is  asymptotically static of order 
$j$ for all $j  \in \mathbb{N}$,

\vspace{.1cm}

\noindent
 {\rm (iii) asymptotically static}  if there exists a static, asymptotically flat  vacuum data set  \\ 
\hspace*{.5cm} $(\tilde{N}, \tilde{k}, v)$ and a diffeomorphism 
$\psi: \tilde{N} \rightarrow \tilde{S}'$ so that 
$ \psi^*\tilde{h} - \tilde{k} = O(|z|^{- j})$   for all  \\ 
\hspace*{.5cm} $j \in \mathbb{N}$ as $|z| \rightarrow \infty$, where $z^a$ denote coordinates as in (i), 

\vspace{.1cm}

\noindent
{\rm (vi) static near space-like infinity}  if there exists a static, asymptotically flat  vacuum  \\ 
\hspace*{.5cm} data set $(\tilde{N}, \tilde{k}, v)$ and a diffeomorphism 
$\psi: \tilde{N} \rightarrow \tilde{S}'$ so that $\psi^*\tilde{h} -  \tilde{k} = 0$. 

\vspace{.1cm}

\noindent
More generally, $(\tilde{S}, \tilde{h})$ will be said to be: 
{\rm (i') conformal to data which are asymptotically static of order $j$}, 
{\rm (ii') conformal to data which are weakly asymptotically static}, 
{\rm(iii') conformal to asymptotically static data}, 
{\rm (iv') conformal to static data near space-like infinity} 
respectively, if  the requirements of {\rm (i)} - {\rm (iv)} hold with the relation $ \psi^*\tilde{h} - \tilde{k}$ replaced by 
$\psi^*\tilde{h} - \theta^2\,\tilde{k}$ where $\theta$ denotes a smooth conformal factor $\theta$ such that  $c \le \theta \le c^{-1}$ on $\tilde{N}$ with some constant $c > 0$.
\end{definition}

In both hierarchies any given condition  implies the preceding one. If $(\tilde{S}, \tilde{h})$ in 
$(iii)$ is such that   $(\tilde{N}, \tilde{k}, v)$ and $\psi$ can be chosen so that 
$\psi^*\tilde{h}$ is real  analytic, it also satisfies $(iv)$.
The discussion in the introduction shows, however, that there exist $C^{\infty}$ spaces 
$(\tilde{S}, \tilde{h})$ satisfying $(iii)$ which are not static in any neighbourhood of space-like infinity.
Even if $(\tilde{S}, \tilde{h})$ in 
$(ii)$ is such that  for any given $j$ the space
$(\tilde{N}, \tilde{k}, v)$ and the map $\psi$ can be chosen so that  $\psi^*\tilde{h}$ is real  analytic and the condition of $(i)$ is satisfied, condition may $(iv)$ not necessarily follow. It has been shown in \cite{friedrich:statconv} that any static vacuum data set $(\tilde{N}, \tilde{k}, v)$ is defined uniquely up to diffeomorphisms by a sequence of `null data'. The map $\psi$ allows one to associate with 
$(\tilde{S}, \tilde{h})$ null data up to some order $j' = j'(j)$. Without further analysis there is, however, no reason to assume that the fact that this can be done for all $j \in \mathbb{N}$ 
implies that there can be associated with $(\tilde{S}, \tilde{h})$  a unique sequence of null data which satisfy the convergence requirement found in   \cite{friedrich:statconv}.

\vspace{.1cm}

Property $(ii')$ is of particular interest here.
Data $(\tilde{S}, \tilde{h})$ which satisfy it are at present the most general ones known to guarantee, possibly
after the suitable conformal rescaling, a smooth extension of the solution jets  
to the critical sets.

\vspace{.1cm}

In the category of conformally compactified  $C^{\infty}$ metrics our problem is a question about 
the structure of the Taylor expansions of the metrics at the point $i$. The direct analysis is not only awkward, however,  but it is also not easy to see how the argument could go. For convenience  we shall  therefore consider at various occasions metrics which are real analytic on some neighbourhood $U$ of $i$. This will allow us to study some of the relevant subproblems in closed form. Once the basic argument is understood the results can then also be obtained by successive order for order arguments  in the $C^{\infty}$ case.

\vspace{1cm}

\section{The basic equation}
\label{d-critrel}

\vspace{.2cm}

To discuss the nature of our problem more closely, we need the following  consequence of the conformal covariance of $L_h$ which slightly generalizes Lemma 2.1 of (\cite{friedrich:confstatic}).

\begin{lemma}
\label{rescfundsol}
Assume that the $C^{\infty}$  initial data set $(S,  h)$ with ADM mass $m$ satisfies conditions (i) - (iii) so that $\mu = m^2/4 > 0$ and  $\rho$ is given on some neighbourhood $U$ of $i$. Let $m'$ a positive number and 
$\vartheta $ be a positive solution of the equation
\begin{equation}
\label{Delta-vartheta-van}
L_h (\vartheta) = 0 \quad \mbox{on} \quad U.
\end{equation}
If we set $h'_{ab} = \vartheta^4\,h_{ab}$
then the Ricci scalar of the metric $h'$ satisfies $R[h'] = 0$ on $U$ and  the functions
$
\rho' = \frac{\mu'}{\mu}\,(\vartheta(i)\,\vartheta)^2\,\rho,
\quad w' =  \sqrt{\frac{\mu}{\mu'}}\,(\vartheta(i)\,\vartheta)^{-1}\,w
$
satisfy 
\[ 
L_{h'}(\frac{1}{\sqrt{\rho'}} + w') = 
\frac{4\,\pi}{\sqrt{\mu'}}\,\delta'_i,
\]
where $\delta'_i$ denotes the density which coincides with the standard Dirac measure $\delta_0$ in i-centered $h'$ normal coordinates. Conditions (\ref{sigmaval}) hold with $\rho$, $\mu$, and $D$ replaced by $\rho'$,$\mu'$ and the derivative operator $D'$ of $h'$, $w'(i) = 0$, and  
$L_{h'}(w')$ vanishes at all orders at $i$. 
\end{lemma}

Assuming that $m' = m$, $\vartheta(i) = 1$ and writing $\omega = \vartheta^{-2}$
we have
\begin{equation}
\label{h-h'}
h'_{ab} = \omega^{-2}\,h_{ab}, \quad 
\rho' =  \omega^{-1}\,\rho,
\end{equation}
\begin{equation}
\label{h-h'-B}
\omega(i) = 1.
\end{equation}
It will be convenient to rewrite equation (\ref{Delta-vartheta-van}), which reads in the new notation
\begin{equation}
\label{U-omega-equ}
L_h(\omega^{-1/2}) = - \frac{1}{8}\,\omega^{-5/2}\,R[h'] = 0,
\end{equation}
in the form
\begin{equation}
\label{B-omega-equ}
0 = \hat{\Sigma}[h, \omega]  \equiv 2\,\omega\,t - D_a\omega\,D^a\omega + \frac{1}{6}\,R[h]\,\omega^2 
\quad \mbox{with} \quad  
t = \frac{1}{3}\,\Delta_h\omega.
\end{equation}
We note that there is still a considerable freedom in choosing the function $\omega$.

To express  
\[
\Sigma_{ab}[h', \rho'] = D'_a D'_b \rho' - s'\,h'_{ab} + \rho' (1 - \rho')\,s'_{ab},
\]
in terms of $h$ and $\rho$, we use
\[
D'_a\,\rho' = \omega^{-1}\,D_a\rho - \omega^{-2}\,\rho\,D_a\omega,
\]
\[
D'_aD'_b\,\rho' = \omega^{-1}\,D_aD_b\,\rho - \omega^{-2}\,\rho\,D_aD_b\,\omega 
- \omega^{-2}\,h_{ab}\,D_c\omega\,D^c\rho 
+ \omega^{-3}\,\rho\,h_{ab}\,D_c\omega\,D^c\omega,
\]
\[
s'\,h'_{ab} = 
\left\{\omega^{-1}\,s - \omega^{-2}\,D_c\omega\,D^c\rho
- \rho\,\left(\frac{1}{3}\,\omega^{-2}\Delta_h\omega -  \omega^{-3}\,D_c \omega\,D^c\omega\right)
\right\}\,h_{ab},
\]
which give with  the general transformation law
\begin{equation} 
\label{sab-conf-trans}
s'_{ab} \equiv s_{ab}[h'] = s_{ab}[h] + \omega^{-1}\,D_aD_b\omega
- \frac{1}{3}\,h_{ab}\,\omega^{-1}\,\Delta_h\omega,
\end{equation}
the relation 
\begin{equation}
\label{Sigma-Sigma'-Pi}
\Sigma_{ab}[h', \rho'] 
= \frac{1}{\omega}\,\Sigma_{ab}[h, \rho] 
- \frac{\rho^2}{\omega^3}\,\hat{\Sigma}_{ab}[h, \omega],
\end{equation}

\noindent
with 
\begin{equation}
\label{DDvan}
\hat{\Sigma}_{ab}[h, \omega] = 
D_aD_b\omega - h_{ab}\,t + \omega\,(1 - \omega)\,s_{ab}.
\end{equation}

\vspace{.2cm}

\noindent
We  note that (\ref{Sigma-Sigma'-Pi}) holds for arbitrary positive conformal factors $\omega$, because equations (\ref{h-h'-B}),  (\ref{B-omega-equ}) have not been used to derive this relation and there has not been assumed a particular conformal scaling of $h$. If $h'' = \vartheta^{-2} h'$, $\rho'' = \vartheta^{-1}\,\rho'$ with some smooth function
$\vartheta > 0$ on $U$ repeated application of the relations above gives
\begin{equation}
\label{repeated-Sigma-Sigma'-Pi}
\Sigma_{ab}[h'', \rho''] = \frac{1}{\vartheta}\,\Sigma_{ab}[h', \rho'] 
- \frac{\rho'^{2}}{ \vartheta^3}\,\hat{\Sigma}[h', \vartheta]
\end{equation}
\[
= \frac{1}{\vartheta\,\omega}\,\Sigma_{ab}[h, \rho] - \frac{\rho^2}{(\vartheta\,\omega)^3}
\left\{ 
\vartheta^2\,\hat{\Sigma}_{ab}[h, \omega] + \omega\,\hat{\Sigma}_{ab}[h', \vartheta]
\right\},
\]
which implies the transformation law
\begin{equation}
\label{hatSigma_{ab}-trafo}
\hat{\Sigma}_{ab}[h, \vartheta\,\omega] =
\vartheta^2\,\hat{\Sigma}_{ab}[h, \omega] + \omega\,\hat{\Sigma}_{ab}[h', \vartheta],
\end{equation}
of the functional $\hat{\Sigma}_{ab}$, which is non-linear in the second argument. Similarly we get the transformation law
\begin{equation}
\label{hatSigma-trafo}
\hat{\Sigma}[h', \vartheta] =
\vartheta^2\,\hat{\Sigma}[h, \omega]
- \frac{1}{6}\,\vartheta^2\,R[h'] +  \frac{1}{6}\,R[h''].
\end{equation}\vspace{.2cm}

 For later use we note some general relations (suppressing $[h, \omega]$). Direct calculations which use (\ref{s-r-Bianchi-id}), (\ref{n=3-Riem-dcomp}) give 
\begin{equation}
\label{D-check-Sigma-ab}
D_c\hat{\Sigma}_{ab} =
D_cD_aD_b\omega - h_{ab}\,D_ct 
+ (1 - 2\,\omega)\,D_c\omega\,s_{ab} 
+ \omega\,(1 - \omega)\,D_cs_{ab} \quad \quad 
\end{equation}
\[
\quad = D_aD_cD_b\omega - R^d\,_{bca}\,D_d\omega
- h_{ab}\,D_ct 
+ (1 - 2\,\omega)\,D_c\omega\,s_{ab} 
+ \omega\,(1 - \omega)\,D_cs_{ab} 
\]
\begin{equation}
\label{div-hat-Sigma}
D^c\hat{\Sigma}_{ca} = 2\,D_at + 2\,(1 - \omega)\,s_{ab}\,D^b\omega + \frac{1}{3}\,R\,D_a\omega
+ \frac{1}{6}\,\omega\,(1 - \omega)\,D_aR,
\end{equation}
\begin{equation}
\label{rot-div-hatSigma}
D_{[c}\,\hat{\Sigma}_{a]b} + \frac{1}{2}\,D^e\,\hat{\Sigma}_{e[c}\,h_{a]b} =
\omega\,\{ (1 - \omega)\,B_{bca} - 2\,D_{[c}\omega\,s_{a]b} - D^e\omega\,s_{e[c}\,h_{a]b}\},
\end{equation}
\begin{equation}
\label{d-hat-scalar-Sima}
D_a\hat{\Sigma} = \omega\,D^c\,\hat{\Sigma}_{ca} - 2\,D^c\omega\,\hat{\Sigma}_{ca} 
+ \frac{1}{6}\,\omega^3\,D_aR.
\end{equation}

\vspace{.1cm}

To show  that smooth data  $(S,  h)$ satisfying conditions (i) - (iii) are on some neighbourhood $U$ of $i$ conformal to static data, it is by (\ref{Sigma-Sigma'-Pi}), (\ref{DDvan})  necessary to show the existence of a smooth  function $\omega$ on $U$ with $\omega(i) = 1$ that satisfies  
(\ref{B-omega-equ}) and the {\it basic equation}
\begin{equation}
\label{add-equ}
D_aD_b\omega - h_{ab}\,t + \omega\,(1 - \omega)\,s_{ab}
- \omega^2\,f_{ab} = 0 \quad \mbox{with} \quad f_{ab} = f_{ab}[h, \rho] = \rho^{-2}\,\Sigma_{ab}[h, \rho].
\end{equation}
If $h$ and $\omega$ where real analytic this would also be sufficient. To show that the data $(S, h)$ are asymptotically conformal to static data in the sense that $h$ could be rescaled so that the field 
$\Sigma_{ab}[h', \rho'] $ in (\ref{Sigma-Sigma'-Pi}) vanishes at all orders at $i$ one would have to find a solution $\omega$ of (\ref{B-omega-equ}) so that the term on the left hand side of 
equation (\ref{add-equ}) vanishes at all orders at $i$.
Even if we ignore the question of analyticity here, these  tasks are complicated by several features of (\ref{B-omega-equ}),  (\ref{add-equ}).

\vspace{.2cm}

\noindent
{\it The system (\ref{add-equ}) is highly singular at $i$}. 

\vspace{.2cm}

In coordinates $x^a$ with origin at $i$ we have 
$\rho^2 = O(|x|^4)$ as $x^a \rightarrow 0$. Thus we can only expect to find a smooth solution $\omega$
if  $f_{ab}$ admits a smooth extension to a whole neighbourhood of $i$. 
As shown below, this can be true only if the Taylor coefficients of $\Sigma_{ab}$ at $i$ satisfy {\it conditions at any order}. Dividing both sides of (\ref{Sigma-Sigma'-Pi}) by $\rho'^2$ and assuming $\omega$ to be an arbitrary smooth  positive  conformal factor on $U$ gives the relation
\begin{equation}
\label{fab-conf-transf}
f_{ab}[h', \rho'] = \omega\,f_{ab}[h, \rho] - \frac{1}{\omega}\,\hat{\Sigma}_{ab}[h, \omega],
\end{equation}
which shows that the possibility to extend $f_{ab}$ as a smooth (real analytic) function to $U$ only depends on the conformal class of $h$.

\vspace{.2cm}

We add a few observations concerning the function $f_{ab}$. Where defined, $f_{ab}$ is symmetric and $h$-trace free. From
\[
D_cf_{ab}= \rho^{-2}\,D_c\Sigma_{ab} - 2\, \rho^{-3}\,D_c\rho\,\Sigma_{ab} =
 \rho^{-2}\,D_c\Sigma_{ab} -  2\, \rho^{-1}\,D_c\rho\,f_{ab}, 
\]
follows with  (\ref{an2fequ}), (\ref{Lambda-def}), (\ref{Cid}) 
\begin{equation}
\label{div-f}
D^a f_{ab} = \frac{2}{\rho^3}\,(\rho\,\Lambda_{a} - D^a \rho\,\Sigma_{ab}) = - \frac{1}{6}\,D_aR.
\end{equation}
The divergence of $f_{ab}$ thus extends smoothly without any assumption.
Moreover, with (\ref{Bcott}) follows
\[
D_{[c}f_{a]b} + \frac{1}{2}\,D^e f_{e[c}\,h_{a]b} =
 \rho^{-2}\left(D_{[c}\Sigma_{a]b} + \frac{1}{2}\,D^e \Sigma_{e[c}\,h_{a]b}\right)
- 2\, \rho^{-1}\left(D_{[c}\rho\,\Sigma_{a]b} + \frac{1}{2}\,D^e \rho\,\Sigma_{e[c}\,h_{a]b}\right)
\]
\[
= \rho^{-1}\left\{(1 - \rho)\,B_{bca} - 2\,D_{[c}\rho\,s_{a]b} - D^e\rho\,s_{e[c}\,h_{a]b}
- 2\,D_{[c}\rho\,f_{a]b} - D^e\rho\,f_{e[c}\,h_{a]b}\right\} 
\]
whence
\begin{equation}
\label{Df-relations}
2\,(1 - \rho)\,B_{bd} - 2\,D_{c}\rho\,(s_{a(b} + f_{a(b})\,\epsilon_{d)}\,^{ca} 
= \rho
\left(D_{c}f_{ab} + \frac{1}{2}\,D^e f_{ec}\,h_{ab}\right)\epsilon_d\,^{ca},
\end{equation}
and, using (\ref{div-f}),
\begin{equation}
\label{2Df-relations}
\rho\,D_{[c}f_{a]b} =
\end{equation}
\[
\frac{1}{12}\,\rho\,D_{[c}R\,h_{a]b}
+ (1 - \rho)\,B_{bca} - 2\,D_{[c}\rho\,s_{a]b} - D^e\rho\,s_{e[c}\,h_{a]b}
- 2\,D_{[c}\rho\,f_{a]b} - D^e\rho\,f_{e[c}\,h_{a]b}. 
\]
If $f_{ab}$ admits an analytic extension to some neighbourhood of $i$  this implies  a relation between $f_{ab}$ and $B_{ab}$. It will be used in section \ref{j-subconds}. 

\vspace{.3cm}

\noindent
{\it The system (\ref{B-omega-equ}), (\ref{add-equ}) is highly overdetermined}. 

\vspace{.2cm}

Assume that $f_{ab}$ extends smoothly to a neighbourhood of $i$.  To some extent the situation is then ameliorated by the following observation.

\begin{lemma}
\label{only-one-equation}
Any solution on $U$ to (\ref{add-equ}) which satisfies $\omega(i) = 1$ and $\hat{\Sigma}(i) = 0$, i.e.
\begin{equation}
\label{t-at-i}
2\,t =  D_a\omega\,D^a\omega - \frac{1}{6}\,R[h] \quad \mbox{at} \quad i,
\end{equation}
also satisfies equation  (\ref{B-omega-equ}) on $U$.
\end{lemma}

\noindent
In fact, (\ref{d-hat-scalar-Sima}), (\ref{add-equ}), (\ref{div-f}) imply
\[
D_a\hat{\Sigma} = \omega\,D^c\,(\omega^2\,f_{ca}) - 2\,D^c\omega\,\hat{\Sigma}_{ca} 
+ \frac{1}{6}\,\omega^3\,D_aR
\]
\[
= \omega^3\,D^cf_{ca} + 2\, D^c\,\omega\,(\omega^2\,f_{ca} 
- \hat{\Sigma}_{ca})
+ \frac{1}{6}\,\omega^3\,D_aR = 0.
\]
$\Box$

Thus apart from requiring $\omega(i) = 1$ and the relation (\ref{t-at-i}), which fixes the value of $t$ at $i$ once 
\begin{equation}
\label{def-ca}
c_a \equiv D_a\omega(i),
\end{equation}
is known, we are left with (\ref{add-equ}). 
It may look somewhat odd that a condition of second order at $i$ has to be specified above but this becomes plausible with the following observation. Applying to both sides of equation (\ref{add-equ}) the  covariant derivative $D_c$, commuting derivatives on the right hand side, contracting, and observing (\ref{div-f}) gives the integrability condition
\begin{equation}
\label{add-equ-int-cond}
0 = \hat{\Lambda}_a[h, \omega] \equiv D_at + (1 - \omega)\,s_{ac}\,D^c\omega 
+ \frac{1}{6}\,R\,D_a\omega + \frac{1}{12}\,\omega\,D_aR - \omega\,f_{ac}\,D^c\omega,
\end{equation}
which may require the specification of $t(i)$.

\vspace{.2cm}

The overdeterminedness does not leave much freedom.
If $]- a, a[ \,\ni \tau \rightarrow \gamma(\tau) \in U$, $a > 0$, denotes a geodesic with unit tangent vector $\dot{\gamma}^a$ and $\gamma(0) = i$ the equations
\[
\dot{\gamma}^a\,\dot{\gamma}^b\,(\hat{\Sigma}_{ab}[h, \omega] - \omega^2\,f_{ab})
= 0, \quad \,\,
\dot{\gamma}^a\,(\hat{\Sigma}_{ab}[h, \omega] - \omega^2\,f_{ab})
= 0, \quad \,\,
\dot{\gamma}^a\,\hat{\Lambda}_{a}[h, \omega] = 0,  
\]
can then be read as a system of ODE's for the quantities $\omega$, $D_a\omega$, $t$
along the geodesic $\gamma$. Solving it along  all such geodesics with initial conditions
\[
\omega(i) = 1, \quad \dot{\gamma}^aD_a\omega(i)  = \dot{\gamma}^a\,c_a, \quad
t(i) = \frac{1}{2}\,c_a\,c^a - \frac{1}{12}\,R[h],
\]
gives smooth fields $\omega$, $D_a\omega$, $t$ on $U$. It follows  that if there exists a positive solution $\omega$ of  (\ref{add-equ}) it is uniquely determined by $c_a$ and it is 
smooth resp. real analytic if $h$ satisfies this requirement.
This does not answer the question, however, whether there does exist a solution to  (\ref{add-equ}). In fact, it is not clear whether the fields denoted here by $D_a\omega$ and $t$ can be obtained from the given function $\omega$ by taking suitable derivatives.
There remains a somewhat subtle problem.

\vspace{.3cm}

\noindent
{\it What determines the value of $c_a$ ?}

\vspace{.2cm}

Assume that $h$ is conformally flat near $i$. Since the question we wish to answer only depends on the conformal structure of $h$ we can assume $h$ to be flat near $i$. Let $x^a$ be $i$-centered
$h$-normal coordinates near $i$ so that 
$h_{ab} = - \delta_{ab}$. Then $\rho = (x, x) \equiv \delta_{ab}\,x^a\,x^b$ and it follows that $\Sigma_{ab} = 0$. The geodesics $\gamma$ are given by the curves 
$\tau \rightarrow x^a(\tau) = \tau\,x_*^a$ with $(x_*, x_*) = 1$ and  the equations above take the form
\[
\frac{d^2}{d\tau^2} \omega + t = 0, \quad 
\frac{d}{d\tau} D^a\omega- t\,k^a = 0, \quad
\frac{d}{d\tau} t = 0.
\]
The solution, given by
\[
t = - 2\,(c, c), \quad 
D^a\omega = 2\,c^a - 2\,(c, c)\,x^a,\quad
\omega = 1 - 2\,(x, c) + (c, c)\,(x, x), 
\]
defines also a solution to  (\ref{add-equ}).
The metric  $h'_{ab} = - \omega^{-2}\,\delta_{ab}$ so obtained is again flat, it holds $h'_{ab} = (f_* h)_{ab}$ where the diffeomorphism $f$, which maps a neighbourhood of $i$ onto another such neighbourhood, is given in our coordiates $x^a$ by
the {\it special conformal transformation}
\[
x^a \rightarrow f^a(x^c) = \frac{1}{\omega}\,\left(x^a - (x, x)\,c^a\right).
\]
In this case the choice of $c_a$ is completely free.

\vspace{.2cm}

In the case where $h$ is not conformally flat near $i$  the situation is more complicated. Assume that $h$ 
satisfies equations (\ref{conf-static-vacuum-equ}). We  could then try again to solve equation (\ref{add-equ}) for some initial datum $c^a \neq 0$. If there existed a solution $\omega$, both metrics, $h$ and $h' = \omega^{-2}\,h$, would satisfy equations  
(\ref{conf-static-vacuum-equ}) with $d\omega \neq 0$ in a neighbourhood of $i$.
It has been shown in \cite{friedrich:confstatic}, \cite{friedrich:exactstat}, however,  that this can 
be the case only for a very restricted class of solutions and that $c_a$ must be related in those cases in a special way to the solution $h$.  In general,  (\ref{add-equ}) is  solvable  only for a unique choice of $c_a$, if it is solvable at all.

\section{The central-harmonic gauge}
\label{h-co-gauge}

Some of the following  considerations will simplify  if the conformal scaling of $h$ is suitably normalized.
There are several  ways to restrict the conformal freedom of $h$ near $i$, but in general  it is not possible to remove it completely.  In general, specifying a conformal gauge leaves the freedom to specify the value of the conformal factor and its differential at $i$. This is related to the fact that dilations and special conformal transformations map certain neighbourhoods of the origin in Minkowski space conformally  onto each other.

The conformal scalings most suited to our purposes seem to be the ones in which the metric $h$ has the property that its coefficients $h_{ab}$  in $i$-centered normal  coordinates satisfy the relation
\begin{equation}
\label{i-central-harmonic-gauge}
 \det(h_{ab}) = - 1 \quad \mbox{near} \quad i.
\end{equation}
The coordinates so obtained have been occasionally referred to as conformal normal coordinates. In \cite{friedrich:schmidt}  the same terminology was used for  coordinates defined in a completely different way and related to a different type of  normalized 
conformal scaling and in section \ref{j-subconds} will be mentioned still another gauge which has also been referred to as a conformal normal gauge. To avoid any confusion we shall not use that name here. 
Following \cite{schouten:1954} (cf. also  
\cite{guenther:1986}) with some simplification,  we shall refer to metrics satisfying (\ref{i-central-harmonic-gauge}) as being given in a
{\it central harmonic (conformal resp. coordinate) gauge}, always assuming that the center referred to is given by the point $i$.

For real analytic Riemannian or pseudo-Riemannian spaces $(S, h')$ of dimension $n \ge 3$ the existence of conformal scalings leading to (\ref{i-central-harmonic-gauge}) with respect to a prescribed  center point  
has been discussed in  \cite{guenther:1986}. It was shown there that for prescribed values $\vartheta(i) > 0$
and $d\vartheta \in T_i^*S$ there exists a unique real analytic function $\vartheta$ near $i$ which assumes these values at $i$ and for which the metric $h = \vartheta^2\,h'$ satisfies in $i$-centered harmonic coordinates the condition (\ref{i-central-harmonic-gauge}).
The existence of such scalings in Riemannian spaces has been discussed in  \cite{cao} in the $C^{\infty}$  and in \cite{guenther:1993}  for suitable values of $k$ and $\alpha$ in the $C^{k, \alpha}$ category. 

As mentioned in the references above,
the central harmonic gauge can be characterized by several useful conditions, which we discuss in the $3$-dimensional Riemannian case.
Condition (\ref{i-central-harmonic-gauge}) is saying that the exponential map $\exp_i$ is volume preserving.
Let $U$ denote a convex normal neighbourhood of the $i$ and  $\Gamma(p)$ the square  of the $h$-distance of a point 
$p  \in U$ from $i$. It represents the unique smooth (real analytic) solution to the conditions
\begin{equation}
\label{i-eikonal}
D_a\Gamma\,D^a\Gamma + 4\,\Gamma = 0
\quad \mbox{on} \quad U, \quad \quad 
\Gamma = \,0, \quad 
D_a\Gamma = 0, \quad 
D_aD_b\Gamma = - 2\,h_{ab}
\quad \mbox{at} \quad i.
\end{equation}
The equation on the left hand side will be referred to as the {\it eikonal equation} (though this is usually used for  the equation which is obtained by rewriting the equation above  in terms of the unknown $\sqrt{\Gamma}$).
In $i$-centered normal coordinates $x^a$, characterized by the condition
$h_{ab}(x)\,x^b = h_{ab}(0)\,x^b = - \delta_{ab}\,x^b$, 
it holds $\Gamma = |x|^2 \equiv \delta_{ab}\,x^a\,x^b$ whence 
$D_a\Gamma = - 2\,h_{ab}\,x^b$ and thus 
\begin{equation}
\label{gen-DeltaGamma + 2n}
D_aD_b\Gamma = - 2\,h_{ab} - h_{ab, \,c}\,x^c, \quad \quad  
\Delta_h \Gamma = - 2\,n - \frac{( \det(h_{ab}))_{,c}\,x^c}{ \det(h_{ab})}.
\end{equation}
This shows that  (\ref{i-central-harmonic-gauge}) is  in our case equivalent  to 
\begin{equation}
\label{A-central-harmonic}
\Delta_h \Gamma = - 6,
\end{equation}
or to
\begin{equation}
\label{B-central-harmonic}
\Delta_h\left(\frac{1}{\sqrt{\Gamma}}\right) = 4\,\pi\,\delta_i.
\end{equation}

On the geodesics sphere  $S_{\epsilon} = \{\sqrt{\Gamma} = \epsilon > 0\}$ with 
small radius $\epsilon$  consider a frame $e_k$, $k = 1, 2, 3$,
with $e^a_1 = (4\,\Gamma)^{- 1/2}\,D^a\Gamma$  the geodesic radial unit vector and 
vectors $e_A$, $A = 2, 3$, tangent to  $S_{\epsilon} $ such that 
$h(e_j, e_k) = - \delta_{jk}$. The second fundamental form on  $S_{\epsilon} $, given in the frame $e_A$ by $\chi_{AB} = h(D_{e_A} e_1, e_B) 
= (4\,\Gamma)^{- 1/2}\,e^a_A\,e^b_B\,D_aD_b\Gamma$,
has then  trace
\[
h^{AB}\,\chi_{AB} =  (4\,\Gamma)^{- 1/2}\,(
 \Delta_h\Gamma + e^a_1\,e^b_1\,D_aD_b\Gamma)
=  (4\,\Gamma)^{- 1/2}\,(\Delta_h\Gamma + 2),
\]
which implies that 
$\Delta_h\Gamma = - 2\,n$ is equivalent to
$h^{AB}\,\chi_{AB}  = - \frac{2}{\sqrt{\Gamma}}$. 

\vspace{.2cm}

A characterization of the central harmonic gauge in terms of fields of higher order which will be used later on  is obtained as follows. Applying two covariant derivatives to  the equation on the left hand side  of (\ref{i-eikonal}), commuting covariant derivatives, and performing a contraction gives with $n = 3$
\[
0 = R_{ab} \,D^a\Gamma\,D^b\Gamma+ D_aD_b\Gamma\,D^aD^b\Gamma - 4\,n
+ D^a\Gamma\,D_a(\Delta_h\Gamma + 2\,n) + 2\,(\Delta_h\Gamma + 2 n).
\]
Because we can write  in $i$ centered normal coordinates $x^a$  
\begin{equation}
\label{gradGamma.D}
D^a\Gamma(\gamma(\tau)) = - 2\,x^a(\gamma(\tau)) = - 2\,\tau\,x^a_*, \quad 
\quad
 x^a_*= const. \neq 0, \,\, \tau \,\,\mbox{an affine parameter},
\end{equation}
along any geodesic $\gamma(\tau)$ passing through $i$,   it follows  for smooth functions $f$ that 
\begin{equation}
\label{BgradGamma.D}
D^a \Gamma\,D_a f = - 2\, \tau \, \frac{d}{d\tau} f
\,\,\,\mbox{along the geodesic $\gamma(\tau)$}, 
\end{equation}
and thus
\[
 D^a\Gamma\,D_a(\Delta_h\Gamma + 2\,n) + 2\,(\Delta_h\Gamma + 2\,n)
 = - 2\,\tau^2\,\frac{d}{d\tau}\left(\frac{\Delta_h\Gamma + 2\,n}{\tau}\right).
 \]
Because
$\Delta_h\Gamma + 2\,n = O(|x|^2)$  in normal coordinates  
as $x^a \rightarrow 0$ by (\ref{gen-DeltaGamma + 2n}),  it follows 

\vspace{.2cm}

\hspace*{1.2cm}{\it $\Delta_h\Gamma = - 2\, n$ holds in normal coordinates near $i$  if and only
if  
\begin{equation}
\label{two-h=-1-consist}
R_{ab} \,D^a\Gamma\,D^b\Gamma = \,4\,n -  D_aD_b\Gamma\,D^aD^b\Gamma 
= - h^{ae}\,h^{bf}\,h_{ab, c}\,x^c\,h_{ef, d}\,x^d,
\end{equation}
\hspace*{1.8cm}in normal coordinates near $i$}. 

\vspace{.2cm}

\noindent
Here equations  
(\ref {i-central-harmonic-gauge}), (\ref{i-eikonal}),  (\ref{gen-DeltaGamma + 2n}) have been used to obtain the expression on right hand side.
The central harmonic gauge thus implies near $i$ the relations
\begin{equation}
\label{consequence:h-is--1}
s_{ab}\,D^a\Gamma\,D^b\Gamma - \frac{4}{3}\,\Gamma\,R
= 4\,n - D_aD_b\Gamma\,D^aD^b\Gamma
= - h^{ae}\,h^{bf}\,h_{ab, c}\,x^c\,h_{ef, d}\,x^d,
\end{equation}
\begin{equation}
\label{Dconsequence:h-is--1}
D_cs_{ab}\,D^a\Gamma\,D^b\Gamma 
+ 2\,s_{ab}\,D^a\Gamma\,D_cD^b\Gamma 
- \frac{4}{3}\,D_c\Gamma\,R
- \frac{4}{3}\,\Gamma\,D_cR
\end{equation}
\[
= - 2\,D_cD_aD_b\Gamma\,D^aD^b\Gamma,
\]
\begin{equation}
\label{DDconsequence:h-is--1}
D_dD_cs_{ab}\,D^a\Gamma\,D^b\Gamma 
+ 2\,D_cs_{ab}\,D_dD^a\Gamma\,D^b\Gamma 
+ 2\,D_ds_{ab}\,D_cD^a\Gamma\,D^b\Gamma 
\end{equation}
\[
+ 2\,s_{ab}\,D_dD_cD^a\Gamma\,D^b\Gamma 
+ 2\,s_{ab}\,D_cD^a\Gamma\,D_dD^b\Gamma 
\]
\[
- \frac{4}{3}\,(D_dD_cR\,\Gamma + D_cR\,D_d\Gamma 
+ D_dR\,D_c\Gamma + R\,D_dD_c\Gamma) 
\]
\[
= - 2\,D_dD_cD_aD_b\Gamma\,D^aD^b\Gamma
- 2\,D_cD_aD_b\Gamma\,D_dD^aD^b\Gamma.
\]

\vspace{.2cm}

\noindent
Relation (\ref{consequence:h-is--1}) implies in particular  that in a central harmonic gauge
$R_{ab} \,x^a\,x^b = O(|x|^4)$ as $|x| \rightarrow 0$
, and thus
\begin{equation}
\label{h-1-R-DR-at-i}
R(i) = 0, \quad s_{ab}(i) = 0, \quad 
D_{a}R(i) = 0, \quad 
D_{(a}s_{bc)}(i) = 0,
\end{equation}
and thus  
\begin{equation}
\label{h-1-R-DR-at-i-and-Cott(i) = 0}
D_aR_{bcde}(i) = 0 \quad \mbox{if} \quad B_{ab}(i) = 0.
\end{equation}

\section{The function $\rho$}
\label{f-hadamard}

In this section we describe a construction of the function $\rho$ which allows us 
to relate it to the properties of the local geometry near $i$ and
to analyse the regularity of $f_{ab}$ at all orders. We assume here $h$ to be real analytic. As pointed out in  \cite{dain:friedrich}, the construction discussed here gives also rise to the construction of a corresponding function $\rho$ in the $C^{\infty}$ case.

\subsection{The Hadamard construction}

Suppose $(S, h)$ is real analytic near the point $i$. We use Hadamard's construction to obtain a parametrix with pole at $i$ for the operator 
\[
L \equiv L_h 
=  \Delta_h - \frac{1}{8}\,R[h] \quad \mbox{with} \quad \Delta_h = D_a\,D^a.
\]
To begin with we keep the scaling of $h$ general. We shall  specialize to the central-harmonic gauge  later. 

Let $x^a$ denote $h$-normal coordinates centered at $i$ and $\Gamma$ the function considered in (\ref{i-eikonal}). Our discussion will be restricted to a sufficiently small convex normal neighbourhood $B(i)$ of $i$ on which $h$ and  the function $\Gamma$ are real analytic. 
A real analytic  function $U(x^{a})$ which satisfies $U(i) = 1$
and $L[U\,\Gamma^{-\frac{1}{2}}] = 0$ on $B(i) \setminus \{i\}$ is obtained  on $B(i)$ as follows. Inserting the ansatz 
\begin{equation}
\label{Uexp}
U = \sum_{p = 0}^{\infty} U_p\,\Gamma^p,
\end{equation} 
with coefficient functions $U_p$   into the equation 
\begin{equation}
\label{fseq}
0 = - \Gamma^{\frac{3}{2}}\,L[U\,\Gamma^{-\frac{1}{2}}]
= D^{a}\,\Gamma\,D_{a}\,U 
+ \frac{1}{2}\,(\Delta_h\,\Gamma + 6)\,U - \Gamma\,L[U],
\end{equation}
writing the resulting expression again as a series in terms of powers of $\Gamma$, and assuming that the coefficients functions vanish seperately  yields the equations 
\begin{equation}
\label{0pqrprop}
D^{a}\Gamma\,D_{a}U_0 = 
- \frac{1}{2}\,(\Delta_h\,\Gamma + 6)\,U_0,
\,\,\,\,\,U(i) = 1,
\end{equation}
\begin{equation}
\label{ppqrprop}
D^{a}\Gamma\,D_{a}U_{p + 1} = 
- \frac{1}{2}\,(\Delta_h\,\Gamma + 2 - 4p)\,U_{p + 1}
- \frac{1}{2p + 1}\,L[U_{p}], \,\,\,\,p = 0, 1,2,\dots \,\,.
\end{equation}

The  solutions  $U_p$ are obtained as follows.
Writing  (\ref{gradGamma.D}) along a given geodesic equation (\ref{0pqrprop})
reduces to an ODE which can be immediately integrated to obtain the solution 
\[
U_0(\tau\,x_*^a) = \exp \left\{ \frac{1}{4}\,\int_0^{\tau}
(\Delta_h\,\Gamma + 6)(s\,x_*^a)\, \frac{ds}{s} \right\},
\]
which can also be expressed in the form
\begin{equation}
\label{P-def}
U_0(x^a) = 
\left\{\frac{1}{4}\,\int_0^1
(\Delta_h\,\Gamma + 6)(s\,x^a)\, \frac{ds}{s}\right\}.
\end{equation}
The remaining equations, rewritten with (\ref{0pqrprop}) in the form
\begin{equation}
\label{Cppqrprop}
D^{a}\Gamma\,D_{a}\left(\frac{U_{p + 1}}{U_0}\right) = 2\,(p + 1)\,\left(\frac{U_{p + 1}}{U_0}\right) 
- \frac{L[U_{p}]}{(2p + 1)\,U_0}, \,\,\,\,p = 0, 1,2,\dots.
\end{equation}
have unique smooth solutions near $\tau = 0$ which are obtained recursively  in the form
\[
U_{p + 1}(\tau\,x_*^a) = \frac{U_0(\tau\,x_*^a)}{(4p + 2)\,\tau^{p + 1}}\,
\int_0^{\tau}\frac{L[U_p]}{U_0}(s\,x_*^a)\,s^p\,ds, \,\,\,\,\,\,p = 0, 1, \dots\,\,.
\]  
They can be written  more concisely
\[
U_{p + 1}(x^a) = \frac{U_0(x^a)}{4p + 2}\,
\int_0^1\frac{L[U_p]}{U_0}(s\,x^a)\,s^p\,ds, \,\,\,\,\,\,p = 0, 1, \dots \,\,.
\]  
The functions $U_p$ are real analytic in a neighbourhood of $i$  if the metric $h_{ab}$  is real analytic there. It has been shown in \cite{garabedian}
that the series defined by (\ref{Uexp}) is absolutely convergent and thus defines a real analytic solution near $i$.

\subsection{An expression for $\rho$}
\label{rho-Drho-DDrho-DDDrho-expr-in-ch-gauge}

The uniqueness of $\rho$, comparison of (\ref{sigmaval}) and (\ref{i-eikonal}) give with  (\ref{Uexp}) 
near $i$ 
\begin{equation}
\label{rho-expansion}
\rho = \mu\,\frac{\Gamma}{U^2} 
= \mu\,\frac{\Gamma}{U_0^2}\,
\left(\sum_{k = 0}^{\infty} (k+1) \left( - \sum_{p = 1}^{\infty} \frac{U_p}{U_0}\,\Gamma^p\right)^k\right)
= \sum_{p = 1}^{\infty}\,V_p\,\Gamma^p,
\end{equation}
where
\[
V_1 = \mu\,\frac{1}{U_0^2} 
\quad 
V_2 = - 2\,\mu\, \frac{U_1}{U^3_0}\
\quad 
V_3 = \mu\, \frac{3\,U^2_1- 2\,U_0\,U_2}{U_0^4}, \quad \dots
\]

\noindent
The following relations will be used later on.
From the expansion above we get with certain analytic functions $F_a$, $F_{ab}$, $F_{abc}$
\begin{equation}
\label{Drho}
D_a\rho = V_1\,D_a\,\Gamma + \Gamma\,F_a,
\end{equation}
\begin{equation}
\label{DDrho}
D_bD_a\,\rho = 
D_bV_1\,D_a\,\Gamma
+ D_aV_1\,D_b\,\Gamma + V_1\,D_b\,D_a\,\Gamma
+  2\,V_2\,D_b\,\Gamma\,D_a\,\Gamma  + \Gamma\,F_{ba},
\end{equation}
\begin{equation}
\label{DDDrho}
D_cD_bD_a\,\rho = V_1\,D_cD_b\,D_a\,\Gamma
+ 3\,D_{(c}D_bV_1\,D_{a)}\,\Gamma 
+ 3\,D_{(c}V_1\,D_bD_{a)}\,\Gamma \quad \quad  \quad \quad 
\end{equation}
\[
+  6\,D_{(c} V_2\,D_b\,\Gamma\,D_{a)}\,\Gamma
+  6\,V_2\,D_{(c}\Gamma\,D_b\,D_{a)}\,\Gamma
+ 6\,V_3\,D_c\,\Gamma\,D_b\,\Gamma\,D_a\,\Gamma + \Gamma\,F_{cba}.
\]

Assuming now a central harmonic gauge, various expressions simplify. We get 
$U_0 = 1$ and thus 
\begin{equation}
\label{V1-V2-centr-harm}
V_1 = \mu, \quad \quad
V_2 = - 2\,\mu\,Q \quad \mbox{with} \quad Q(x^a)  = U_1(x^a) = - \frac{1}{16}\,\int_0^1 R(s\,x^a)\,ds,
\end{equation}
\begin{equation}
\label{V3-centr-harm}
V_3 = \mu\,(3\,Q^2 - 2\,P) \quad 
 \mbox{with} \quad P(x^a)  = U_2(x^a) = \frac{1}{6}\,\int_0^1 s\,(\Delta_hQ - \frac{1}{8}\,R\,Q)(s\,x^a)\,ds.
\end{equation}
In some calculations it is useful to write with the notation of (\ref{gradGamma.D}) 
\begin{equation}
\label{Q-P-geod-expr}
Q(\tau\,x_*^a)  = - \frac{1}{16\,\tau}\,\int_0^{\tau} R(s\,x_*^a)\,ds, \quad 
P(\tau\,x_*^a)  = \frac{1}{6\,\tau^2}\,\int_0^{\tau} s\,(\Delta_hQ - \frac{1}{8}\,R\,Q)(s\,x_*^a)\,ds.
\end{equation}
Moreover, equations (\ref{ppqrprop}) imply in particular
\begin{equation}
\label{Q-equ}
D^{a}\Gamma\,D_{a}Q = 2\,Q + \frac{1}{8}\,R, \quad \quad
D^{a}\Gamma\,D_{a}P = 4\,P
- \frac{1}{3}\,(\Delta_hQ - \frac{1}{8}\,Q).
\end{equation}

\section{The holomorphic extension}
\label{g-holomorphic}

Assume $h$ to be real analytic and all fields to be given in $i$-centered normal coordinates $x^a$
on some open convex normal neighbourhood $O$ of $i$. If we consider 
$\mathbb{R}^3$ and thus $O$ as being embedded  in the usual  way in
$\mathbb{C}^3$, all real analytic fields considered so far extend in a unique way as holomorphic functions to some connected open neighbourhood  ${\cal O}$ of $O$ in $\mathbb{C}^3$ where the extension of $h_{ab}$ is non-degenerate. 
The differential geometric relations satisfied on $O$ will be preserved on ${\cal O}$ and we can assume the extended functions $x^a$ to define normal coordinates in the sense that $x^a\,h_{ab}(x^c) = x^a\,h_{ab}(0)$ on ${\cal O}$.

\vspace{.2cm}

We are only interested in properties which hold on some unspecified  connected open neighbourhood of $i$. In various of the following statements it is understood that 
${\cal O}$ or other neighbourhoods of the point $x^a = 0$ are chosen sufficiently small and `close to $i$' should be read as a reminder of this.  Though it  would lead to logically more satisfactory statements, we shall not introduce the language   of  {\it germs of analytic functions} (cf. \cite{gunning:1990}), 
hoping our simple applications of complex analysis to be obvious enough.

\vspace{.2cm}

The extension into the complex domain will allow us to analyse certain relations by differential geometric techniques which otherwise would have  to be discussed in terms  of formal expansions.
The symmetric tensor  $h_{ab}$ defines at  each point of ${\cal O}$ a cone of null vectors. Of particular interest to us will be  the subset  ${\cal N}_i$ of  ${\cal O}$ which is generated by the complex null geodesics passing through $i$. In terms of the normal coordinates $x^a$ or the extended function $\Gamma$, which satisfies (\ref{i-eikonal}) on ${\cal O}$, we have
\begin{equation}
\label{calN-i}
{\cal N}_i = \{x^a \in  {\cal O}|\,\delta_{ab}\,x^a\,x^b = 0\} = \{x^a \in  {\cal O}|\,\Gamma(x^a) = 0\}.
\end{equation}

Because null geodesics considered as point sets are conformal invariants, the set ${\cal N}_i$ is a conformal invariant as well. The set ${\cal N}_i \setminus \{i\}$ represents an analytic null hypersurface
and ${\cal N}_i$ is a holomorphic subvariety  of  ${\cal O}$ with singular point $i$   (\cite{gunning:1990}).

\vspace{.3cm}

Equation (\ref{add-equ}) can only admit smooth solutions if the tensor field $ f_{ab} = \rho^{-2}\,\Sigma_{ab}$ extends smoothly to $i$. This requirement implies at all orders restrictions on the coefficients of the Taylor expansion at $i$ of $\Sigma_{ab}$. The details of this fact are most easily discussed if $h$ is assumed to be real analytic. The field $f_{ab}$ then extends as a real analytic field to $i$ if and only if it admits a holomorphic extension to ${\cal O}$.  Because $\rho = \mu\,\Gamma\,U^{-2}$, where $\mu\, U(i) \neq 0$, this then  implies that  the fields $\Sigma_{ab}$ and  $D_c\Sigma_{ab}$ are holomorphic and  vanish on ${\cal N}_i$. The infinite sequence of conditions on the Taylor coefficients of these fields at $i$ follow because $i$ is a singular point of ${\cal N}_i$.

\begin{lemma}
\label{ten-on-N-vanish}
A holomorphic tensor field $T$ on some neighbourhood ${\cal O}$ of $i$
vanishes on ${\cal N}_i$ if and only if its covariant derivatives at $i$ satisfy in space spinor notation  the sequence of conditions
\begin{equation}
\label{spin-T-vanishes-on-N}
\,D_{(C_pD_p} \ldots D_{ C_1D_1)}T(i) = 0
\quad p = 0, 1, 3, \ldots \,, 
\end{equation}
where the brackets denote symmetrization. The equivalent conditions  in tensor notation read
\begin{equation}
\label{tensor-T-vanishes-on-N}
D_{ \{ c_p} \ldots D_{ c_1\}}\,T(i) = 0,
\quad p = 0, 1, 3, \ldots \,, 
\end{equation}
where the curved brackets denote the symmetric trace free part with respect to the indices in brackets.

\end{lemma}

\noindent
{\bf Proof}: Since we consider tensor relations we can use coordinates and a frame field which are well adapted to the situation.
Let  $e_{\bf a}$, ${\bf a} = 1, 2, 3$, be
an $h$-orthonormal frame field on ${\cal O}$ near $i$ which is parallelly transported along
the $h$-geodesics through $i$ and let $x^a$ denote normal coordinates
centered at $i$ so that $e^b\,_{\bf a} \equiv \,<dx^b, e_{\bf a}>\, =
\delta^b\,_{\bf a}$ at $i$. 
At the point with coordinates $x^a$ the coefficients of
the frame then satisfy
\begin{equation}
\label{normalformchar}
e^{b}\,_{\bf a}\,x^a = \delta^{b}\,_{\bf a}\,x^a,
\quad \quad 
x_b\,e^{b}\,_{\bf a} = x_b\,\delta^b\,_{\bf a} \quad \mbox{where} \quad 
x_a = \delta_{ab}\, x^b, 
\end{equation}
where it is assumed, as will be done  in the following,
that the summation rule does not distinguish between bold face and
other indices. 
In the following all tensor fields, except the frame
field $e_{\bf a}$ are
expressed in terms of this frame field, so that the metric is given by 
$h_{\bf ab} \equiv h(e_{\bf a}, e_{\bf c})  = - \delta_{\bf ab}$.
With $D_{\bf a} \equiv D_{e_{\bf a}}$ the connection
coefficients with respect to $e_{\bf a}$ are defined by
$D_{\bf a}\,e_{\bf c} = \Gamma_{\bf a}\,^{\bf b}\,_{\bf c}\,e_{\bf b}$. 
Let $X$ denote the vector field near $i$ 
which has in normal coordinates the expansion $X(x) = x^b\,\delta^{\bf
a}\,_b\,e_{\bf a}$ so that
$X = - 1/2\,\,grad_h \,\Gamma$ and thus tangential to the geodesics passing through $i$.

Suppose $T$ is a tensor field of rank $(r,s)$ near $i$ which has
components  $T^{{\bf a}_1 \ldots {\bf a}_r}\,_{{\bf b}_1 \ldots {\bf b}_s}$ with respect to the frame
$e_{\bf c}$. Since $D_X\,e_{\bf c} = 0$, we find for $x^s$ sufficiently  small and $|\tau| \le 1$   
\[
\frac{d}{d \tau}(T^{{\bf a}_1 \ldots {\bf a}_r}\,_{{\bf b}_1 \ldots {\bf b}_s}(\tau\,x^{e}))
= x^f\,(\frac{\partial}{\partial x^f}T^{{\bf a}_1 \ldots {\bf a}_r}\,_{{\bf b}_1 \ldots {\bf b}_s})
(\tau\,x^{e}) = X^{\bf c}(x^{e})
(D_{\bf c}T^{{\bf a}_1 \ldots {\bf a}_r}\,_{{\bf b}_1 \ldots {\bf b}_s}) (\tau\,x^{e}). 
\]
Observing that such formulae also hold for the covariant differentials of $T$,
we get for $ p = 0,1,2, \dots$ by induction
\[
\frac{d^p}{d \tau ^p}T^{{\bf a}_1 \ldots {\bf a}_r}\,_{{\bf b}_1 \ldots {\bf b}_s}  
(\tau\,x^{e})
= X^{ {\bf c}_p}(x^{e}) \ldots X^{ {\bf c}_1}(x^{e})
D_{ {\bf c}_p} \ldots D_{ {\bf c}_1}(T^{{\bf a}_1 \ldots {\bf a}_r}\,_{{\bf b}_1 \ldots {\bf b}_s}
)(\tau\,x^{e}).
\]
This implies a Taylor expansion of the form
\begin{equation}
\label{T-taylor}
T^{{\bf a}_1 \ldots {\bf a}_r}\,_{{\bf b}_1 \ldots {\bf b}_s}(x^a) = 
\sum_{p = 0}^{\infty}
\frac{1}{p\,!}\,X^{ {\bf c}_p}(x^a) \ldots X^{ {\bf c}_1}(x^a)
\,D_{ {\bf c}_p} \ldots D_{ {\bf c}_1}T^{{\bf a}_1 \ldots {\bf a}_r}\,_{{\bf b}_1 \ldots {\bf b}_s}(i),
\end{equation}
which is absolutely convergent  for sufficiently small values of $x^a$.

Let $\gamma(\tau)$ a null geodesic on ${\cal N}_i$ with $\gamma(0) = i$. In the normal coordinates $x^a$  it has a representation $\tau \rightarrow \tau\,x^a_*$ with some $x^a_* \neq 0$ which satisfies $\delta_{ab}\,x^a_*\,x^b_* = 0$. 
In the space spinor formalism, in which the frame is written $e_{AB}$  with $e_{AB} = e_{(AB)}$,  the vector field $X = X^{AB}\,e_{AB}$ is null along $\gamma$ so that we can write $X^{AB}(\gamma(\tau)) = \tau\,\iota^A\,\iota^B$ with a spinor field $\iota^A$ which satisfies  $D_{\dot{\gamma}} \iota^A = 0$. With this notation the expansion 
(\ref{T-taylor}) implies along $\gamma$
\begin{equation}
\label{gamma-T-taylor}
T^{{\bf a}_1 \ldots {\bf a}_r}\,_{{\bf b}_1 \ldots {\bf b}_s}(\gamma(\tau)) = 
\sum_{p = 0}^{\infty}
\frac{1}{p\,!}\,\tau^p\,\iota^{C_p}\,\iota^{D_p} \ldots \iota^{C_1}\,\iota^{D_1}\,
\,D_{C_pD_p} \ldots D_{ C_1D_1}T^{{\bf a}_1 \ldots {\bf a}_r}\,_{{\bf b}_1 \ldots {\bf b}_s}(i).
\end{equation}
We can symmetrize here over the indices $C_p \ldots D_1$ and conclude that $T$ vanishes along $\gamma$ if and only if
\[
0 = \iota^{C_p}\,\iota^{D_p} \ldots \iota^{C_1}\,\iota^{D_1}\,
\,D_{(C_pD_p} \ldots D_{ C_1D_1)}T^{{\bf a}_1 \ldots {\bf a}_r}\,_{{\bf b}_1 \ldots {\bf b}_s}(i)
\quad p = 0, 1, 3, \ldots 
\]
Since  $\gamma$ was arbitrary these equations must hold for arbitrary $\iota^A$, which implies  
for all $p \ge 0$ that
$D_{(C_pD_p} \ldots D_{ C_1D_1)}T^{{\bf a}_1 \ldots {\bf a}_r}\,_{{\bf b}_1 \ldots {\bf b}_s}(i) = 0$. Relation (\ref{tensor-T-vanishes-on-N}) is just the translation of this equation into tensor notation.\\
$\Box$

\vspace{.1cm}

We note that the singular nature of the point $i$ with respect to ${\cal N}_i$ comes into play only in the last step of the argument, where it is used that all null directions at $i$ are tangent to ${\cal N}_i$.
In the analytic case a similar  argument implies: 

\vspace{.1cm}

\noindent
 {\it Condition (\ref{regcond}) with $p_* = \infty$  is equivalent to the condition that
 \begin{equation}
 \label{B(Drho,Drho)-on-N-vanishes}
 B_{ab}\,D^a\rho\,D^b\rho = 0 \quad \mbox{on} \quad {\cal N}_i
 \quad \mbox{near} \quad i, 
 \end{equation}
which follows from  (\ref{Drho-Drho-contr-stat-vac-dual-Xi-def})}.

\vspace{.1cm}

In fact, the gradient $D^{AB}\rho$ is by (\ref{rho-expansion}) proportional to 
$D^{AB}\Gamma$ on ${\cal N}_i$ whence tangential to the null directions on 
${\cal N}_i$ and thus also proportional to $\iota^A\,\iota^B$ along  $\gamma$.
On $\gamma$ the relation  (\ref{Drho-Drho-contr-stat-vac-dual-Xi-def}) is thus equivalent to
$B_{ABCD}\,\iota^A\,\iota^B\,\iota^C\,\iota^D = 0$ . The conclusion then follows 
with the type of argument above.

Observing that $D^a\rho$ is tangential to the null directions of ${\cal N}_i$, which are conformal invariants, and that $B_{ab}$ is a conformal density, condition 
(\ref{B(Drho,Drho)-on-N-vanishes}) and thus condition (\ref{regcond}) with $p_* = \infty$  is seen  to be conformally invariant.

\vspace{.1cm}

To control the smooth extensibility of $f_{ab}$ we would like to make use of 
 condition (\ref{regcond}) with $p_* = \infty$. Sorting out in terms of Taylor coefficients whether this condition implies relations like (\ref{tensor-T-vanishes-on-N}) with $T$ corresponding to the fields $\Sigma_{ab}$ and $D_c\Sigma_{ab}$ would require some  awkward algebra. 
 The following result will allow us to discuss the question in a more geometric way.

\begin{lemma}
\label{reg-ext}
Let $g$ be a real analytic function on $O$. Then the real analytic function $f = \frac{g}{\Gamma}$  on $O \setminus \{i\}$ extends as a real analytic function to $i$ if and only if the holomorphic extension of $g$  to ${\cal O}$ vanishes  on ${\cal N}_i$.
\end{lemma}

\noindent
{\bf Proof}: If $f$ admits the desired extension the relation 
$g = \Gamma\,f$  then satisfied by  the three holomorphic  functions to ${\cal O}$
implies that $g = 0$ on ${\cal N}_i$.
 
Assume that $g \neq 0$ on ${\cal O}$ and $g = 0$ on ${\cal N}_i$. Denoting the normal coordinates $x^a$ by $x$, $y$, $z$  we have $\Gamma = x^2 + y^2 + z^2$. By the Weierstrass Division Theorem 
 (\cite{gunning:1990}) there exist then  holomorphic functions $k = k(x, y, z)$, $a = a(x, y)$, and $b = b(x, y)$ such that 
 \begin{equation}
 \label{WDT-decomp}
 g = k\,\Gamma + a\,z + b \quad \mbox{near} \quad i.
 \end{equation}
Our assumption implies that  $a(x, y)\,z + b(x, y) = 0$ if $x^2 + y^2 + z^2 = 0$ and thus
\begin{equation}
\label{non-equ}
b^2 = q\,a^2 \,\,\, \mbox{with} \,\,\, q = - (x^2 + y^2), 
\end{equation}
for $(x, y)$ close to $(0, 0)$.

Consider the ring $H_0$ of functions which are defined and holomorphic on some connected open neighbourhood of the origin in $\mathbb{C}^2$. We consider these functions as representing their Taylor  series at the origin  which converge on some neighbourhood of the origin.  A function $m \in  H_0$ is called a {\it unit}  if it has a multiplicative inverse  $m^{-1}$ close to the origin, that is $m(0, 0) \neq 0$. A non-vanishing function $m \in H_0$ is called a {\it nonunit} if $m(0, 0) = 0$. A nonunit  $m \in H_0$ is called {\it irreducible} over $H_0$ if it cannot be written as a product 
$m = m_1\,m_2$ of two nonunits $m_1$, $m_2 $ in $H_0$.

To show that (\ref{non-equ}) leads to contradictions unless
$a$ and $b$ vanish near $(0, 0)$ we use the fact 
 that {\it every nonunit in $H_0$ can be written near the origin as a product of a finite number of  irreducible factors which is unique up to reorderings and insertions of  factors $\epsilon\cdot \epsilon^{-1}$  with units $\epsilon$}  (\cite{gunning:1990}). An example of this situation is given by the relation
\[
q = q_+\,\,q_- = \epsilon\,\,q_+\,\epsilon^{-1}\,\,q_- \,\,\,\,\mbox{with} \,\,\,\,
q_{\pm}  =  i\,x \pm y \,\,\,\,\mbox{and} \,\,\,\,\epsilon\,\,\,\,\mbox{a unit in}\,\,H_0.
\]
Most important for us is the observation  that $\epsilon$ can not be chosen here such that $\epsilon\,\,q_+ = \epsilon^{-1}\,\,q_- $,  in other words, $q$ can not be written in the form $q = c^2$ with some $c \in H_0$.
It follows then immediately from (\ref{non-equ}) that $a$ cannot be a unit. 
Thus, if $a \neq 0$, $a$ and $b$ admit factorizations in terms of irreducible factors so that  (\ref{non-equ}) takes the form
\[
b^2_1\cdot b^2_2 \cdot \,\ldots \,\cdot b^2_k 
= q\,\,a^2_1\cdot \,\ldots\,\cdot \,a_i^2.
\]
But the uniqueness of the factorization implies that 
this is in conflict with the observation that $q$ cannot be written as a square.
It follows  that $a =0$, $b = 0$ in equation  (\ref{WDT-decomp}) which  implies that $g = 0$ on ${\cal N}_i$.\\
$\Box$

\vspace{.3cm}

\subsection{Some relations on ${\cal N}_i$}
\label{relations-on-Ni}

For the analysis of fields near  ${\cal N}_i$ it will be convenient to consider a neighbourhood $W$ of a given null geodesic $\gamma(\tau)$ in ${\cal N}_i$, 
$\gamma(0) = i$, and assume $W$ to be ruled by null geodesics. Let 
\begin{equation}
\label{A-frame-norm}
n^a, \,\,p^a, \,\,m^a \quad \mbox{with} \quad n^a = D^a\Gamma,
\end{equation}
be a smooth frame on $W \setminus\{i\}$ satisfying
\begin{equation}
\label{B-frame-norm}
n_a\,n^a = p_a\,p^a = n_a\,m^a =p_a\,m^a = 0, \quad \quad
n_a\,p^a = 1,  \quad \quad m_a\,m^a = - 1/2.
\end{equation}
Assuming that $m^a$ is parallelly propagated long the null geodesics, it follows 
\begin{equation}
\label{A-frame-on-N-prop}
n^aD_an^b = -2\,n^b, 
\quad \quad
n^aD_am^b = 0, 
\quad \quad
n^aD_ap^b = 2\,p^b
\quad \quad \mbox{on}\,\,\,\, W \setminus\{i\}.
\quad \quad
\end{equation}
In the notation of (\ref{gradGamma.D}), (\ref{BgradGamma.D})
along the null geodesic $\gamma$, where $\delta_{ab}\,x^a_*\,x^b_* =0$,
it follows 
\begin{equation}
\label{n-p--near-i}
n^a(\gamma(\tau)) = - 2\,\tau\,x^a_*, \quad \quad
p^a(\gamma(\tau)) = O(|\tau|^{-1}) \quad \mbox{as} \quad 
\tau \rightarrow 0.
\end{equation}
The frame is fixed by these conditions  up to transformations of the form
\begin{equation}
\label{frame-transf}
m^a \rightarrow m'^a = \pm m^a + c\,n^a, \quad \quad 
p^a \rightarrow p'^a = p^a \pm 2\,c\,m^a + c^2\,n^a,
\end{equation}
where $c$ is a function on $W \setminus\{i\}$ which satisfies 
there $D^a\Gamma\,D_ac = 2\,c$.

The vectors $n^a$, $m^a$ span the tangent space of ${\cal N}_i \setminus\{i\}$ at  points of $W \setminus\{i\}$ and it holds 
\begin{equation}
\label{h-on-W-minus-i}
h_{ab} = n_{a}\,p_{b} + n_{b}\,p_{a} - 2\, m_a\, m_b.
\end{equation}
Furthermore, it follows, possibly after replacing $m^a$ by $- m^a$, 
\begin{equation}
\label{epsilon-on-N}
\epsilon_{abc} = \alpha\,n_{[a}\,m_b\,p_{c]} \quad \mbox{with} \quad 
\alpha = i\,6\,\sqrt{2},
\end{equation}
and consequently
\begin{equation}
\label{epsilon-on-N-contr}
n^a\,\epsilon_{abc} = \frac{\alpha}{3}\,\,n_{[b}\,m_{c]}, \quad
m^a\,\epsilon_{abc} = \frac{\alpha}{6}\,\,n_{[b}\,p_{c]}, \quad
p^a\,\epsilon_{abc} = \frac{\alpha}{3}\,\,m_{[b}\,p_{c]}.
\end{equation}
In the following we assume (\ref{epsilon-on-N}), which removes the freedom 
to choose the sign in (\ref{frame-transf}).

\vspace{.5cm}

Observing that 
$D^a\rho = \mu\,U_0^{-2}\,D^a\Gamma$ on ${\cal N}_i$ and 
$ \mu\,U_0^{-2} \neq 0$ near $i$ by  (\ref{rho-expansion})
we see that  (\ref{Drho-Drho-contr-stat-vac-dual-Xi-def}) implies 
\begin{equation}
\label{B(n,n) = 0} 
D^a\Gamma\,D^b\Gamma\,B_{ab} = 0 \quad \mbox{on} \quad {\cal N}_i. 
\end{equation}
The following consequence of this relation  will be important for us.

\begin{lemma}
\label{BGG=0-cons}
If $D^a\Gamma\,D^b\Gamma\,B_{ab} = 0$ and  
$D^a\Gamma\,D^b\Gamma\,s_{ab} = 0$ hold on ${\cal N}_i$ then     
$D^a\Gamma\,m^{d}\,s_{ad} = 0$ holds  on ${\cal N}_i$.
\end{lemma}

\noindent
{\bf Proof}:  We have 
\[
n^e\,D_e\,n^a = D^e\Gamma\,D_eD^a\Gamma = 1/2\,\,D^a(D_e\Gamma\,D^e\Gamma) = - 2\,D^a\Gamma
 = - 2\,n^a,
\]
and thus with suitable coefficient functions 
\[
m^e\,D_e\,n^a = \gamma\,n^a + \delta\,m^a, \quad 
n^e\,D_e\,m^a = \rho\,n^a + \pi\,m^a, 
\] 
because the members  on the left hand sides contract to zero with $n^a$. It holds 
\[
\gamma\,n^a + \delta\,m^a = 
m^b\,D_bD^a\Gamma \rightarrow - 2\,m^a 
\,\,\, \mbox{whence} \,\,\, 
\gamma \rightarrow 0, \,\,\, 
\delta \rightarrow - 2 \,\,\, \mbox{as} \,\,\, x^a \rightarrow 0.
\]
For the Cotton tensor we have
\begin{equation}
\label{Cott-form}
B_{de} =  
 \frac{1}{2}\,D_{a}\,s_{bd} \,\epsilon_{e}\,^{ab} + \frac{1}{24}\,D_aR\,\epsilon_{de}\,^a
=  \frac{1}{2}\,D_{a}\,s_{b(d} \,\epsilon_{e)}\,^{ab},
\end{equation}
where  the Bianchi identity has been used in the second step.
With (\ref{epsilon-on-N-contr}) it   follows on ${\cal N}_i$ 
\begin{equation} 
\label{B(n,n)-on-N-expr}
n^a\,n^b\,B_{ab} =  \frac{\alpha}{12}\,(n^a\,m^b - m^a\,n^b)\,n^d\,D_{a}\,s_{bd} 
\end{equation}
\[
=  \frac{\alpha}{12}\,
\left\{ n^{c}\,D_{c}\,(s_{da}\, m^{d}\,n^a)  - m^{c}\,D_{c}\,(s_{da}\, n^{d}\,n^a)
- \left(n^{c}\,D_{c}(m^{d}\,n^a) - m^{c}\,D_{c}\,(n^{d}\,n^a)\right)s_{da} \right\}
\]
\[
= \frac{\alpha}{12}\,
\left\{ n^{c}\,D_{c}\,(s_{da}\, m^{d}\,n^a) - m^{c}\,D_{c}\,(s_{da}\, n^{d}\,n^a)
- (\rho - 2\,\gamma)\,n^a\,n^b\,s_{ab} - (\pi - 2 - 2\,\delta)\,n^a\,m^a\,s_{ab}\right\},
\]
which implies with our assumptions the ODE
\[
0 = 2\,\tau\,\frac{d}{d\tau}(s_{ab}\, m^{a}\,n^b) + (\pi - 2 - 2\,\delta)\,(s_{ab}\,m^a\,n^b),
\]
for $s_{ab}\, m^a\,n^b$ along the null geodesic $\gamma$ on ${\cal N}_i$.
Because  $\pi  - 2 - 2\,\delta  \rightarrow 2 >  0$ as $\tau \rightarrow 0$ and 
because $\gamma$ is arbitrary, the result follows. \\
$\Box$

\section{The smoothness of $f_{ab}$}
\label{analyticity of f}

In this section we show that  condition (\ref{regcond}) with $p_* = \infty$
ensures the regularity of our basic equation.

\begin{proposition}
\label{f-is-analytic-if-B(bb)-vanishes-on-N}
Suppose the metric $h$ is real analytic near $i$. The field $f_{ab} = \rho^{-2}\,\Sigma_{ab}$ extends then as an analytic tensor field to $i$ if and only if the Cotton tensor satisfies condition   (\ref{regcond}) with 
$p_* = \infty$ or, equivalently,
condition (\ref{B(Drho,Drho)-on-N-vanishes}).
\end{proposition}

\noindent
{\bf Proof}: Lemma \ref{ten-on-N-vanish}
shows that  the Taylor expansion coefficients of $\Sigma_{ab}$ at $i$ must satisfy conditions at  any order if  $ f_{ab}$ is to extend smoothly to $i$. Instead of analysing the expansion coefficients we shall study the holomorphic extension of $\Sigma_{ab}$ on  ${\cal N}_i$. 
 The requirement that $f_{ab} = \rho^{-2}\,\Sigma_{ab}$ extends as a real analytic  field to $i$ translates in view of the Lemma  (\ref{reg-ext}) into the condition that $\Sigma_{ab}$ extends to a holomorphic tensor field near $i$ which satisfies
\begin{equation}
 \label{Sigma-dSigma-on N}
\Sigma_{ab} = 0, \quad D_c \Sigma_{ab} = 0
\quad \mbox{on} \quad {\cal N}_i \quad \mbox{near} \quad i.
\end{equation}

\vspace{.1cm}

\noindent
In fact, $\Sigma_{ab} = \rho^{2} f_{ab}$ and (\ref{rho-expansion}) imply equations (\ref{Sigma-dSigma-on N}) near $i$ if $f_{ab}$ extends as a real analytic  function to $i$.  Conversely, (\ref{Sigma-dSigma-on N}) implies with  (\ref{rho-expansion})  and  Lemma  (\ref{reg-ext})  the existence of holomorphic functions  $l_{ab}$ and $l_{cab}$ on ${\cal O}$ near $i$ such that
$ \Sigma_{ab} = \rho\,l_{ab}$ and $D_c \Sigma_{ab} = \rho\,l_{cab}$. It follows that
$D_c \rho\,\,l_{ab} = \rho\,(l_{cab} - D_c l_{ab})$ and thus  $l_{ab} = 0$ on ${\cal N}_i$
because $D_c\rho \neq 0$ on ${\cal N}_i \setminus \{i\}$.
By Lemma   (\ref{reg-ext}) it follows  that 
$l_{ab} = \rho\,f_{ab}$ whence $\Sigma_{ab} = \rho^{2} f_{ab}$
with some holomorphic function $f_{ab}$ near $i$. 

With the formulation  (\ref{Sigma-dSigma-on N}) of the problem
the assertion of the Proposition follows now as
a consequence of Lemmas \ref{Sigma-vanishes-iff-B(bb)-vanishes-on-N} and 
\ref{DSigma-vanishes-if-B(bb)-vanishes-on-N}
proven below. \\
$\Box$

 \vspace{.5cm}

The behaviour of conditions (\ref{Sigma-dSigma-on N}) under conformal  rescalings is of interest here. We have seen that a rescaling of type (\ref{h-h'}) leads to a transformation of the form 
\[
\Sigma_{ab} \rightarrow \Sigma'_{ab} 
= \frac{1}{\omega}\,\Sigma_{ab} 
- \frac{\rho^2}{\omega^3}\,\hat{\Sigma}_{ab},
\]
and thus
\[
D_c\,\Sigma_{ab} \rightarrow D'_c\,\Sigma'_{ab} = \frac{1}{\omega^2}\,\left\{2\,\Sigma_{c(a}\,D_{b)}\omega + D_c \omega \,\Sigma_{ab}
- 2\,h_{c(a}\,\Sigma_{b)e}\,D^e\omega\right\} 
+ \frac{1}{\omega}\,D_c\Sigma_{ab} - \frac{2\,\rho}{\omega^3}\,D_c\rho\,\hat{\Sigma}_{ab}
\]
\[
- \frac{\rho^2}{\omega^4}\left\{\omega\,D_c\hat{\Sigma}_{ab}
+ 2\,\hat{\Sigma}_{c(a}\,D_{b)}\omega - D_c\omega\,\hat{\Sigma}_{ab} - 2\,h_{c(a}\,\hat{\Sigma}_{b)e}\,D^e\omega
\right\},
\]
which implies the transformations
\[
\Sigma_{ab}|_{{\cal N}_i} \rightarrow \Sigma'_{ab}|_{{\cal N}_i}  = \frac{1}{\omega}\,\Sigma_{ab}|_{{\cal N}_i}, 
\] 
\[
D_c\Sigma_{ab}|_{{\cal N}_i} \rightarrow
D'_c\,\Sigma'_{ab}|_{{\cal N}_i}  = 
\left(\frac{1}{\omega^2}\,\left\{2\,\Sigma_{c(a}\,D_{b)}\omega + D_c \omega \,\Sigma_{ab}
- 2\,h_{c(a}\,\Sigma_{b)e}\,D^e\omega\right\} 
+ \frac{1}{\omega}\,D_c\Sigma_{ab} \right)|_{{\cal N}_i} .
\]
While in the first case we find a transformation behaviour as satisfied by conformal densities, the 
behaviour is more difficult in the second case. Nevertheless, the pair of conditions 
(\ref{Sigma-dSigma-on N}) is invariant under conformal rescalings. This allows us to analyse these conditions in a convenient conformal gauge

Not all of the conditions (\ref{Sigma-dSigma-on N}) imply restrictions on the conformal structure of $h$.
In fact, it follows from (\ref{Bcott}) that the relation 
\begin{equation}
\label{0-onN-Bcott}
\left(D_{[c}\, \Sigma_{a]b} + 
\frac{1}{2}\,D^e\,\Sigma_{e[c}\,h_{a]b}\right)|_{{\cal N}_i} = 0,
\end{equation}
is satisfied identically for any metric $h$.  Observing that $\Sigma_{ab}$ is symmetric and trace free we get the decomposition
\begin{equation}
\label{DSigma-decomp}
D_a\Sigma_{bc} = D_{\{a}\Sigma_{bc\}} + \frac{2}{3}\,(D_{[a}\,\Sigma_{b]c} + D_{[a}\,\Sigma_{c]b})
+ \frac{2}{5}\,D^e\,\Sigma_{e(a}\,h_{bc)}.
\end{equation}
It follows that (\ref{Sigma-dSigma-on N}) holds if and only if 
\begin{equation}
\label{reduced-Sigma-dSigma-on N}
\Sigma_{ab} = 0, \quad D^a \Sigma_{ab} = 0, \quad D_{\{c} \Sigma_{ab\}} = 0
\quad \mbox{on} \quad {\cal N}_i \quad \mbox{near} \quad i.
\end{equation}

Assume now a central harmonic conformal gauge and associated normal coordinates 
$x^a$.  With the expressions given in subsection \ref{rho-Drho-DDrho-DDDrho-expr-in-ch-gauge} a direct calculation using (\ref{gen-DeltaGamma + 2n}) gives
\[
\frac{1}{\mu}\,\Sigma_{ab}|_{{\cal N}_i} 
= (D_aD_b\Gamma + 2\,h_{ab}  - 4\,Q\,D_a\Gamma\,D_b\Gamma)|_{{\cal N}_i}.
\]
Observing (\ref{Lambda-def}), we need to analyse $\Lambda_a|_{{\cal N}_i}$, which is given in our gauge by 
\[
\frac{1}{\mu}\,\Lambda_a|_{{\cal N}_i} 
= s_{ah}\,D^h\Gamma
+ (8\,Q - \frac{1}{6}\,R)\,D_a\Gamma.
\]
Finally, we get in our gauge
\[
\frac{1}{\mu}\,D_{\{c}\Sigma_{ba\}}|_{{\cal N}_i} 
= D_aD_bD_c\Gamma
+  \frac{1}{6}\,R\,D_{a}\Gamma\,h_{bc} 
+ D_a\Gamma\,s_{bc}
+ D^d\Gamma\,s_{da}   \,h_{bc}. 
\]
\[
- 12\,D_{(a}Q\,D_b\Gamma\,D_{c)}  \Gamma 
- 12\,Q\,D_{(a}\Gamma\,D_bD_{c)} \Gamma
+ (18\,Q^2 - 12\,P)
\,D_{a}\Gamma\,D_b\Gamma\,D_{c} \Gamma
\]
\[
+ \left(\frac{3}{10}\,R - \frac{72}{5}\,Q\right)\,D_{(a}\Gamma\,h_{bc)}
- \frac{9}{5}\,D^d\Gamma\,s_{d(a}\,h_{bc)}.
\]

\noindent
Conditions (\ref{reduced-Sigma-dSigma-on N}) thus read, in the same order, in a central harmonic gauge

\begin{equation}
\label{Bh-1gauge:Sigma-van-on N}
D_aD_b\Gamma + 2\,h_{ab}  = 4\,Q\,D_a\Gamma\,D_b\Gamma
\quad \mbox{on} \quad {\cal N}_i \quad \mbox{near} \quad i,
\end{equation}

\begin{equation}
\label{Bh-1gauge:Lambda-van-on-N}
s_{ah}\,D^h\Gamma =  \left(\frac{1}{6}\,R - 8\,Q\right)\,D_a\Gamma
\quad \mbox{on} \quad {\cal N}_i \quad \mbox{near} \quad i,
\end{equation}

\begin{equation}
\label{Bh-1gauge:-DSigma-s-tf-van-on N}
D_aD_bD_c\Gamma = 
(12\,P -18\,Q^2)D_{a}\Gamma\,D_b\Gamma\,D_{c} \Gamma
+ 12\,D_{(a}Q\,D_b\Gamma\,D_{c)}  \Gamma 
\end{equation}
\[
+ 12\,Q\,D_{(a}\Gamma\,D_bD_{c)} \Gamma
-  \frac{1}{6}\,R\,D_{a}\Gamma\,h_{bc} 
- D_a\Gamma\,s_{bc}
- D^d\Gamma\,s_{da}   \,h_{bc}
\]
\[
+ \left( \frac{72}{5}\,Q   - \frac{3}{10}\,R \right)\,D_{(a}\Gamma\,h_{bc)}
+ \frac{9}{5}\,D^d\Gamma\,s_{d(a}\,h_{bc)}
\quad \mbox{on} \quad {\cal N}_i \quad \mbox{near} \quad i.
\]

\begin{lemma}
\label{Sigma-vanishes-iff-B(bb)-vanishes-on-N}
The conformally invariant condition $\Sigma_{ab}|_{{\cal N}_i} = 0$ is equivalent to the conformally invariant condition $D^a\rho\,D^b\rho\,B_{ab}|_{{\cal N}_i} = 0$.
\end{lemma}

\noindent
{\bf Proof}: The central harmonic gauge implies the following important 
the relation
\begin{equation}
\label{s(n,n) = 0-on-N}
s_{ab}\,D^a\Gamma\,D^b\Gamma = 0 \quad \mbox{on} \quad {\cal N}_i.
\end{equation}
In fact, taking a derivative on the left hand side of the eiconal equation in
 (\ref{i-eikonal}) shows that the field $\Pi_{ab} = D_aD_b\Gamma + 2\,h_{ab}$ satisfies $D^a\Gamma\,\Pi_{ab} = 0$.  Observing that 
$\Pi_a\,^a = \Delta_h\Gamma + 6  = 0$ by (\ref{A-central-harmonic}) it follows that $\Pi_{ab}$ has in a frame 
satisfying (\ref{B-frame-norm}) the expansion 
\[
\Pi_{ab} = \kappa\,n_a\,n_b + \beta\,m_{(a}\,n_{b)}
\quad  \mbox{with} \quad \kappa = p^a\,p^b\,\Pi_{ab}, 
\quad \beta = - 4\,p^a\,m^b\,\Pi_{ab}.
\]
This implies that 
\[
0 = \Pi_{ab}\,\Pi^{ab} = D_aD_b\Gamma\,D^aD^b\Gamma - 12
\quad \mbox{on} \quad {\cal N}_i.
\]
Restricting (\ref{consequence:h-is--1}) to ${\cal N}_i$ and observing the relation above gives 
(\ref{s(n,n) = 0-on-N}).

\vspace{.3cm}

Taking a second derivative on the left hand side of the eiconal equation and commuting derivatives gives with (\ref{n=3-Riem-dcomp}) 
\begin{equation}
\label{B-DGD3G}
D^a\Gamma\,D_a\Pi_{bc} + \Pi_{ba}\,\Pi_c\,^a - 2\,\Pi_{bc}
\end{equation}
\[
= D^f\Gamma\,s_{fb} \, D_c\Gamma 
+  D^f\Gamma\,s_{fc} \, D_b\Gamma 
- D^h\Gamma\,D^f\Gamma\,s_{hf}\,h_{bc} 
+ \frac{R}{6}\,D_b\Gamma\,D_c\Gamma  
+ 4\,\Gamma\,\{s_{bc}  + \frac{R}{6}\,h_{cb}\}.
\]
Contracting with $p^a\,m^b$  and observing (\ref{A-frame-on-N-prop}), we get the equation 
$n^aD_a\beta - 4\,\beta = n^a\,m^b\,s_{ab}$,
which reads with (\ref{gradGamma.D}), (\ref{BgradGamma.D}) 
\[
\frac{d}{d\tau}(\tau^2\,\beta) = - \frac{\tau}{2}\,n^a\,m^b\,s_{ab}.
\]
Since $\tau^2\,\beta \rightarrow 0$ as $\tau \rightarrow 0$ by (\ref{h-1-R-DR-at-i}) and (\ref{n-p--near-i}),  it implies in view of (\ref{B(n,n)-on-N-expr}), (\ref{s(n,n) = 0-on-N}) and Lemma \ref{BGG=0-cons} that 
{\it the function $\beta$ vanishes along a given null generator of ${\cal N}_i$ if and only if $B_{ab}\,D^a\Gamma\,D^b\Gamma$ vanishes along that generator}.

\vspace{.3cm}

This shows that $\Sigma_{ab}|_{{\cal N}_i}= 0$, i.e. (\ref{Bh-1gauge:Sigma-van-on N}), implies $D^a\rho\,D^b\rho\,B_{ab}|_{{\cal N}_i} = 0$. The converse will follow if it can be shown that 
$\kappa = 4\,Q$ if $\beta = 0$.

\vspace{.3cm}

To derive the relevant equation we use the eiconal equation repeatedly and observe that
\[
D_cD_bD_a\Gamma\,D^a\Gamma\,D^b\Gamma 
= D^a\Gamma\left\{ D_c(D_bD_a\Gamma\,D^b\Gamma) - D_bD_a\Gamma\,D_cD^b\Gamma\right\}
\] 
\[
= D^a\Gamma\left\{ - 2\,D_cD_a\Gamma - D_bD_a\Gamma\,D_cD^b\Gamma\right\} = 0
\quad \mbox{near } \quad i.
\] 
Assume now that $\beta = 0$, so that
\[
D_aD_b\Gamma = - 2\,h_{ab} + \kappa\,n_a\,n_b 
\quad \mbox{and} \quad n^a\,m^b\,s_{ab} = 0 
\quad \mbox{on} \quad W \setminus \{i\}.
\]
This implies that
\begin{equation}
\label{frame-derivs-on-N}
m^aD_an^b = - 2\,m^b, \quad 
m^aD_am^b = - p^b - \nu\,n^b, \quad
m^aD_ap^b = - 2\,\nu\,m^b
\quad \mbox{on} \quad W \setminus\{i\},
\end{equation}
\[
\mbox{with} \quad \nu = - m^a\,p^b\,D_am_b = m^a\,m^bD_a\,p_b.
\]

Contracting (\ref{Dconsequence:h-is--1}) on $W \setminus\{i\}$  with $p^c$ and observing  (\ref{A-central-harmonic}),  (\ref{s(n,n) = 0-on-N}) and
the equation above gives 
\begin{equation}
\label{B-Dconsequence:h-is--1}
p^c\,n^a\,n^b\,D_cs_{ab} 
- 4\,n^a\,p^b\,s_{ab} - \frac{4}{3}\,R = 0.
\end{equation}
With our assumption, the Bianchi identity, (\ref{h-on-W-minus-i}), 
(\ref{s(n,n) = 0-on-N}),  (\ref{A-frame-on-N-prop}), (\ref{frame-derivs-on-N}) and the relation 
$0 = s_a\,^a = 2\,(n^a\,p^b - m^a\,m^b)s_{ab} $  follows 
\[
p^c\,n^a\,n^b\,D_cs_{ab} = (h^{ca} - n^c\,p^a + 2\,m^c\,m^a)\,n^b\,D_cs_{ab} 
\]
\[
= \frac{1}{6}\,n^a D_aR - n^cD_c(p^a\,n^b\,s_{ab}) + n^cD_c(p^a\,n^b)s_{ab}
+ 2\,m^cD_c(m^a\,n^b\,s_{ab})  - 2\,m^cDc(m^a\,n^b)s_{ab} 
\]
\[
= \frac{1}{6}\,n^a D_aR - n^cD_c(p^a\,n^b\,s_{ab}) + 6\,n^a\,p^bs_{ab},
\]
which implies with (\ref{BgradGamma.D}), (\ref{Dconsequence:h-is--1})
\[
0 = \frac{1}{6}\,n^a D_aR - \frac{4}{3}\,R - n^aD_a(n^b\,p^c\,s_{bc}) + 2\,n^b\,p^c\,s_{bc}
\]
\[
= 2\,\frac{d}{d\tau}(\tau\,n^a\,p^bs_{ab}) - \frac{1}{3}\left(\frac{d}{d\tau}(\tau\,R) + 3\,R\right),
\]
and thus the important relation
\begin{equation}
\label{s(n,p)=1/6R-8Q}
n^a\,p^bs_{ab} = \frac{1}{6}\,R - 8\,Q \quad \mbox{on} \quad W \setminus \{i\}.
\end{equation}
Transvecting (\ref{B-DGD3G}) with $p^b\,p^c$ gives
\[
n^aD_a\kappa - 6\,\kappa = 2\,n^a\,p^bs_{ab} + \frac{1}{6}\,R,
\]
and thus with the equation above 
\[
\frac{d}{d\tau}(\tau^3\,\kappa) = \tau^2\,\left(8\,Q - \frac{1}{4}\,R\right).
\]
With the first of equations (\ref{Q-equ}) this implies
\[
\frac{d}{d\tau}\left(\tau^3\,(\kappa - 4\,Q)\right) 
= \tau^2\left(8\,Q - \frac{1}{4}\,R\right) - 12\,\tau^2\,Q 
+ 4\, \tau^2\left(Q + \frac{1}{16}\,R\right) = 0,
\]
and thus $\kappa = 4\,Q$ on $W \setminus \{i\}$. Since $W$ is a neighbourhood of an arbitrary geodesic $\gamma$, the desired  result follows. \\
$\Box$

\begin{lemma}
\label{DSigma-vanishes-if-B(bb)-vanishes-on-N}
If any of the equivalent conditions of Lemma  \ref{Sigma-vanishes-iff-B(bb)-vanishes-on-N}
is satisfied then 
\[
D_c\Sigma_{ab} = 0 \quad \mbox{on} \quad {\cal N}_i \quad \mbox{near } \,\,\, i.
\]
\end{lemma}

\noindent
{\bf Proof}: The relations 
\begin{equation}
\label{s-comps-on-N}
n^a\,n^bs_{ab} = 0, \quad n^a\,m^bs_{ab} = 0, \quad n^a\,p^bs_{ab} = \frac{1}{6}\,R - 8\,Q,
\quad \mbox{on} \quad W \setminus \{i\},
\end{equation}
which we have seen in the proof above to be a consequence of our gauge conditions and the  conditions of Lemma \ref{Sigma-vanishes-iff-B(bb)-vanishes-on-N}, immediately  imply 
(\ref{Bh-1gauge:Lambda-van-on-N}) and thus $D^a\Sigma_{ab} = 0$ on 
${\cal N}_i$. 

To show that $D_{\{a} \Sigma_{bc\}} = 0$ on ${\cal N}_i$ we consider equation 
(\ref{Bh-1gauge:Sigma-van-on N}). Because $n^a$ and $m^a$ are tangential to $W \setminus \{i\}$, we can take derivatives in these directions which give on $W \setminus \{i\}$ with equations  (\ref{Q-equ}) and   (\ref{s(n,p)=1/6R-8Q})
\begin{equation}
\label{n-D3Omega}
n^a\,D_aD_bD_c \Gamma = \left(\frac{1}{2}\,R - 8\,Q\right)n_b\,n_c,
\end{equation}
\begin{equation}
\label{m-D3Omega}
m^a\,D_aD_bD_c \Gamma = 4\,m^aD_aQ\,n_b\,n_c - 16\,Q\,n_{(b}\,m_{c)} 
\end{equation}
and, after commuting derivatives,
\begin{equation}
\label{D3Omega-n}
n^c\,D_aD_bD_c \Gamma = \,8\,Q\,n_a\,n_b,
\end{equation}
\begin{equation}
\label{D3Omega-m}
m^c\,D_aD_bD_c \Gamma = (4\,m^cD_cQ - m^c\,p^d\,s_{cd})\,n_a\,n_b
- 8\,Q\,(2\,n_a\,m_b + n_b\,m_a).
\end{equation}
Corresponding contractions of (\ref{Bh-1gauge:-DSigma-s-tf-van-on N}) with $n^a$, $m^a$ give precisely the same results. This implies that with the possible exception of
$p^c\,p^b\,p^a\,D_{\{c}\Sigma_{ba\}}$ all components of
$D_{\{c}\Sigma_{ba\}}$ in our frame vanish on $W$.   

We set 
\[
\zeta =  p^c\,p^b\,p^aD_cD_bD_a\Gamma,
\]
and  denote the contraction of the right hand side of 
(\ref{Bh-1gauge:-DSigma-s-tf-van-on N}) with $p^c\,p^b\,p^a$ by $\hat{\zeta}$. Then
\begin{equation}
\label{zeta-hat}
\hat{\zeta} = 12\,p^a\,D_a Q + 30\,Q^2 + 12\,P - p^a\,p^c\,s_{ac},
\end{equation}
and the proof will be complete when it can  be shown that 
\begin{equation}
\label{D3Gamma(ppp)}
\zeta = \hat{\zeta} \quad \mbox{on} \quad W \setminus \{i\}  \quad \mbox{near} \quad i.
\end{equation}
Because  (\ref{Bh-1gauge:Sigma-van-on N})  it is only known so far to hold on  ${\cal N}_i$,
this equation does not allow us  to calculate a threefold derivative of $\Gamma$ in a direction transverse 
to ${\cal N}_i$. It can therefore not be used to  derive (\ref{D3Gamma(ppp)}) and we need a different representation of $\Gamma$.

\vspace{.3cm}

\noindent
{\it Calculation of $\zeta$ and comparison with $\hat{\zeta}$.}

\vspace{.3cm}

Let $\gamma(\tau)$ with $\gamma(0) = i$ denote one of the null geodesics ruling $W$ and let 
$k$, $m$, $p$  denote a frame at $i$ with $k = \gamma'(0)$ and 
\[
h(k, q) = 1, \quad
h(m, m) = - \frac{1}{2}, \quad
h(k, k) = h(q, q) = h(k, m) = h(q, m) = 0.
\]
Assume the frame to be  parallelly transported along $\gamma$ so that 
\[
D_k k = 0, \quad D_k m = 0, \quad D_k q = 0.
\] 
In the normal coordinates $x^a$  the geodesic $\gamma$ is then given near $i$ by the curve
\[
B_{\epsilon} \equiv \{z \in \mathbb{C}|\,|z| < \epsilon\}  \ni \tau
 \rightarrow  x^a(\tau)  = \tau\,x^a_*  \in {\cal N}_i
\quad \mbox{ with} \quad x^a_* = k^a, \quad \epsilon > 0.
\]
and 
\[
n^a(\gamma(\tau))  = D^a\Gamma(\gamma(\tau)) = - 2\,\tau\,k^a  \quad \mbox{along} \quad \gamma.
\]
We assume that the vector field $m^a$ considered here  coincides on $\gamma$ with the field $m^a$ introduced in section \ref{relations-on-Ni}.  Because the fields $q^a$ and $p^a$ are determined by the normalization conditions  uniquely once $k^a$,$m^a$ and $n^a$, $m^a$ are given respectively, it follows that  
\begin{equation}
\label{p-q-rel}
p^a = - \frac{1}{2\,\tau}\,q^a  \quad \mbox{along} \quad \gamma.
\end{equation}

For $(\tau, \lambda) \in B_{\epsilon'} \times B_{\epsilon'}$ with some 
$\epsilon' > 0$  a 1-parameter family  $x^a(\tau, \lambda)$  of geodesics with affine parameter 
 $\lambda$ and family parameter $\tau$ is defined by the following conditions. It holds
 \[
x^a(\tau, 0) = \gamma^a(\tau) \quad \mbox{so that} \quad x^a(0, 0) = x^a(i) = 0,
\]
and for given value of $\tau$ the curve 
\[
\lambda \rightarrow x^a(\tau, \lambda),
\]
is the geodesic which has  tangential  vector $q$ at  $\gamma(\tau)$. 

We extend the fields $k$ and $q$ given  
on $\gamma$ to the points $x^a(\tau, \lambda)$ by setting 
\[
k^a(\tau, \lambda) = \frac{d}{d\tau}x^a(\tau, \lambda), \quad \quad  
q^a(\tau, \lambda)  = \frac{d}{d\lambda}x^a(\tau, \lambda).
\]
Then $q^a$ is geodesic and the field $k^a$ is a Jacobi field which satisfies
the equation
\[
D_q^2\,k^a = R^a\,_{bcd}\,q^b\,q^c\,k^d
\quad \mbox{along the geodesic} \quad \lambda \rightarrow x(\tau, \lambda), 
\]
and the initial conditions
\[
k^a(\tau, 0) =  \gamma'^{a}(\tau), \quad \quad
q^cD_c\,k^a(\tau, 0) = k^cD_c\,q^a(\tau, 0)  =  k^cD_c\,q^a|_{\gamma(\tau)} = 0.
\]

\vspace{.2cm}

By construction, the curve $\lambda \rightarrow x(0, \lambda)$ is a generator of ${\cal N}_i$. 
Therefore  
$\Gamma(x(0, \lambda)) = 0$ whence
\[
\Gamma(x(\tau, \lambda)) 
= \int_0^{\tau} \frac{d}{d\tau'} \Gamma(x(\tau', \lambda))\,d\tau'
= \int_0^{\tau} k^aD_a\Gamma (x(\tau', \lambda))\,d\tau'.
\]
Along the geodesic  $\lambda \rightarrow x(\tau, \lambda)$ 
it follows then
\[
q^a\,q^b\,q^c\,D_aD_bD_c\Gamma|_{x(\tau, \lambda)} = 
\frac{d^3}{d\lambda^3}(\Gamma(x(\tau, \lambda)))
= \int_0^{\tau} \frac{d^3}{d\lambda^3}\left(k^aD_a\Gamma (x(\tau', \lambda))\right)\,d\tau'
\]
\[
= \int_0^{\tau} q^a\,q^b\,q^c\,D_aD_bD_c \left(k^aD_a\Gamma (x(\tau', \lambda))\right)\,d\tau'
=  \int_0^{\tau} J \,d\tau'.
\]
With  the Jacobi equation the integrand $J$ can be written
\[
J  = q^e\,n^a\,q^b\,q^c\,k^d\,D_eR_{abcd}
+n^a\,q^b\,q^c \,q^eD_ek^d\,R_{abcd}
\]
\[
+ 3\, q^eD_eD^a\Gamma\,q^b\,q^c\,k^d\,R_{abcd}
+ 3\, q^eD_ek^a\,q^c\,q^bD_cD_bD_a\Gamma
+ k^a\,q^b\,q^c\,q^d\,D_bD_cD_dD_a\Gamma.
\]
Restricting the equations above to $\gamma$ and observing 
(\ref{p-q-rel}) and (\ref{Bh-1gauge:Sigma-van-on N})
gives
\[
- 8\,\tau^3\,\zeta = \int_0^{\tau} J \,d\tau',
\]
where we have along $\gamma$ 
\[
\frac{1}{4\,\tau^2}\, J
= p^e\,n^a\,p^b\,p^c\,n^d\,D_eR_{abcd}
+ 12\,Q\,n^a\,p^b\,p^c\,n^d\,R_{abcd}
+ p^b\,p^c\,p^d\,D_bD_cD_dD_a\Gamma\,D^a\Gamma.
\]

The last term on the right hand side can be simplified by
using the eiconal equation and (\ref{Bh-1gauge:Sigma-van-on N}). It holds 
 \begin{equation}
 \label{DDDGamma.DGamma}
D_cD_dD_a\Gamma\,D^a\Gamma
= D_c(D_dD_a\Gamma\,D^a\Gamma) - D_dD_a\Gamma\,D_cD^a\Gamma
\end{equation}
\[
= - 2\,D_cD_d\Gamma 
+ 2\, D_dD_c\Gamma
+ 8\,Q\,n_c\,n_d = 8\,Q\,n_c\,n_d \quad \mbox{on} \quad {\cal N}_i,
\]
which implies with (\ref{A-central-harmonic})  and (\ref{Bh-1gauge:Sigma-van-on N})
\begin{equation}
\label{DDDGamma.DDGamma}
D_cD_dD_a\Gamma\,D^dD^a\Gamma = 0
\quad \mbox{on} \quad {\cal N}_i,
\end{equation}
and 
 \begin{equation}
 \label{DDDDGamma.DGamma}
D_bD_cD_dD_a\Gamma\,D^a\Gamma
= D_b(D_cD_dD_a\Gamma\,D^a\Gamma) - D_cD_dD_a\Gamma\,D_bD^a\Gamma
\end{equation}
\[
= D_b\left\{D_c(D_dD_a\Gamma\,D^a\Gamma) - D_dD_a\Gamma\,D_cD^a\Gamma\right\} - D_cD_dD_a\Gamma\,D_bD^a\Gamma
\]
\[
= - 2\,D_bD_cD_d\Gamma 
- D_bD_dD_a\Gamma\,D_cD^a\Gamma
- D_dD_a\Gamma\,D_bD_cD^a\Gamma
 - D_cD_dD_a\Gamma\,D_bD^a\Gamma
\]
\[
= - 2\,D_bD_cD_d\Gamma 
+ 2\, D_bD_dD_c\Gamma
+ 2\, D_bD_cD_d\Gamma
+ 2\, D_cD_dD_b\Gamma
\]
\[
- 4\,Q\,(n_c\,D_bD_dD_a\Gamma
+ n_d\,D_bD_cD_a\Gamma
+ n_b\,D_cD_dD_a\Gamma\,)D^a\Gamma
\]
\[
= 2\, D_bD_cD_d\Gamma + 2\, D_cD_bD_d\Gamma
- 96\,Q^2\,n_b\,n_c\,n_d  \quad \mbox{on} \quad {\cal N}_i,
\]
whence 
 \begin{equation}
 \label{DDDDGamma.DDGamma}
D_bD_cD_dD_a\Gamma\,D^dD^a\Gamma = 2^7\,Q^2\,n_b\,n_c
 \quad \mbox{on} \quad {\cal N}_i.
\end{equation}
With (\ref{n=3-Riem-dcomp}), (\ref{s(n,p)=1/6R-8Q}), (\ref{DDDGamma.DGamma})
and (\ref{DDDDGamma.DGamma}) follows 
\[
\frac{1}{4\,\tau^2}\, J(\gamma(\tau))
=
2\,p^c\,n^b\,p^a\,D_c\,s_{ba}  
+ \frac{1}{6}\,p^c D_cR 
+ 6\,R\,Q 
 - 288\,Q^2 +  4\,\zeta,
\]
and thus
\begin{equation}
\label{zeta-int-equ}
\tau^3\,\zeta = \int_0^{\tau} \tau'^2\,(\eta - 2\,\zeta) \,d\tau',
\end{equation}
with 
\[
\eta \equiv - p^c\,n^b\,p^a\,D_c\,s_{ba}  
- \frac{1}{12}\,p^c D_cR 
- 3\,R\,Q 
+ 144\,Q^2
\]
\[
= - (h^{cb} - n^c\,p^b + 2\,m^c\,m^b)\,p^a\,D_c\,s_{ba}  
- \frac{1}{12}\,p^e D_eR 
- 3\,R\,Q 
+ 144\,Q^2,
\]
whence
\begin{equation}
\label{eta-expr}
\eta = n^c\,p^b\,p^a\,D_c\,s_{ba}  
- 2\,m^c\,m^b\,p^a\,D_c\,s_{ba} 
- \frac{1}{4}\,p^e D_eR 
- 3\,R\,Q 
+ 144\,Q^2.
\end{equation}
Taking  in (\ref{zeta-int-equ}) derivatives with respect to $\tau$ gives 
\[
\tau^3\,\frac{d}{d\tau} \zeta  + 3\,\tau^2\,\zeta = \tau^2\,(\eta - 2\,\zeta),
\]
whence
\[
\frac{d}{d\tau}(\tau^5\,\zeta) =
\tau^5\,\frac{d}{d\tau} \zeta  + 5\,\tau^4\,\zeta = \tau^4\,\eta,
\]
and thus finally 
\begin{equation}
\label{p-p-p-DGamma}
\zeta =
\frac{1}{\tau^5} \int_0^{\tau} \tau'^4\,\eta\,d\tau'
\quad \mbox{along} \quad \gamma.
\end{equation}

Because $\hat{\zeta}(\tau)  = O(1)$ as  $\tau \rightarrow 0$ by 
(\ref{h-1-R-DR-at-i-and-Cott(i) = 0}) and (\ref{p-q-rel}), we can write the function 
$\hat{\zeta}$ given by (\ref{zeta-hat}) in the form
\[
\hat{\zeta} =
\frac{1}{\tau^5} \int_0^{\tau} \frac{d}{d\tau'}(\tau'^5\,\hat{\zeta})\,d\tau'.
\]
It follows then with (\ref{p-p-p-DGamma}) that
\[
\zeta = \hat{\zeta}\,\,\,on\,\,\gamma\,\,near\,\,i\,\,\,if \,\,and\,\,only\,\,if \,\,K = 0\,\,\,there,
\,\,where
\]
\begin{equation}
\label{K-def}
K \equiv \frac{2}{\tau^4} \left( \frac{d}{d\tau}(\tau^5\,\hat{\zeta}) - \tau^4\,\eta \right)
= - n^aD_a\,\hat{\zeta} + 10\,\hat{\zeta} - 2\,\eta.
\end{equation}

\noindent
The right hand side of (\ref{K-def}) reads  more explicitly 
\[
K = - n^cD_c\left(- p^a\,p^b\,s_{ab} + 12\,P + 12\,p^aD_aQ + 30\,Q^2\right)
\]
\[
+ 10\left(- p^a\,p^b\,s_{ab} + 12\,P + 12\,p^aD_aQ + 30\,Q^2\right)
\]
\[
- 2\left(n^cD_c(p^a\,p^b\,s_{ab}) - 4\,p^a\,p^b\,s_{ab} - 2\,m^c\,m^b\,p^a\,D_cs_{ab}
- \frac{1}{4}\,p^aD_aR + 144\,Q^2 - 3\,R\,Q\right)
\]
\[
= - n^cD_c(p^a\,p^b\,s_{ab}) - 2\,p^a\,p^b\,s_{ab} + 4\,m^c\,m^b\,p^a\,D_cs_{ab}
\]
\[
- 12\left\{4\,P - \frac{2}{3}\,(n^a\,p^b - m^a\,m^b)\,D_aD_bQ + \frac{1}{24}\,R\,Q\right\}
+ 120\,P
- 24\,p^aD_aQ 
\]
\[+ 120\,p^aD_aQ + \frac{1}{2}\,p^aD_aR 
- 12\,n^a\,p^bD_aD_bQ
- 60\,Q\left(\frac{R}{8} + 2\,Q\right) + 12\,Q^2 + 6\,R\,Q
\]
\[
= - n^cD_c(p^a\,p^b\,s_{ab}) - 2\,p^a\,p^b\,s_{ab} + 4\,m^c\,m^b\,p^a\,D_cs_{ab} + 72\,P
-8\,m^a\,m^bD_aD_bQ 
\]
\[
- 4\,p^a\,D^b\Gamma\,D_aD_bQ 
- 2\,R\,Q
+ 96\,p^aD_aQ 
+ \frac{1}{2}\,p^aD_aR - 108\,Q^2
\]
\[
= - n^cD_c(p^a\,p^b\,s_{ab}) - 2\,p^a\,p^b\,s_{ab} + 4\,m^c\,m^b\,p^a\,D_cs_{ab} + 72\,P
-8\,m^a\,m^bD_aD_bQ 
\]
\[
- 4\,\left\{  p^a\,D_a(D^b\Gamma\,D_bQ) - p^a\,D_aD^b\Gamma\,D_bQ \right\}
- 2\,R\,Q
+ 96\,p^aD_aQ 
+ \frac{1}{2}\,p^aD_aR - 108\,Q^2,
\]
where we used that equations (\ref{Q-equ}) hold in a full neighbourhood of $i$.
It follows 
\begin{equation}
\label{fin-K-expr}
K = - n^cD_c(p^a\,p^b\,s_{ab}) - 2\,p^a\,p^b\,s_{ab} + 4\,m^c\,m^b\,p^a\,D_cs_{ab}
\quad 
\end{equation}
\[
\quad  + 72\,P - 8\,m^a\,m^bD_aD_bQ 
+ 80\,p^aD_aQ - 76\,Q^2.
\]

To make use of the second of equations (\ref{Q-equ}) we consider the ODE
\[
n^aD_aK - 4\,K = M,
\]
on $\gamma$, which is such that the function $M$ on the right hand side does not contain the function $P$. Because
\[
n^aD_aK - 4\,K = - 2 \left(\tau\,\frac{d}{d\tau}K + 2\,K\right) 
= - \frac{2}{\tau}\,\frac{d}{d\tau}(\tau^2\,K),
\]
and $\tau^2\,K \rightarrow 0$ as $\tau \rightarrow 0$, it follows that 
\[
K \,\,vanishes \,\,along \,\,\gamma \,\,near \,\,i \,\,if \,\,and \,\,only\,\, if \,\,M \,\,vanishes \,\,there.
\]

With the expression (\ref{fin-K-expr}) for $K$, a useful form of $M$ is obtained as follows.
\[
M = n^dD_d\left\{
- n^cD_c(p^a\,p^b\,s_{ab}) - 2\,p^a\,p^b\,s_{ab} + 4\,m^c\,m^b\,p^a\,D_cs_{ab} + 72\,P
\right.
\]
\[
\left.
-8\,m^a\,m^bD_aD_bQ 
+ 80\,p^aD_aQ - 76\,Q^2
\right\}
- 4\left\{
- n^cD_c(p^a\,p^b\,s_{ab}) - 2\,p^a\,p^b\,s_{ab}\right.
\]
\[
\left. 
+ 4\,m^c\,m^b\,p^a\,D_cs_{ab} + 72\,P
-8\,m^a\,m^bD_aD_bQ 
+ 80\,p^aD_aQ - 76\,Q^2
\right\}
\]
\[
= - n^dD_d\left(n^cD_c(p^a\,p^b\,s_{ab})\right) + 2\,n^cD_c(p^a\,p^b\,s_{ab})
+ 8\,p^a\,p^b\,s_{ab} + 4\,n^dD_d(m^c\,m^b\,p^a\,D_cs_{ab})
\]
\[
 - 16\,m^c\,m^b\,p^a\,D_cs_{ab} - 8\,m^a\,m^b\,n^cD_cD_aD_bQ 
 + 72\,\left\{ - \frac{1}{3}\,\Delta_h Q + \frac{1}{24}\,R\,Q\right\}
+ 160\,p^aD_aQ 
\]
\[+ 80\,n^b\,p^a\,D_bD_aQ - 152\,Q \left(\frac{R}{8} + 2\,Q\right)
+ 32\,m^a\,m^b\,D_aD_bQ - 320\,p^aD_aQ + 304\,Q^2
\]
\[
= - n^dD_d\left(n^cD_c(p^a\,p^b\,s_{ab})\right) + 2\,n^cD_c(p^a\,p^b\,s_{ab})
+ 8\,p^a\,p^b\,s_{ab} 
\]
\[
+ 4\,n^dD_d(m^c\,m^b\,p^a\,D_cs_{ab})
 - 16\,m^c\,m^b\,p^a\,D_cs_{ab} - 8\,m^a\,m^b\,n^cD_cD_aD_bQ 
 \]
 \[
+ 32\,n^a\,p^bD_aD_bQ 
+ 80\,m^a\,m^b\,D_aD_bQ - 160\,p^aD_aQ - 16\,R\,Q.
\]
Using here 
\[
n^a\,p^bD_aD_bQ = p^bD_b(D^a\Gamma\,D_aQ) -  p^bD_bD^a\Gamma\,D_aQ
= \frac{1}{8}\,p^aD_aR + 4\,p^aD_aQ - \frac{1}{2}\,R\,Q - 8\,Q^2,
\]
and 
\[
m^a\,m^b\,n^c D_aD_bD_c Q 
\]
\[
= m^aD_a\left\{m^bD_b(n^cD_cQ) - m^bD_bn^c\,D_cQ\right\}
- \left\{m^aD_am^b\,n^c + m^b\,m^aD_an^c\right\}D_bD_cQ
\]
\[
= 
m^a\,m^bD_aD_b(n^cD_cQ) - (p^b + \nu\,n^b)D_b(n^cD_cQ) 
+ 2\,m^a\,m^c\,D_aD_cQ 
\]
\[
- 2\,(p^c + \nu\,n^c)\,D_cQ
+ \left\{ (p^b + \nu\,n^b)\,n^c + 2\,m^b\,m^c\right\}D_bD_cQ
\]
\[
= m^a\,m^b\,D_aD_b\left(\frac{R}{8} + 2\,Q\right) + 4\,m^am^b\,D_aD_bQ - \frac{1}{2}\,R\,Q - 8\,Q^2,
\]
which gives with
\[
R^f\,_{bca}\,D_fQ\,m^a\,m^b\,n^c = 
 \left(m^a\,m^b\,s_{ab} - \frac{R}{24}\right)n^cD_cQ 
- \frac{1}{2} \left(n^a\,p^b\,s_{ab} + \frac{R}{12}\right)n^cD_cQ 
\]
\[
= - \frac{1}{2}\,R\,Q - 8\,Q^2,
\]
the expression
\[
m^a\,m^b\,n^cD_cD_aD_bQ  = m^a\,m^b\,n^c\left(D_aD_bD_c Q - R^f\,_{bca}\,D_fQ\right)
\]
\[
= \frac{1}{8}\,m^a\,m^bD_aD_bR + 6\,m^a\,m^bD_aD_bQ,
\]
we finally get

\begin{equation}
\label{fin-M-expr}
M = - n^dD_d\left(n^cD_c(p^a\,p^b\,s_{ab})\right) + 2\,n^cD_c(p^a\,p^b\,s_{ab})
+ 8\,p^a\,p^b\,s_{ab} 
\end{equation}

\[
+ 4\,n^dD_d(m^c\,m^b\,p^a\,D_cs_{ab})
- 16\,m^c\,m^b\,p^a\,D_cs_{ab}  - m^a\,m^bD_aD_bR 
\]

 \[
 + 4\,p^aD_aR  + 2^5 \cdot (m^a\,m^b\,D_aD_bQ - p^aD_aQ - R\,Q)
- 2^8 \cdot Q^2.
\]

\vspace{.3cm}

\noindent
Lemma \ref{DSigma-vanishes-if-B(bb)-vanishes-on-N} will be proven when it can be shown that this function vanishes on $\gamma$ near $i$.

\vspace{.3cm}

\noindent
{\it Proof that $M = 0$ on $\gamma$ near $i$.}

\vspace{.3cm}

To make use of equations (\ref{Dconsequence:h-is--1}) and  (\ref{DDconsequence:h-is--1}) we observe (\ref{DDDGamma.DDGamma}), (\ref{DDDDGamma.DDGamma})
and the fact that equations (\ref{n-D3Omega}), (\ref{m-D3Omega}), (\ref{D3Omega-n}), 
(\ref{D3Omega-m}) allow us to derive  the expansion
\[
D_aD_bD_c\Gamma   
= n_a\,n_b\,n_c\,(p, p, p) 
+ \left(\frac{1}{2}\,R - 8\,Q \right)p_a\,n_b\,n_c  
\]
\[
+ 8\,Q\,n_a\,(p_b\,n_c  + n_b\,p_c)
- \left(8\,m^aD_aQ + 16\,Q - \frac{1}{3}\,R\right) n_a (n_b\,m_c + m_b\,n_c)
\]
\[
+ 32\,Q\,n_a\,m_b\,m_c
- 8\,m^aD_aQ\, m_a\,n_b\,n_c
+16\,Q \,m_a (n_b\,m_c + m_b\,n_c),
\]
which implies
\begin{equation}
\label{D3Gamma-D3Gamma}
D_cD_aD_b\Gamma\,D_dD^aD^b\Gamma = 3 \cdot 2^7 \cdot Q^2\,n_c\,n_d
\quad \mbox{on} \quad W.
\end{equation}
The restrictions of 
(\ref{Dconsequence:h-is--1}) and (\ref{DDconsequence:h-is--1}) to $W$ are then obtained in the form
\begin{equation}
\label{C-Dconsequence:h-is--1}
0 = D_cs_{ab}\,D^a\Gamma\,D^b\Gamma 
+ 2\,s_{ab}\,D^a\Gamma\,D_cD^b\Gamma 
- \frac{4}{3}\,D_c\Gamma\,R,
\end{equation}
\begin{equation}
\label{h=-1-2nd-order-on-W}
0 = D_dD_cs_{ab}\,n^a\,n^b 
- 4\,(D_cs_{db}\,n^b + D_ds_{cb}\,n^b) 
\end{equation}
\[
+ 8\,Q\,(n_c\,D_ds_{ab}\,n^a\,n^b + n_d\,D_cs_{ab}\,n^a\,n^b) 
+ 8\,s_{cd}
\]
\[
- \frac{4}{3}\,(D_cR\,n_d + D_dR\,n_c  
- 2\,h_{cd}\,R) 
- 8\,R\,Q\,n_c\,n_d
+ 9 \cdot 2^{7}\,Q^2\,n_c\,n_d.
\]

\noindent
Contraction of (\ref{B-Dconsequence:h-is--1}) with $p^c$ gives 
\[
p^c\,n^a\,n^b\,D_cs_{ab} 
= 4\,n^b \,p^c\,s_{cb} + \frac{4}{3}\,R 
= 2\,R - 32\,Q 
\quad \mbox{on} \quad W. 
\]
Contracting (\ref{h=-1-2nd-order-on-W}) with $p^d\,p^c$ and observing  the equation above we get
\[
0 = p^d\,p^c\,D_dD_cs_{ab}\,n^a\,n^b 
- 8\,p^d\,p^c\,D_cs_{db}\,n^b   
+ 16\,Q\,p^d\,D_ds_{ab}\,n^a\,n^b 
\]
\[
+ 8\,p^c\,p^d\,s_{cd}
- \frac{8}{3}\,p^c\,D_cR
- 8\,R\,Q
+ 9 \cdot 2^{7}\,Q^2
\]
\[
= p^d\,p^c\,D_dD_cs_{ab}\,n^a\,n^b 
- 8\,p^d\,p^c\,D_cs_{db}\,n^b   
+ 16\,Q\, (2\,R - 32\,Q)  
\]
\[
+ 8\,p^c\,p^d\,s_{cd}
- \frac{8}{3}\,p^c\,D_cR
- 8\,R\,Q
+ 9 \cdot 2^{7}\,Q^2
\]
whence, finally, 
\begin{equation}
\label{ppnnDDs}
0 = p^d\,p^c\,D_dD_cs_{ab}\,n^a\,n^b 
- 8\,p^c\,n^b\,p^a\,D_cs_{ba}   
+ 8\,p^c\,p^d\,s_{cd}
\end{equation}
\[
- \frac{8}{3}\,p^c\,D_cR
+ 24\,R\,Q
+ 5 \cdot 2^{7}\,Q^2
\quad \mbox{on} \quad
W.
\]

To bring the expression on the right hand side of  (\ref{ppnnDDs}) into a form similar to 
(\ref{fin-M-expr}), we need ways to swap in the first term  the positions of the vectors $n$ and 
$p$. To achieve this we shall repeatedly use the representation (\ref{h-on-W-minus-i})
of the metric and the Bianchi identity. This  gives for the first three terms of (\ref{ppnnDDs})
\[
p^d\,p^c\,D_dD_cs_{ab}\,n^a\,n^b 
- 8\,p^c\,n^b\,p^a\,D_cs_{ba}   
+ 8\,p^c\,p^d\,s_{cd}
\]
\[
= p^d\,n^a\,p^c\,n^b \,D_dD_cs_{ab} 
- 8\,p^c\,n^b\,p^a\,D_cs_{ba}   
+ 8\,p^c\,p^d\,s_{cd}
\]
\[
= (h^{da} - n^d\,p^a + 2\,m^d\,m^a)
\,(h^{cb} - n^c\,p^b + 2\,m^c\,m^b)\,D_dD_cs_{ab} 
\]
\[
- 8\,(h^{cb} - n^c\,p^b + 2\,m^c\,m^b)\,p^a\,D_cs_{ba}   
+ 8\,p^c\,p^d\,s_{cd}
\]
\[
= D^aD^bs_{ab}  - n^d\,p^a \,D_dD^bs_{ab} + 2\,m^d\,m^a\,D_dD^bs_{ab} 
\]
\[
+ (- n^c\,p^b + 2\,m^c\,m^b)\,(D_cD^as_{ab} - 2\,s_{f(a}\,R^f\,_{b)}\,^a\,_c)
+ n^d\,n^c\,p^b \,p^a\,D_dD_cs_{ab} 
\]
\[
- 2\, n^d\,m^c\,p^a\,m^b\,D_dD_cs_{ab} 
- 2\,m^d\,n^c\,m^a\,p^b\,D_dD_cs_{ab} 
+ 4\,m^d\,m^c\,m^b\,m^a\,D_dD_cs_{ab} 
\]
\[
- 8\,p^a\,D^bs_{ba} + 8\, n^c\,p^b\,p^a\,D_cs_{ba}  - 16\,m^c\,m^b\,p^a\,D_cs_{ba}   
+ 8\,p^c\,p^d\,s_{cd}
\]
\[
= n^d\,n^c\,p^b \,p^a\,D_dD_cs_{ab} 
- 4\, n^d\,m^c\,p^a\,m^b\,D_dD_cs_{ab} 
+ 4\,m^d\,m^c\,m^b\,m^a\,D_dD_cs_{ab} 
\]
\[
+ \frac{1}{6}\,\Delta_h R  -  \frac{1}{6}\,n^d\,p^a \,D_dD_aR 
+  \frac{1}{3}\,m^d\,m^a\,D_dD_aR
- \frac{1}{6}\,n^c\,p^b\,D_cD_bR
\]
\[
+ \frac{1}{3}\,\,m^c\,m^b\,D_cD_bR
- 4\,m^d\,n^c\,m^a\,p^b\,s_{f(a}\,R^f\,_{b)cd} 
+ (2\,n^c\,p^b - 4\,m^c\,m^b)\,s_{f(a}\,R^f\,_{b)}\,^a\,_c
\]
\[
-  \frac{4}{3}\,\,p^a\,D_aR + 8\, n^c\,p^b\,p^a\,D_cs_{ba}  - 16\,m^c\,m^b\,p^a\,D_cs_{ba}   
+ 8\,p^c\,p^d\,s_{cd}
\]
\[
= n^d\,n^c\,p^b \,p^a\,D_dD_cs_{ab} 
- 4\, n^d\,m^c\,p^a\,m^b\,D_dD_cs_{ab} 
+ 4\,m^d\,m^c\,m^b\,m^a\,D_dD_cs_{ab} 
\]
\[
+ 8\, n^c\,p^b\,p^a\,D_cs_{ba}  - 16\,m^c\,m^b\,p^a\,D_cs_{ba}   
+ \frac{1}{3}\,m^d\,m^a\,D_dD_aR
-  \frac{4}{3}\,\,p^a\,D_aR 
\]
\[
+ 8\,p^c\,p^d\,s_{cd}
- 4\,m^d\,n^c\,m^a\,p^b\,s_{f(a}\,R^f\,_{b)cd} 
- (2\,n^c\,p^b - 4\,m^c\,m^b)\,s_{f(a}\,R^f\,_{b)c}\,^a
\]
\[
= 
n^dD_d(n^c\,p^b \,p^a\,D_cs_{ab}) 
- n^dD_d(n^c\,p^b \,p^a)\,D_cs_{ab} 
- 4\, n^dD_d(m^c\,p^a\,m^b\,D_cs_{ab}) 
\]
\[
+ 4\, n^dD_d(m^c\,p^a\,m^b)\,D_cs_{ab} 
+ 4\,m^dD_d(m^c\,m^b\,m^a\,D_cs_{ab}) 
- 4\,m^d D_d(m^c\,m^b\,m^a)\,D_cs_{ab} 
\]
\[
+ 8\, n^c\,p^b\,p^a\,D_cs_{ba}  - 16\,m^c\,m^b\,p^a\,D_cs_{ba}   
+ \frac{1}{3}\,m^d\,m^a\,D_dD_aR
\]
\[
-  \frac{4}{3}\,\,p^a\,D_aR 
+ 8\,p^c\,p^d\,s_{cd}
- 6\,(s_{ab}\,n^a\,p^b)^2 
+ R\,\,s_{ab}\,n^a\,p^b,
\]
where we use the relations
\[
s^c\,_a n^a s_{cb} \,p^b = (n^a \,p^b s_{ab})^2, \quad
s^c\,_a m^as_{cb}\,m^b = - 2\,(n^a\,p^b s_{ab})^2, \quad
s_{ab}\,s^{ab} = 6\,(n^a\,p^b s_{ab})^2, 
\]
and
\[
- 4\,m^d\,n^c\,m^b\,p^a\,s_{f(a}\,R^f\,_{b)cd} = - 3\,(s_{ab}\,n^a\,p^b)^2 
+ \frac{1}{2}\,R\,\,s_{ab}\,n^a\,p^b,
\]
\[
- (2\,n^c\,p^b - 4\,m^c\,m^b)\,s_{f(a}\,R^f\,_{b)c}\,^a
= - 3\,(s_{ab}\,n^a\,p^b)^2 
+ \frac{1}{2}\,R\,\,s_{ab}\,n^a\,p^b.
\]

It follows 
\[
p^d\,p^c\,D_dD_cs_{ab}\,n^a\,n^b 
- 8\,p^c\,n^b\,p^a\,D_cs_{ba}   
+ 8\,p^c\,p^d\,s_{cd}
\]
\[
= 
n^dD_d\left\{n^cD_c(p^b \,p^a\,s_{ba}) 
- 4\,p^b \,p^a\,s_{ab}\right\} 
- 2\,n^c\,p^b \,p^a\,D_cs_{ba} 
- 4\, n^dD_d(m^c\,p^a\,m^b\,D_cs_{ba}) 
\]
\[
+ 8\,m^c\,p^b\,m^a\,D_cs_{ba} 
+ 4\,m^dD_d\left\{m^cD_c(m^b\,m^a\,s_{ab})  
+ 2\,(p^b + \nu\,n^b)\,m^a\,s_{ab} \right\} 
\]
\[
+ \left\{ 4\,(p^c + \nu\,n^c)\,m^b\,m^a
+ 8\,m^c\,(p^b + \nu\,n^b)\,m^a\right\} D_cs_{ab} 
+ 8\, n^c\,p^b\,p^a\,D_cs_{ba}  
\]
\[
- 16\,m^c\,m^b\,p^a\,D_cs_{ba}   
+ \frac{1}{3}\,m^d\,m^a\,D_dD_aR
-  \frac{4}{3}\,\,p^a\,D_aR 
+ 8\,p^c\,p^d\,s_{cd}
\]
\[
- 6\,(s_{ab}\,n^a\,p^b)^2 
+ R\,\,s_{ab}\,n^a\,p^b
\]
\[
= 
n^dD_d\left\{n^cD_c(p^b \,p^a\,s_{ba}) \right\}  
- 4\,n^dD_d(p^b \,p^a\,s_{ab}) 
+ 6\, n^cD_c(p^b\,p^a\,s_{ba})  
- 24\,p^b\,p^a\,s_{ba}  
\]
\[
- 4\, n^dD_d(m^c\,p^a\,m^b\,D_cs_{ba}) 
- 8\,m^c\,m^b\,p^a\,D_cs_{ba}   
+ 4\,m^dD_d\left\{m^cD_c(m^b\,m^a\,s_{ab})\right\}  
\]
\[ 
+ 8\,m^dD_d(p^b\,m^a\,s_{ab}) 
+ 8\,m^dD_d(\nu\,n^b\,m^a\,s_{ab}) 
+ 4\,p^c\,m^b\,m^a\,D_cs_{ab} 
\]
\[
+ 4\,\nu\,n^c\,m^b\,m^a\,D_cs_{ab} 
+ 8\,m^c\,p^b\,m^a\,D_cs_{ab} 
+ 8\,\nu\,m^c\,n^b\,m^a\,D_cs_{ab} 
\]
\[
+ \frac{1}{3}\,m^d\,m^a\,D_dD_aR
-  \frac{4}{3}\,\,p^a\,D_aR 
+ 8\,p^c\,p^d\,s_{cd}
- 6\,(s_{ab}\,n^a\,p^b)^2 
+ R\,\,s_{ab}\,n^a\,p^b
\]
\[
= 
n^dD_d\left\{n^cD_c(p^b \,p^a\,s_{ba}) \right\}  
+ 2\, n^cD_c(p^b\,p^a\,s_{ba})  
- 24\,p^b\,p^a\,s_{ba}  
\]
\[
- 4\, n^dD_d(m^c\,p^a\,m^b\,D_cs_{ba}) 
+ 4\,m^dD_d\left\{m^cD_c(m^b\,m^a\,s_{ab})\right\}   
\]
\[
+ 8\,m^d\,p^b\,m^a\,D_ds_{ab}
+ 4\,p^c\,m^b\,m^a\,D_cs_{ab} 
+ 4\,\nu\,n^cD_c(m^b\,m^a\,s_{ab}) 
\]
\[
+ \frac{1}{3}\,m^d\,m^a\,D_dD_aR
-  \frac{4}{3}\,\,p^a\,D_aR 
- 6\,(s_{ab}\,n^a\,p^b)^2 
+ R\,\,s_{ab}\,n^a\,p^b. 
\]
With
\[
4\,p^c\,m^b\,m^a\,D_cs_{ba} = 2\,p^c\,(- h^{ba} + p^b\,n^a\, + n^b\,p^a)\,D_cs_{ba} 
= 4\,p^c\,n^b\,p^a\,D_cs_{ba} 
\]
\[
= 4\,(h^{cb} -  n^c\,p^b  + 2\,m^c\,m^b)\,p^a\,D_cs_{ba} 
= \frac{2}{3}\,p^a\,D_aR  -  4\,n^c\,p^b\,p^a\,D_cs_{ba}   + 8\,m^c\,m^b\,p^a\,D_cs_{ba}, 
\]
\[
= \frac{2}{3}\,p^a\,D_aR  
-  4\,n^cD_c(p^b\,p^a\,s_{ba})  
+  16\,p^b\,p^a\,s_{ba}  
 + 8\,m^c\,m^b\,p^a\,D_cs_{ba}, 
\]
this implies
\[
p^d\,p^c\,D_dD_cs_{ab}\,n^a\,n^b 
- 8\,p^c\,n^b\,p^a\,D_cs_{ba}   
+ 8\,p^c\,p^d\,s_{cd}
\]
\[
= 
n^dD_d\left\{n^cD_c(p^b \,p^a\,s_{ba}) \right\}  
- 2\, n^cD_c(p^b\,p^a\,s_{ba})  
- 8\,p^b\,p^a\,s_{ba}  
- 4\, n^dD_d(m^c\,m^b\,p^a\,D_cs_{ba})
\]
\[ 
+ 16\,m^c\,p^b\,m^a\,D_cs_{ab}
+ 4\,m^dD_d\left\{m^cD_c(m^b\,m^a\,s_{ab})\right\}   
+ 4\,\nu\,n^cD_c(m^b\,m^a\,s_{ab}) 
\]
\[
+ \frac{1}{3}\,m^d\,m^a\,D_dD_aR
-  \frac{2}{3}\,\,p^a\,D_aR 
- 6\,(s_{ab}\,n^a\,p^b)^2 
+ R\,\,s_{ab}\,n^a\,p^b 
\]
\[
= 
n^dD_d\left\{n^cD_c(p^b \,p^a\,s_{ba}) \right\}  
- 2\, n^cD_c(p^b\,p^a\,s_{ba})  
- 8\,p^b\,p^a\,s_{ba}  
\]
\[
- 4\, n^dD_d(m^c\,m^b\,p^a\,D_cs_{ba}) 
+ 16\,m^c\,p^b\,m^a\,D_cs_{ab}
\]
\[
+ 4\,m^dD_d\left\{m^cD_c\left(\frac{1}{6}\,R - 8\,Q\right)\right\}   
+ 4\,\nu\,n^cD_c\left(\frac{1}{6}\,R - 8\,Q\right)
\]
\[
+ \frac{1}{3}\,m^b\,m^a\,D_bD_aR
-  \frac{2}{3}\,\,p^a\,D_aR 
- 6\,(s_{ab}\,n^a\,p^b)^2 
+ R\,\,s_{ab}\,n^a\,p^b 
\]
\[
= 
n^dD_d\left\{n^cD_c(p^b \,p^a\,s_{ba}) \right\}  
- 2\, n^cD_c(p^b\,p^a\,s_{ba})  
- 8\,p^b\,p^a\,s_{ba}  
\]
\[
- 4\, n^dD_d(m^c\,m^b\,p^a\,D_cs_{ba}) 
+ 16\,m^c\,p^b\,m^a\,D_cs_{ab}
+  \frac{2}{3}\,m^b\,m^a\,D_bD_a\,R
\]
\[   
-  \frac{2}{3}\,p^a\,D_a\,R   
-  \frac{2}{3}\,\nu\,n^a\,D_a\,R   
- 32\,m^b\,m^a\,D_bD_a\,Q 
+ 32\,p^a\,D_a\,Q 
\]
\[
+ 32\,\nu\,n^a\,D_a\,Q 
+ \frac{2}{3}\,\nu\,n^cD_c\,R 
- 2^6 \cdot \,\nu\,Q 
- 4\,\nu\,R
\]
\[
+ \frac{1}{3}\,m^b\,m^a\,D_bD_aR
-  \frac{2}{3}\,\,p^a\,D_aR 
- 6\,(s_{ab}\,n^a\,p^b)^2 
+ R\,\,s_{ab}\,n^a\,p^b,
\]
whence
\[
p^d\,p^c\,D_dD_cs_{ab}\,n^a\,n^b 
- 8\,p^c\,n^b\,p^a\,D_cs_{ba}   
+ 8\,p^c\,p^d\,s_{cd}
\]
\[
= 
n^dD_d\left\{n^cD_c(p^b \,p^a\,s_{ba}) \right\}  
- 2\, n^cD_c(p^b\,p^a\,s_{ba})  
- 8\,p^b\,p^a\,s_{ba}  
\]
\[
- 4\, n^dD_d(m^c\,m^b\,p^a\,D_cs_{ba}) 
+ 16\,m^c\,p^b\,m^a\,D_cs_{ab}
+ m^b\,m^a\,D_bD_aR
\]
\[
-  \frac{4}{3}\,\,p^aD_aR 
- 32\,m^b\,m^a\,D_bD_a\,Q 
+ 32\,p^aD_a\,Q 
+ 8\,R\,Q - 3 \cdot 2^7 \cdot Q^2.
\]
With this result equation (\ref{ppnnDDs}) takes the form
\begin{equation}
\label{B-ppnnDDs}
0 = n^dD_d\left\{n^cD_c(p^b \,p^a\,s_{ba}) \right\}  
- 2\, n^cD_c(p^b\,p^a\,s_{ba})  
- 8\,p^b\,p^a\,s_{ba}  
\end{equation}

\[
- 4\, n^dD_d(m^c\,m^b\,p^a\,D_cs_{ba}) 
+ 16\,m^c\,p^b\,m^a\,D_cs_{ab}
+ m^b\,m^a\,D_bD_aR
\]

\[
-  4\,p^aD_aR 
- 2^5\,(m^b\,m^a\,D_bD_a\,Q 
- p^aD_a\,Q  - R\,Q) 
+ 2^8 \cdot Q^2 = - M,
\]
with $M$ as given by (\ref{fin-M-expr}). \\
$\Box$

\subsection{Remarks on the $C^{\infty}$ case}
\label{smoothness of f}

Let $x^a$ denote $i$-centered $h$-normal coordinates. With the notation of (\ref{rho-expansion}) Proposition  \ref{f-is-analytic-if-B(bb)-vanishes-on-N} says that the field $f_{ab}$ defined by the relation $\,\mu^{-2}\,U^{4}\,\Sigma_{ab} = |x|^4\,f_{ab}$ is analytic if $h$ is analytic.  The discussion following equation (\ref{an2fequ}) shows that the expressions of the Taylor coefficients of $\,\mu^{-2}\,U^{4}\,\Sigma_{ab}$ in terms of $\mu$ and quantities derived from $h$ are independent of $h$ being smooth or analytic. By Taylor's theorem  one has thus  in the $C^{\infty}$ case for each $N \in \mathbb{N}$ a representation  
\[
\mu^{-2}\,U^{4}\,\Sigma_{ab} = |x|^4\,p^N_{ab} + R^{N+1}_{ab},
\]
where the  components of  $p^N_{ab}(x)$  are polynomials of order $N$ and the remainder term satisfies $R^{N+1}_{ab}(x) = O(|x|^{N+1})$ as $|x| \rightarrow 0$. The $p^N_{ab}$ define the partial sums of a formal power series which, in general, will not converge near $i$ and if it does it defines an analytic function which may not be related  to  
$\mu^{-2}\,U^{4}\,\Sigma_{ab}$ away from $i$.  By Borel's theorem (\cite{dieudonne:I}) there exists, however, a $C^{\infty}$  field $\hat{f}_{ab}$ near $i$ such that we have 
$\hat{f}_{ab} = p^N_{ab} + \bar{R}^{N+1}_{ab}$ for any $N$ with a smooth remainder term such that 
$\bar{R}^{N+1}_{ab}(x) = O(|x|^{N+1})$ as $|x| \rightarrow 0$. This implies that 
\[
\Sigma_{ab} = \rho^2\,f_{ab},
\]
with $f_{ab} = \hat{f}_{ab} + \check{f}_{ab}$,
where $ \check{f}_{ab}(0)  = 0$, $ \check{f}_{ab}(x) = |x|^{-4}\,R^{N+1}_{ab}  - \bar{R}^{N+1}_{ab}$ if $x \neq 0$
for arbitrary $N \in \mathbb{N}$. It follows  that $\check{f}_{ab}(x) = |x|^{-4}\,\tilde{f}_{ab}$ with a smooth function $\tilde{f}_{ab}$ that vanishes at $i$ together with its derivatives of any order. The same property follows then for 
$\check {f}_{ab}$, which implies that $f_{ab}$ is smooth.

\section{The overdeterminedness}
\label{j-subconds}

In the following we assume that the field $f_{ab}$ extends  smoothly to a neighbourhood  of $i$.  It remains to study  the problems arising from the overdeterminedness of the basic equation and the need to find $D_a\omega(i)$. It turns out that the analysis simplifies if we use instead of $f_{ab}$ the field
\begin{equation}
\label{def-t}
t_{ab} = f_{ab} + s_{ab}= \frac{1}{\rho^2}\,(D_aD_b\rho - s\,h_{ab} + \rho\,s_{ab}),
\end{equation}
which also is smooth and has  various important properties.
Equations  (\ref{sab-conf-trans}) and (\ref{fab-conf-transf}) show that $t_{ab}$ is 
as a conformal density of conformal weight $-1$,  
\begin{equation}
\label{t-conf-density}
t_{ab}[\vartheta^{-2}\,h] = \vartheta\,t_{ab}[h],
\end{equation}
and it follows from  (\ref{div-f}) and the Bianchi identity that, independent of the scaling of $h$, 
\begin{equation}
\label{div-tab=0}
D^a t_{ab} = 0.
\end{equation}
The tensors $t_{ab}$ and  $B_{ab} = \frac{1}{2}\,B_{acd}\,\epsilon_b\,^{cd}$ share these properties and they are in fact related by  (\ref{2Df-relations}), which can be written
\begin{equation}
\label{Bab-tab-relation}
B_{abc} = \rho\,D_{[b}t_{c]a} 
+ D^e\rho\,(t_{e[b}\,h_{c]a} + 2\,h_{e[b}\,t_{c]a}),
\end{equation}
or in space spinor notation
\begin{equation}
\label{spin-Bab-tab-relation}
b_{ABCD} = \rho\,D_A\,^H\,t_{BCDH} + 2\,D_{(A}\,^H\rho\,\,t_{BCD)H},
\end{equation}
where we write
\[
B_{ABCDEF} = - \frac{1}{2}\,(b_{ABCE}\,\epsilon_{DF} + b_{ABDF}\,\epsilon_{CE}),
\]
so that  
\begin{equation}
\label{B-s-rel}
b_{ABCE} = - B_{ABCFE}\,^F = D_{(A}\,^Hs_{BCE)H}.
\end{equation}
(The spinor $b_{ABCD}$ differs by a constant factor from the spinor  obtained by directly translating the frame form of $B_{ab} = \frac{1}{2}\,B_{acd}\,\epsilon_b\,^{cd}$ with the van der Waerden symbols into a spinor field). Another property which $t_{ab}$ has in common with $B_{ab}$ is the following.

\vspace{.3cm}

\noindent
{\it The metric  $h$ is locally conformally flat near $i$ if and only if 
$t_{ab}$ vanishes on some neighbourhood of $i$}.

\vspace{.3cm}

\noindent
In fact, it follows from  (\ref{Bab-tab-relation}) that $B_{abc}$ vanishes on open sets on which $t_{ab}$ vanishes. Conversely, on a sufficiently small neighbourhood of $i$ on which $B_{abc} = 0$ there exists a conformal gauge such that $h_{ab}$ is flat near $i$. Then $s_{ab} = 0$ there and   
also $f_{ab} = 0$  because the flat data are Schwarzschild data whence static  in the preferred conformal gauge. $\Box$

\vspace{.5cm}

The relation between $B_{abc}$ and $t_{ab}$ can be reversed at the point $i$ in the sense that equation  (\ref{Bab-tab-relation}) allows us to obtain an expression for $t_{ab}$ and its derivatives at $i$ in terms of derivatives of $B_{abc}$ and lower order terms at $i$. At the lowest orders this is seen as follows.
Taking a derivative of  (\ref{Bab-tab-relation}) at $i$ gives
\begin{equation}
\label{DB-t}
D_dB_{abc}(i) = - 2\,\mu\,(t_{d[b}\,h_{c]a} +2\,h_{d[b}\,t_{c]a}),
\end{equation}
whence
\begin{equation}
\label{tab-at-i}
 t_{ac}(i) = - \frac{1}{3\,\mu}\,D^bB_{abc}(i)
 \quad \mbox{resp.} \quad 
 t_{ABCD}(i) = \frac{1}{3\,\mu}\,D_A\,^HB_{BCDH}(i).
 \end{equation}
The relation  
\[
\left\{D_cB_{ba} +
\frac{2}{3}\,(D_{[a}B_{c]b} + D_{[b}B_{c]a})\right\}(i)= D_{(c}\,B_{ba)}(i) = D_{\{c}\,B_{ba\}}(i)  = 0,
\] 
implies that
\begin{equation}
\label{t-DB}
D_cB_{ba}(i) \neq 0 
\quad \mbox{iff} \quad D^bB_{abc}(i)  \neq 0
\quad \mbox{iff} \quad t_{ab}(i) \neq 0.
\end{equation}

\noindent
Taking a second derivative  of  (\ref{Bab-tab-relation}) at $i$ gives
\[
D_eD_dB_{abc}(i) 
\]
\[
= - 2\,\mu\,\{h_{ed}\,D_{[b}\,t_{c]a}
+ D_e\,(t_{d[b}\,h_{c]a} +2\,h_{d[b}\,t_{c]a}) 
+ D_d(t_{e[b}\,h_{c]a} +2\,h_{e[b}\,t_{c]a})\}.
\]
The only non-trivial contractions are 
 \begin{equation}
 \label{rot(t)}
D_fD^fB_{abc}(i) =   - 14\,\mu\,D_{[b}t_{c]a}(i),
\end{equation}
and
\[
D_eD_fB_a\,^f\,_c(i) = D_fD_eB_a\,^f\,_c(i) 
= - \mu\,(4\,D_{e}\,t_{ca} + 2\,D_{[e}\,t_{c]a} + 2\,D_{[e}\,t_{a]c}), 
\]
which implies
\begin{equation}
\label{Dt-DDB}
D_e\,t_{ac}(i) = - \frac{1}{4\,\mu}D_eD_fB_a\,^f\,_c
- \frac{1}{14\,\mu}\,D_fD^fB_{(ac)e}.
\end{equation}
Because 
\[
D^fB_{afc} 
= \frac{1}{2}\,D^fD_{f}s_{ca} 
-  \frac{1}{8}\,D_aD_c R 
- \frac{1}{2}\,s^h\,_cs_{ha} 
- \frac{1}{6}\,R\,s_{ac} 
+ \frac{1}{2}\,R^h\,_a\,^f\,_cs_{fh} 
+ \frac{1}{24}\,\Delta_hR\,h_{ac},
\] 
whence
\begin{equation}
\label{DB-props}
D_fB_{a}\,^f\,_{c} = D_fB_{(a}\,^f\,_{c)}, \quad 
D_fB_{h}\,^{fh} = 0, 
\end{equation}
the right hand side of (\ref{Dt-DDB}) reflects indeed  the algebraic properties on $t_{ab}$.

More generally, observing (\ref{sigmaval}) and the fact that the equation for $\rho$ implies 
that $D_aD_bD_c\rho(i) = 0$, it follows with  (\ref{Bab-tab-relation}) for $q \ge 2$ 
\begin{equation}
\label{gen-t-B-at-i-rel}
D_{d_{q + 2}} \ldots D_{d_1}B_{abc}(i) = 
\end{equation}
\[
= - 2\,\mu\left\{
\sum_{1 \le i < j \le q + 2} h_{d_i d_j}
D_{d_{q + 2}} \ldots \hat{D}_{d_j} \ldots  \hat{D}_{d_i} \ldots D_{d_1}
D_{[b}\,t_{c]a}
\right.
\]
\[
\quad \quad \quad \quad \left. +
\sum_{1 \le i \le q + 2}D_{d_{q + 2}} \ldots \hat{D}_{d_i} \ldots D_{d_1}
(t_{d_i[b}\,h_{c]a} + 2\,h_{d_i [b}\,t_{c]a})
\right\} +L_{q - 1}, 
\]
where $L_{q - 1}$ denotes an expression which is linear in the derivatives 
$D_{c_j} \ldots D_{c_1}\,t_{ab}(i)$, $ 0 \le j \le q - 1$, and linear  in the terms
$D_{e_l} \ldots D_{e_1}\,\rho(i)$, $ 0 \le l \le q + 3$. The structure of the term in curly brackets suggests that (\ref{tab-at-i}), (\ref{Dt-DDB})
and the equations above can be used  successively  to obtain formulas for
$D_{c_{q + 1}} \ldots D_{c_1}\,t_{ab}(i)$ in terms of a linear expression in 
$D_{d_{q + 2}} \ldots D_{d_1}B_{abc}(i)$ and terms in involving lower order derivatives of $B_{abc}$ and and derivatives of $\rho$ at $i$. We do not work out the details here.

\vspace{.3cm}

Equation (\ref{Bab-tab-relation})  implies furthermore   
\[
D^a\rho\,D^d\rho\,B_{ad} 
= \frac{1}{2}\,\rho\,D^a\rho\,D^d\rho\,D_{b}t_{ca}\,\epsilon_d\,^{bc},
\]
so that, if $h$ is real analytic,  the holomorphic extension satisfies
\[
D^a\rho\,D^d\rho\,B_{ad}|_{{\cal N}_i} = 0.
\]
{\it It follows that for real analytic metrics equation (\ref{Bab-tab-relation})  holds if and only if the metric satisfies condition (\ref{regcond}) with $p_* = \infty$}. We omit a  discussion of the smooth case.

In terms of  $t_{ab}$  the basic equation (\ref{add-equ}) assumes the form
\begin{equation}
\label{A-t-form-add-equ}
D_aD_b\omega - t\,h_{ab} + \omega\,s_{ab} = 
\omega^2\,t_{ab}[h].
\end{equation}
If we set 
\[
 \Upsilon_{ab}[ \omega, h]  = D_aD_b\omega - \frac{D_c\omega\,D^c\omega}{2\,\omega}\,h_{ab} + \omega\,L_{ab}[h] - \omega^2\,t_{ab}[h],
 \]
equations (\ref{B-omega-equ}) and (\ref{A-t-form-add-equ}) are combined in the relation 
\begin{equation}
\label{comb-crit-equ}
\Upsilon_{ab}[ \omega, h] = 0.
\end{equation}
Equation  (\ref{Sigma-Sigma'-Pi}) takes in terms of $t_{ab}$ the form 
\begin{equation}
\label{new-crit-form}
\Sigma_{ab}[h', \rho'] = - \frac{\rho^2}{\omega^3}\, \Upsilon_{ab}[h, \omega].
\end{equation}
This follows by using equations (\ref{an2fequ}), (\ref{n1fequ}), (\ref{def-t}) to write
$\Sigma_{ab}[h, \rho] = \rho^2\,(t_{ab} - s_{ab})$ and equations
(\ref{B-omega-equ}), (\ref{comb-crit-equ}) to obtain 
 (\ref{DDvan}) in the form $\hat{\Sigma}[h, \omega]  = \Upsilon_{ab}[h, \omega]  + \omega^2\,(t_{ab} - s_{ab})$.
Equation (\ref{comb-crit-equ}) implies a  conformal staticity criterion  in terms of the conformal density $t_{ab}$.

\begin{lemma}
\label{f=l-crit}
If the metric $h$ is analytic it is conformal to a static datum if and only if there exists a conformal factor satisfying  $\omega(i) = 1$ and 
\[
R_{ab}[\omega^{-2}\,h] = \omega\,t_{ab}[h].
\]
\end{lemma}

\noindent
{\bf Proof}:
As an immediate consequence of  the general transformation law
\[
L_{ab}[\omega^{-2}\,h] = L_{ab}[h] + \omega^{-1}\,D_aD_b\omega
- \frac{1}{2}\omega^{-2}\,D_c\omega\,D^c\omega\,h_{ab},
\]
of the Schouten tensor and the behaviour (\ref{t-conf-density}) of $t_{ab}$ the relation (\ref{comb-crit-equ}) is seen to be equivalent to 
$\omega\,t_{ab}[h] = t_{ab}[\omega^{-2}\,h] = L_{ab}[\omega^{-2}\,h]$.\\
$\Box$
 
\vspace{.3cm}

If $h$ were real analytic and there existed a solution to (\ref{comb-crit-equ}) with $\omega(i) = 1$ and some value of the differential $D_a\omega(i)$, it would then be given in
$i$-centered normal coordinates $x^a$ by the function 
\[
\hat{\omega} = 1 + \sum_{p \ge 1} \frac{1}{p!}\,x^{a_p} \ldots x^{a_1}\,
D_{a_p} \ldots D_{a_1}\omega(i),
\]
where the $D_a$ denote covariant derivatives in the directions of an orthonormal frame which is parallely propagated along the geodesics through $i$ and  the covariant derivatives of $\omega$ are obtained sucessively by taking formal derivatives of equation (\ref{comb-crit-equ}) and restricting to  $i$. 
One might think of using this procedure to construct a solution. The function $\hat{\omega}$ is determined, however,  by the symmetrized coefficients  $D_{(a_p} \ldots D_{a_1)}\omega(i)$ only and without further information it is not clear whether the covariant derivatives $D_{a_p} \ldots D_{a_1}\hat{\omega}(i)$ coincide with the coefficients 
$D_{a_p} \ldots D_{a_1}\omega(i)$ obtained by taking formal derivatives of  (\ref{comb-crit-equ}). This corresponds to the fact that apart from the choice of the scaling factor and its differential at a given point a conformal gauge is fixed in general uniquely at all orders at  that point  by 
 imposing conditions on the {\it symmetrized} covariant derivatives of the Ricci tensor (\cite{fefferman:graham}, \cite{lee:parker}) or the Schouten tensor (\cite{gover:2001}) at that point.
Whether the assumption (\ref{regcond}) with $p_* = \infty$ allows us to say more is not clear and we proceed along different lines.

With  the expression  
$\Upsilon_{ab}$ in (\ref{comb-crit-equ}) we get
\[
D_c\Upsilon_{ab} 
= D_cD_aD_b\omega 
- \omega^{-1}\,\Upsilon_{cd}\,D^d\omega\,h_{ab}  
+ D^d\omega\,( L_{cd}\,h_{ab}  + h_{dc}\,L_{ab})
\]
\[
+ \omega\,D_cL_{ab} - \omega^2\,D_ct_{ab}
- \omega\,D^d\omega\,(t_{dc}\,h_{ab} 
+ 2\,h_{dc}\,t_{ab}),
\]
whence
\begin{equation}
\label{rot-Pi-rel}
 \omega^{-1}\,D_{[c}\Upsilon_{a]b}  
+ \omega^{-2}\,D^d\omega\,\Upsilon_{d[c}\,h_{a]b}  
\end{equation}
\[
= B_{bca}  - \omega\,D_{[c}t_{a]b}
- D^d\omega\,(t_{d[c}\,h_{a]b} 
+2\,h_{d[c}\,t_{a]b}).
\]
It follows that equation (\ref{comb-crit-equ}) can only hold if the function $\omega$ satisfies the 
 {\it compatibility condition}

\begin{equation}
\label{comp-cond}
B_{bca}  = \omega\,D_{[c}t_{a]b}
+ D^d\omega\,(t_{d[c}\,h_{a]b} 
+2\,h_{d[c}\,t_{a]b}).
\end{equation}

\vspace{.3cm}

The relations (\ref{tab-at-i}), (\ref{Dt-DDB}), (\ref{gen-t-B-at-i-rel}) and equation
(\ref{comp-cond}) suggest that the requirement of conformal staticity induces restrictions on the conformal structure of the metric $h$ in terms of
a sequence of {\it differential relations  on the Cotton tensor} at the point $i$, whose lowest order member would be given by 
 \[
B_{bca} = - \frac{1}{14\,\mu}\,\omega\,D_fD^fB_{bca}
- \frac{1}{3\,\mu}\,D^d\omega(D^fB_{df[c}\,h_{a]b} + 2\,h_{d[c}\,D^fB_{a]fb})
\quad \mbox{at} \quad i.
\]
These relations involve besides  the expansion coefficients of the Cotton tensor also those of $\omega$, however,  and one would have to determine those in accordance with 
(\ref{comb-crit-equ}) and (\ref{comp-cond}) to obtain conditions expressed entirely in terms of $h$ and its derived structures.

 We shall concentrate instead on analyzing equation (\ref{comp-cond}), considered  as a differential equation for $\omega$. We note that this equation is  implicit, highly overdetermined and a priori a solution to it need not even satisfy the equation
$\Upsilon_{ab} = 0$ because (\ref{comp-cond}) and (\ref{rot-Pi-rel}) only  imply
\[
 \omega\,D_{[c}\Upsilon_{a]b}  + D^d\omega\,\Upsilon_{d[c}\,h_{a]b} = 0. 
\]

\subsection{Analysis of the compatibility condition.} 

Using (\ref{Bab-tab-relation}) we can rewrite (\ref{comp-cond}) near $i$  in the form
\begin{equation}
\label{B-comp-cond}
\zeta^e\,(t_{e[c}\,h_{a]b} 
+2\,h_{e[c}\,t_{a]b}) = - D_{[c}t_{a]b},
\end{equation}
with 
\[
\zeta_a = D_a\chi, \quad \chi = \log(\omega - \rho).
\]
We need to understand now the conditions under which equation (\ref{B-comp-cond}) can be solved for a smooth 1-form 
$\zeta_a$, the  conditions which ensure this 1-form to be  closed so that we can write 
$\zeta_a = D_a\chi$ with a function $\chi$ satisfying $\chi(i) = 1$, and finally 
whether the function $\omega = e^{\chi} + \rho$ so obtained does indeed satisfy equation (\ref{comb-crit-equ}).
We shall discuss these questions in two different ways. The first method, which imposes right at the beginning a  non-degeneracy condition, gives concise expressions and illustrates the overall argument. The second method, discussed in section \ref{det-on-stat-crit} gives  more detailed information but also requires more detailed information on the underlying structure.

Because $t_{ab}$ is symmetric there exist at each point orthogonal eigenvectors $\xi^a$, $\lambda^a$, $\rho^a$ 
with corresponding real eigenvalues $\alpha$, $\beta$ $\gamma$ (satisfying $\alpha + \beta  + \gamma = 0$ because
$t_a\,^a = 0$) so that
\[
t^a\,_b\,\xi^b = \alpha\,\xi^a, \quad 
t^a\,_b\,\lambda^b = \beta\,\lambda^a, \quad 
t^a\,_b\,\rho^b = \gamma\,\rho^a.
\]
We  assume that 
\begin{equation}
\label{simple-ev's}
\alpha \neq \beta \neq \gamma \neq \alpha,
\end{equation}
at the point $i$. This condition will then also be satisfied and 
$t_{ab}\,t^{ab} = \alpha^2 + \beta^2 + \gamma^2 \neq 0$
on some neighbourhood $U$ of $i$. 
This requirement  is not particularly restrictive. By (\ref{t-DB}) it reduces to the condition that the derivative $D^bB_{abc}$ of the Cotton tensor symmetrizes with three different eigenvalues at $i$.  As discussed in more detail below, the assumption (\ref{simple-ev's}) excludes in particular  the situations in which the map of static data onto their conformal classes is
not $1 : 1$. Moreover, it fixes our problem uniquely. 

On $U$ the equation resulting from the  contraction of (\ref{B-comp-cond}) with $t^{ab}$ can then be written 
\[
D_a\chi\,(\delta^a\,_b - T^a\,_b) 
= - \frac{t^{cd}\,D_{[b}t_{c]d}}{t_{ef}\,t^{ef}\, } \quad \mbox{with}
\quad T^a\,_b = \frac{3}{2}\,\frac{t^a\,_c\,t^c\,_b}{t_{de}\,t^{de}}.
\]
We note that these three equations are not necessarily equivalent to the original five equations.
Any solution to (\ref{B-comp-cond})  will solve the equation above but the latter may admit solutions $\chi$ which do not solve  (\ref{B-comp-cond}).

The tensor $T^a\,_b$, with the given index position, is conformally invariant,
\[
T^a\,_b[\vartheta^{-2}\,h] = T^a\,_b[h].
\] 
The matrix $T = (T^a\,_b)$ is symmetric with respect to the symmetric form defined by $h$ and
with the notation above we have  
\[
T^a\,_b\,\xi^a = u\,\xi^a, \quad
T^a\,_b\,\lambda^a = v\,\lambda^a, \quad
T^a\,_b\,\rho^a = w\,\rho^a, 
\]
with 
\[
u = \frac{3\,\alpha^2}{2\,(\alpha^2 + \beta^2 + \gamma^2)}, \quad
v = \frac{3\,\beta^2}{2\,(\alpha^2 + \beta^2 + \gamma^2)} \quad
w = \frac{3\,\gamma^2}{2\,(\alpha^2 + \beta^2 + \gamma^2)}. 
\]
The eigenvalues $u$, $v$, $w$ are independent of the scaling of $h$.
The relations
\[
1 - u = \frac{(\beta - \gamma)^2}{2\,(\alpha^2 + \beta^2 + \gamma^2)}, \quad 
1 - v = \frac{(\gamma - \alpha)^2}{2\,(\alpha^2 + \beta^2 + \gamma^2)}, \quad 
1 - w =  \frac{(\alpha - \beta)^2}{2\,(\alpha^2 + \beta^2 + \gamma^2)},
\]
imply  
\[
0 \le u, \,v, \,w < 1, \quad \mbox{if} \quad \alpha \neq \beta \neq \gamma \neq \alpha,
\]
\[
u = 1, \quad 0 < v, \, w < 1  \quad \mbox{if} \quad \beta = \gamma = - \frac{1}{2}\,\alpha \neq 0.
\]
In the first case, which is considered here,   the matrix $1 - T$ is invertible on $U$ with an inverse whose entries are given by 
\begin{equation}
\label{M-def}
M^a\,_b = (1 + \sum_{k = 1}^{\infty} T^k)^a\,_b,
\end{equation}
where the series is normally convergent with respect to the operator norm implied by the standard Euclidean norm on $\mathbb{R}^3$. The invertibility being given, it follows from Cramer's rule that  the functions $M^a\,_b$  and thus the components of $\zeta_a$ are smooth resp. real analytic on $U$ if $h$ is.
Our equation can thus be written
\begin{equation}
\label{t-contr-B-comp-cond} 
D_a\chi = - \frac{t^{cd}\,D_{[b}t_{c]d}}{t_{ef}\,t^{ef}\, }\,M^b\,_a.
\end{equation}

\subsection{Conformal staticity and asymptotic conformal \\ staticity criteria.} 

The relation above gives rise to a criterion which characterizes initial data  satisfying the requirement (iv') of  Definition \ref{as-stat}.

\begin{theorem}
\label{conf-stat-cond}
Suppose the metric $h_{ab}$ is real analytic near $i$ and the dual $B_{ab}$ of its Cotton tensor satisfies at the point $i$ the condition 
\begin{equation}
\label{X-B-regcond}
D_{\{a_1} \cdots D_{a_p}\,B_{ab\}}(i) = 0 \quad \mbox{for} \quad
p = 0, 1, 2, 3, \ldots \,\,.
\end{equation}
Suppose furthermore that $h$ is generic in the sense that  the map  associated with the conformal density $t_{ab}$  defined by (\ref{def-t}) has three simple eigenvalues at $i$.
Then $h$ is conformal to asymptotically flat  
static vacuum data near $i$  if and only if 

\vspace{.1cm}

\noindent
(i) the 1-form
\[
\kappa = \kappa_a\,dx^a \quad \mbox{with} \quad \kappa_a = \frac{t^{cd}\,D_{[b}t_{c]d}}{t_{ef}\,t^{ef}\, }\,M^b\,_a,
\]
\hspace*{.6cm}with $M^a\,_b$ given by (\ref{M-def}), is closed, i.e.
\begin{equation}
\label{B-first-restr-on-h}
D_{[a}\kappa_{b]} = 0,
\end{equation}
\hspace*{.6cm}and

\vspace{.1cm}

\noindent
(ii) the integral $\chi$ defined by 
\begin{equation}
\label{int-chi}
D_a \chi  = - \kappa_a, \quad \chi(i) = 0,
\end{equation}
\hspace*{.7cm}satisfies the equation
\begin{equation}
\label{X-critical-equ}
0 = D_aD_b\chi + D_a\chi \,D_b\chi + L_{ab} -  (e^{\chi} + 2\,\rho)\,t_{ab}
\quad \quad \quad \quad \quad \quad \quad \quad \quad 
\end{equation} 
\[
\quad \quad \quad \quad \quad \quad 
+ \frac{1}{e^{\chi} + \rho}
\left\{
\frac{1}{3}\,\Delta_h\rho + \frac{1}{12}\,\rho\,R
- \frac{1}{2}\,e^{\chi}\,D_c\chi\,D^c\chi
- D_c\chi\,D^c\rho
\right\}h_{ab}.
\]

\end{theorem}

\vspace{.2cm}

\noindent
{\bf Proof:} 
Equation (\ref{t-contr-B-comp-cond}) implies that  (\ref{B-first-restr-on-h}) must be required. This being satisfied we can integrate $\chi$ and set
$\omega= e^{\chi} + \rho$ so that $\omega(i) = 1$.
Using then the equation
\begin{equation}
\label{combined-rho-equs}
D_aD_b\rho - \frac{D_c\rho\,D^c\rho}{2\,\rho}\,h_{ab} + \rho\,L_{ab} - \rho^2\,t_{ab} = 0,
\end{equation}
which combines (\ref{an2fequ}) (where $\check{\rho} = 0$ by the analyticity assumption) and (\ref{def-t}), a direct calculation shows
that the function $\omega$ does in fact satisfy the critical equation in the form
(\ref{comb-crit-equ}) and, as a consequence, the compatibility condition 
(\ref{comp-cond}). This calculation also shows that (\ref{X-critical-equ}) must be required because the right hand  side of (\ref{X-critical-equ}) is just a rewrite of 
$e^{-\chi}\,\Upsilon_{ab}$  in terms of $\chi$.\\
$\Box$

\vspace{.2cm}

It is not clear to what extent conditions (\ref{X-B-regcond}) and (\ref{B-first-restr-on-h}) are independent of each other. This requires further analysis.
Similarly, it is not clear whether the conformally invariant conditions (\ref{X-B-regcond}) and  (\ref{B-first-restr-on-h}) suffice 
to characterize $h$ as conformally static vacuum data so that  (\ref{X-critical-equ}) would just be a consequence.

If (\ref{B-first-restr-on-h}) holds, there remains no freedom. The problem of 
{\it solving}  the overdetermined non-linear PDE  (\ref{comb-crit-equ}) for $\omega$ is replaced here, however,  by  the 
problem of  {\it integrating} the linear equation (\ref{int-chi}) for $\chi$ along the geodesics through the point $i$, and to {\it checking} whether $\chi$ does indeed solve (\ref{X-critical-equ}).
The integration could be avoided if equation  (\ref{X-critical-equ}) could be expressed directly  in terms of $\kappa_a$. This equation contains, however,  $\chi$ explicitly and this makes sense because only one of the potentials for $\kappa_a$ can  possibly represent  the desired solution $\omega$ to  (\ref{comb-crit-equ}).

There is a further reason which suggest that a check of  (\ref{comb-crit-equ}) is needed.
With the assumption (\ref{simple-ev's}) equation (\ref{comp-cond}) determines $D_a\omega(i)$ uniquely {\it  if there exits a solution at all} and in that case it must hold 
$D_a\omega(i) = D_a\chi(i)$ with the differential of $\chi$ determined (uniquely) by  (\ref{t-contr-B-comp-cond}). However, because the latter equation is obtained by a contraction of 
(\ref{B-comp-cond}) with $t^{ab}$ it is a priori not clear that a solution to 
(\ref{t-contr-B-comp-cond}) provides in fact  a solution to  (\ref{comp-cond}) resp.
(\ref{comb-crit-equ}).

\vspace{.2cm}

 We note that if $h$ is conformally static and given  in the conformal gauge in which  $t_{ab} = s_{ab}$ a direct calculation using equation (\ref{Xi-def}) with $ \Xi_{bca} = 0$ gives 
in fact $\kappa_a = - D_a(\log(1 - \rho))$ so that $\omega = 1$, as to be expected.

\vspace{.2cm}

The field  $\kappa_a$ does not satisfy a homogeneous transformation law under conformal rescalings, it holds 
\[
\kappa_a[\vartheta^{-2} h] = \kappa_a[h] +D_a (\log \vartheta).
\]
With $\vartheta(i) = 1$ this gives the  expected transformation behaviour
\[
\omega[\vartheta^{-2} h] = \vartheta^{-1} \, \omega[h].
\]
The transformation law of $\kappa_a$ implies
\[
D_{[a}\kappa_{b]}[\vartheta^{-2} h]  = D_{[a}\kappa_{b]}[h].
\]
{\it This  conformal invariance shows  that (\ref{B-first-restr-on-h}) imposes in fact only a restriction  on the conformal structure of $h$}. 

\vspace{.2cm}

To simplify the criterion and to understand the nature of its conditions  it would be desirable to obtain a simple expression for $D_{[a}\kappa_{b]}$. Writing 
\[
L_b = \frac{t^{cd}\,D_{[b}t_{c]d}}{t_{ef}\,t^{ef}\, },
\]
and contracting $D_{[a}\kappa_{b]}$ twice with $\delta^a\,_b - T^a\,_b$ we obtain 
(\ref{B-first-restr-on-h}) in the form
\[
0 = D_{[a}L_{b]} - T_{c[a}\,D^cL_{b]}
+ L_c\,M^c\,_d\,(D_{[a}\,T^d\,_{b]} - T_{e[a}D^e\,T^d\,_{b]}).
\]
This must be understood as a {\it differential relation for $t_{ab}$} because the field $L_a$ reads more explicitly
\[
L_a = \frac{1}{4}\,D_a \log(t_{cd}\,t^{cd})
-  \frac{1}{3}\,D_e \log(t_{cd}\,t^{cd})\,T^e\,_a - \frac{1}{3}\,D_e\,T^e\,_a,
\]
and there is no way to express $t_{cd}\,t^{cd}$ in terms of $T^a\,_b$.

\vspace{.3cm}

In the case of $C^{\infty}$- data an analysis related to the one above supplies a characterization of  data that satisfy the requirements of part  (ii') of Definition \ref{as-stat}.

\begin{theorem}
\label{as-conf-stat-cond}
Suppose the metric $h_{ab}$ is smooth near $i$, its Cotton tensor satisfies at the point $i$  condition (\ref{X-B-regcond}) and $h$ is generic in the sense that  the map  associated with the conformal density $t_{ab}$  defined by (\ref{def-t}) has three simple eigenvalues at $i$. Then the metric $h$ is conformal to vacuum data which are weakly asymptotically static,
if  for all $k, j \in \mathbb{N}$:   

\vspace{.1cm}

\noindent
(i) the 1-form $\kappa$ of Theorem \ref{conf-stat-cond} satisfies
\begin{equation}
\label{O-B-first-restr-on-h}
D_{[a}\kappa_{b]} = O(|x|^k), 
\end{equation}

\noindent
(ii) the $C^{\infty}$ function $\chi$ which is given near $i$ in $i$-centered, $h$-normal coordinates $x^a$ by
\begin{equation}
\label{O-chi-expression}
\chi(x) = - \int_0^1 f(\tau x) \,\frac{d \tau}{\tau} \quad \mbox{with}\quad f(x) = x^a\,\kappa_a(x),
\end{equation}
\hspace*{.7cm}satisfies
\begin{equation}
\label{O-X-critical-equ}
D_aD_b\chi + D_a\chi \,D_b\chi + L_{ab} -  (e^{\chi} + 2\,\rho)\,t_{ab}
\quad \quad \quad \quad \quad \quad \quad \quad \quad 
 \quad \quad \quad \quad \quad 
\end{equation} 
\[
\quad \quad \quad \,
+ \frac{1}{e^{\chi} + \rho}
\left\{
\frac{1}{3}\,\Delta_h\rho + \frac{1}{12}\,\rho\,R
- \frac{1}{2}\,e^{\chi}\,D_c\chi\,D^c\chi
- D_c\chi\,D^c\rho
\right\}h_{ab} = O(|x|^j),
\]
\hspace*{.7cm}as $|x| \rightarrow 0$

\vspace{.1cm}

\end{theorem}

\vspace{.2cm}

\noindent
{\bf Proof}: For any $C^{\infty}$ function $f(x)$ defined for $x^a$ close to $x^a(i) = 0$ 
the notation $f = {\cal O}^{\infty}_{x_*}$ will mean in the following that 
$f(x) = O(|x - x_*|^k)$ as $x^a \rightarrow x^a_*$ for all $k \in \mathbb{N}$.

By equation (\ref{O-chi-expression}) the function $\chi$ satisfies $\chi(i) = 0$ and for $|\tau|$ and $|x|$ sufficiently small 
\[
0 = - 2\,\left(\frac{d}{d \tau}\chi(\tau\,x) + x^a\kappa_a(\tau\,x)\right) =
D^a\Gamma\,(D_a\chi + \kappa_a)(\tau\,x).
\]
With (\ref{O-B-first-restr-on-h}) this allows us to derive the relation
\[
D_bD^a\Gamma\,(D_a\chi + \kappa_a) + 
D^a\Gamma\,D_a\,(D_b\chi + \kappa_b)  = {\cal O}^{\infty}_{0}.
\]
Restricting the left hand side and its derivatives of higher order to $i$ thus  implies
\[
D_a\chi + \kappa_a =  {\cal O}^{\infty}_{0},
\]
so that the necessary condition (\ref{t-contr-B-comp-cond}) for conformal staticity is  satisfied asymptotically.

In terms of $\omega = e^{\chi} + \rho$ relation (\ref{O-X-critical-equ}) reads 
$e^{-\chi}\,\Upsilon[\omega, h] = {\cal O}^{\infty}_{0}$ which is equivalent to the asymptotic versions 
\[
2\,\omega\,t - D_a\omega\,D^a\omega + \frac{1}{6}\,R[h]\omega^2 = {\cal O}^{\infty}_{0},
\quad \quad
D_aD_b\omega - t\,h_{ab} + \omega\,s_{ab} - \omega^2t_{ab}
= {\cal O}^{\infty}_{0},
\]
of (\ref{B-omega-equ}) and (\ref{add-equ}) resp. (\ref{A-t-form-add-equ}). 

\vspace{.1cm}

\noindent
{\it Because no particular conformal gauge has be employed so far, we can now conveniently assume that the scaling of $h$ has been chosen  such that 
$\omega - 1 = {\cal O}^{\infty}_{0}$}.

\vspace{.1cm}

\noindent
The equations above then reduce to
\[
R[h] = {\cal O}^{\infty}_{0}, \quad \quad \Sigma_{ab}[h, \rho] = {\cal O}^{\infty}_{0},
\]
and imply by (\ref{dual-Xi-def}) the relation $\Xi_{ab} =  {\cal O}^{\infty}_{0}$,
which reads in terms of space spinors as a relation  for the trace free part $s_{ABCD}$ of the Ricci spinor
\[
(1 - \rho)\,D_A\,^F\,s_{BCDF} - 2\,s_{F(ABC}\,D_{D)}\,^F \rho =  {\cal O}^{\infty}_{0}.
\]
The following conclusions follow from the analysis of \cite{friedrich:statconv}.
 For given $j \in \mathbb{N}$, $j \ge 2$, the relation above allows us to express
 the covariant derivatives of $s_{ABCD}$ up to order $j$, and thus the derivatives of the fields $h_{ab}$ and $\rho$ up to order $j+2$, uniquely  in terms of the {\it formal null data} 
\[ 
\psi_{ABCD} = s_{ABCD}(i), \,\,\,\psi_{A_nB_n} \ldots _{A_1B_1ABCD} =
D_{(A_nB_n} \ldots D_{A_1B_1} s_{ABCD)}(i), 
\,\,\, n = 1, 2, \ldots j.
\]
If these data satisfy a certain decay condition as $j \rightarrow \infty$, the metric is real analytic, represents in fact a (conformal) static vacuum field with mass $m$ near $i$, and  there is nothing further to show. If the decay condition is not satisfied we argue as follows. The null data above can be complemented  (in a rather arbitrary way) by symmetric spinors
\[ 
\psi_{A_nB_n} \ldots _{A_1B_1ABCD}, 
\quad j + 1 \le n < \infty,
\]
such that the decay condition  is satisfied. The complete set of null data then defines a (conformal) static vacuum field $h^*_{ab}$ with mass $m$ near $i$ so that 
the $h^*$-covariant derivatives of $s^*_{ABCD}$  of order $\le j$ coincide at $i$ with the $h$-covariant derivatives of $s_{ABCD}$ at $i$ and 
\[ 
\psi_{A_nB_n} \ldots _{A_1B_1ABCD} =
D^*_{(A_nB_n} \ldots D^*_{A_1B_1} s^*_{ABCD)}(i), 
\quad j + 1 \le n < \infty.
\]
It follows that $h_{ab} - h^*_{ab}$ and $\rho - \rho^*$ vanish at orders $\le j + 2$  at $i$ and the fields (cf. (\ref{conf-vac-transition}))
\[
\tilde{h}_{ab} = \frac{\mu^2\,(1 + \sqrt{\rho})^4}{\rho^2}\,h_{ab}, \quad \quad 
v_* = \frac{1 - \sqrt{\rho_*}}{1 + \sqrt{\rho_*}} \quad \mbox{and} \quad
\tilde{h^*}_{ab} =  \frac{\mu^2\,(1 + \sqrt{\rho_*})^4}{\rho_*^2}\,h^*_{ab},
\]
satisfy $v^* \rightarrow 1$ as $|z| \rightarrow \infty$ and 
\[
R_{ab}[\tilde{h^*}] - \frac{1}{v_*}\,\tilde{D^*}_a\tilde{D^*}_b v_* = 0,
\quad \quad 
\Delta_{\tilde{h^*}} v^* = 0, \quad 
|h_{ab} - h^*_{ab}| = O    \left(   \frac{1}{|z|^{j + 2}}     \right)
 \quad \mbox{as} \quad |z| \rightarrow \infty,
\]
where the fields on the left hand side are given in the coordinates
$z^a = \frac{x^a}{|x|^2}$ in  which $\tilde{h}_{ab}$, $\tilde{h^*}_{ab}$ explicitly  satisfy the asymptotic flatness condition.
$\Box$

\subsection{Some facts underlying the  staticity criterion.} 
\label{det-on-stat-crit}

 To exhibit some of the conditions on the conformal structure implicit in   (\ref{B-first-restr-on-h}),
 to understand better the role of the non-degeneracy condition (\ref{simple-ev's}), and to derive a more explicit form of (\ref{B-first-restr-on-h}) we discuss the relevant equations now in more detail.
It will be convenient to write (\ref{B-comp-cond}) in terms of space spinors.
In view of  (\ref{div-tab=0}) it  reads then 
\begin{equation}
\label{spin-B-comp-cond}
2\,\zeta_{(A}\,^Ht_{BCD)H} = - D_A\,^H\,t_{BCDH},
\end{equation}
with 
\[
\zeta_{AB} = D_{AB} \chi    , \quad \chi = \log(\omega - \rho).
\]
To analyse the pointwise restrictions induced by this equation on the derivatives of $t_{ABCD}$
we consider at a given point a covector  $\zeta_{AB} \neq 0$ and analyse the map 
$t_{ABCD}  \rightarrow \zeta_{(A}\,^H t_{BCD)H}$ acting on symmetric spinors.
It holds
\begin{equation}
\label{zeta-t-sym}
 \zeta_{(A}\,^H t_{BCD)H} = \frac{1}{4}\,(
  \zeta_{A}\,^H t_{BCDH} +
 \zeta_{D}\,^H t_{ABCH} +
 \zeta_{C}\,^H t_{DABH} +
 \zeta_{B}\,^H t_{CDAH})
 \end{equation}
\[
=  \zeta_{A}\,^H t_{BCDH}  + \frac{3}{4}\,\epsilon_{A(B}\,t_{CD)HK}\,\zeta^{HK}.
\]
Thus
\begin{equation}
\label{A-zeta-t-sym=0}
 \zeta_{(A}\,^H t_{BCD)H} = 0,
 \end{equation}
if and only if 
\[
\zeta_{A}\,^H t_{BCDH}  = - \frac{3}{4}\,\epsilon_{A(B}\,t_{CD)HK}\,\zeta^{HK}.
\]
With  
$\zeta_{AB}\,\zeta^{CB} = \frac{1}{2}\,\zeta_{HK}\,\zeta^{HK}\, \epsilon_A\,^C$,
this is seen to be equivalent to 
\begin{equation}
\label{B-zeta-t-sym=0}
 \frac{1}{2}\,\zeta_{HK}\,\zeta^{HK}\,t_{BCDE} =
- \zeta^A\,_E \,\zeta_{A}\,^H t_{BCDH}  = \frac{3}{4}\, \zeta_{E(B}\,t_{CD)HK}\,\zeta^{HK}.
 \end{equation}
This implies
\[
0 =  \zeta_{HK}\,\zeta^{HK}\,t_{ECD}\,^E = \zeta^E\,_{(C}\,t_{D)EHK}\,\zeta^{HK},
\]
which  is equivalent to the existence of a real factor $f$ such that 
\[
 \zeta^E\,_{C}\,t_{DEHK}\,\zeta^{HK} = \zeta_{HK}\,\zeta^{HK}\,f\,\epsilon_{CD},
 \]
whence
\[
 t_{ABHK}\,\zeta^{HK} = - 2\,f\,\zeta_{AB}.
 \]
Observing this in (\ref{B-zeta-t-sym=0}) we get the representation
\begin{equation}
\label{zeta-map-kernel}
t_{BCDE} =
 - \frac{3\,f}{\zeta_{HK}\,\zeta^{HK}}\, \zeta_{(EB}\,\zeta_{CD)},
\end{equation}
which implies in turn (\ref{A-zeta-t-sym=0}).

Let  $\lambda_{AB} = \lambda_{(AB)}  \neq 0$ be a real spinor field with $\zeta_{AB}\,\lambda^{AB} = 0$. 
Then 
$\zeta_{BA}\,\lambda_C\,^{B}\, = \zeta_{B(A}\,\lambda_{C)}\,^{B}$ whence
\[
t_{BCDE}\,\lambda^{DE} =
 \frac{f}{\zeta_{HK}\,\zeta^{HK}}\, 
\, (\zeta^E\,_{B}\,\zeta_{CD}\,\lambda_E\,^{D} 
+ \zeta^E\,_{C}\,\zeta_{DB}\,\lambda_E\,^{D} )
\]
\[
=
 \frac{f}{\zeta_{HK}\,\zeta^{HK}}\, 
\, (\zeta^E\,_{B}\,\zeta_{ED}\,\lambda_C\,^{D} 
+ \zeta^E\,_{C}\,\zeta_{DE}\,\lambda_B\,^{D} )
= f\,\lambda_{BC}.
\]

\vspace{.2cm}

\noindent
{\it It follows that $t_{ABCD} \neq 0$ satisfies (\ref{A-zeta-t-sym=0}) if and only if it is of the form 
(\ref{zeta-map-kernel}) so that the map $\xi^{AB} \rightarrow t^{AB}\,_{CD}\,\xi^{CD}$ has a simple  eigenvalue $\alpha = - 2\,f \neq 0$ with eigenvector $\zeta^{AB}$ and  eigenvalues $\beta = \gamma = f$  with eigenvectors  orthogonal to $\zeta^{AB}$}.

\vspace{.2cm}

Consider now (\ref{spin-B-comp-cond}) as an equation for the covector $\zeta_{AB}$
with $t_{ABCD}$  and $D_{A}\,^H\,t_{BCDH}$ given at a fixed point $q$.
Depending on the (real) eigenvalues $\alpha$, $\beta$, $\gamma$ of  the map $\xi^{AB} \rightarrow t^{AB}\,_{CD}(p)\,\xi^{CD}$ three different cases can occur. 

\vspace{.1cm}

\noindent
$-$ The map has no simple eigenvalue. Because $t_{ab}$ is trace free his happens only if 
$t_{ab}(q) = 0$.  The formulas (\ref{tab-at-i}), (\ref{Dt-DDB}), (\ref{gen-t-B-at-i-rel}) and the subsequent discussion show that this case requires a quite detailed analysis of situations in which $t_{ab}$ and the Cotton tensor vanish up to some given orders $p$ resp. $p + 1$  at $i$.  The discussion in \cite{friedrich:statconv} shows that there do in fact exist conformally static data for which  $p$ is arbitrarily large. We shall not consider this case any further here.

\vspace{.1cm}

\noindent
$-$ $t_{ab}(q) \neq 0$ and the map has eigenvalues  
\begin{equation}
\label{double-eigenvalue}
\beta =  \gamma = - \frac{1}{2} \alpha = f \neq 0.
\end{equation}
Assume that the covectors $\xi^*_{AB}$ and $\xi_{AB}$ 
solve $2\,\xi_{(A}\,^H\,\,t_{BCD)H} = - D_A\,^H\,t_{BCDH}$ at $q$.
Then $\zeta_{AB} =  \xi^*_{AB} - \xi_{AB}$ satisfies (\ref{A-zeta-t-sym=0}) and we can have the situation where  $\zeta_{AB} \neq 0$ and thus 
$\zeta^{AB}\,t_{ABCD} = \alpha\,\zeta_{BC}$. All solutions to
$2\,\xi_{(A}\,^H\,\,t_{BCD)H} = - D_A\,^H\,t_{BCDH}$ at the point $q$
are then  of the form $\xi_{AB} = c\,\zeta_{AB} + \xi^*_{AB}$  
where  $\xi^*_{AB}$ is a given solution and $c \in \mathbb{R}$.
Such situations occur at the point $q = i$ in particular  in the case of the exceptional static data, 
which admit non-trivial conformal rescalings that yield again  static data
(\cite{friedrich:confstatic}, \cite{friedrich:exactstat}). This case requires a detailed analysis because the number of simple eigenvalues may change with near $i$. We shall make some observations about this case but not analyse it in detail.

\vspace{.1cm}

\noindent
$-$ $t_{ab}(q) \neq 0$ and the map has simple eigenvalues, 
\begin{equation}
\label{simple-eigenvalues}
\alpha \neq \beta \neq \gamma \neq \alpha \quad \mbox{with} \quad  
\alpha + \beta + \gamma = 0.
\end{equation}
This condition will then also be satisfied on some neighbourhood of $q$.
The set of metrics with this property is open
in any topology on the set of metrics under consideration  
in which  $B_{abc}$ is  $C^1$-tensor field.
It is non-empty because, as we have seen above, 
$t_{ab}(i) = s_{ab}(i)$ for static data, for which $s_{ab}(i)$  
can be prescribed arbitrarily as part of their 
null data (\cite{friedrich:statconv}).

\vspace{.1cm}

In the following we shall mainly be concerned with the case 
where (\ref{simple-eigenvalues}) holds on some neighbourhood of $i$.
The discussion of  (\ref{A-zeta-t-sym=0}) shows that if there exists a solution 
$\zeta_{AB}$  to $2\,\zeta_{(A}\,^H\,\,t_{BCD)H} = - D_A\,^H\,t_{BCDH}$ at $i$, it is unique.

\vspace{.1cm}

\noindent
{\it The question about $D_a\omega(i)$ raised in section \ref{d-critrel} thus finds 
a complete answer if (\ref{simple-eigenvalues}) holds at $i$. If for a given metric $h$ satisfying this condition there exists a solution to our problem near $i$ then  it is unique.
In the case (\ref{double-eigenvalue}) the possible values of  $D_a(i)$ are restricted to the extent to which this is consistent with the results of  \cite{friedrich:confstatic}, \cite{friedrich:exactstat}. In the cases in which $t_{ab}(i) = 0$ we may expect restrictions on $D_a\omega(i)$, depending on the structure of the non-vanishing derivative of $B_{abc}$ of lowest order at $i$.}

\vspace{.3cm}

Denote by $\xi_{AB}$, $\lambda_{AB}$, $\eta_{AB}$ an orthogonal frame of eigenvectors of
$t_{ABCD}$ with corresponding eigenvalues $\alpha$, $\beta$, $\gamma$ which are
normalized so that
\begin{equation}
\label{frame-norm}
\xi_{AB}\,\xi^{AB} = \lambda_{AB}\,\lambda^{AB} = \eta_{AB}\,\eta^{AB} = - 2,
\end{equation}
whence
\[
\xi_{AB}\,\xi^{CB} = \lambda_{AB}\,\lambda^{CB} = \eta_{AB}\,\eta^{CB} = - \,\epsilon_A\,^C,
\] 
and assume the frame to be oriented such that 
\begin{equation}
\label{frame-orientation}
\eta_{AB} = i\,\,\xi_{EA}\,\lambda_B\,^E, \quad  
\xi_{AB} = i\,\,\lambda_{EA}\,\eta_B\,^E, \quad \quad 
\lambda_{AB} = i\,\,\eta_{EA}\,\xi_B\,^E.
\end{equation}

We consider first the case (\ref{double-eigenvalue}) so that 
$t_{ABCD}\,\xi^{CD} = - 2\,f\,\xi_{AB}$. Then 
\begin{equation}
\label{double-ev-t}
t_{ABCD} = \frac{f}{2} \left(2\,\xi_{AB}\,\xi_{CD} - \lambda_{AB}\,\lambda_{CD}
- \eta_{AB}\,\eta_{CD}\right)
=
 \frac{f}{2} \,(\xi_{AB}\,\xi_{CD}
+ \xi_{AD}\,\xi_{BC} + \xi_{AC}\,\xi_{DB}),
\end{equation}
where second equation is obtained by using the different representations 
\[
\epsilon_{AC}\,\epsilon_{BD} + \epsilon_{AD}\,\epsilon_{BC} = 2\,h_{ABCD}
=  - \,(\xi_{AB}\,\xi_{CD} + \lambda_{AB}\,\lambda_{CD} + \eta_{AB}\,\eta_{CD}), 
\]
of the metric. It holds
\[
\xi_{(A}\,^E\,t_{BCD)E} = 0, 
\,\,\,
\lambda_{(A}\,^E\,t_{BCD)E} = - \frac{3}{2}\,i\,f\,\eta_{(AB}\,\xi_{CD)}, \,\,\, 
\eta_{(A}\,^E\,t_{BCD)E} =  \frac{3}{2}\,i\,f\,\xi_{(AB}\,\lambda_{CD)}.
\]
The expansion
\begin{equation}
\label{zeta-expr}
\zeta_{AB} = x\,\xi_{AB} + y\,\lambda_{AB} + z\,\eta_{AB},
\end{equation}
thus implies
\[
- 2\,\zeta_{(A}\,^E\,t_{BCD)E} = 3\,i\,f\,\left(y\,\eta_{(AB}\,\xi_{CD)}
- z\,\xi_{(AB}\,\lambda_{CD)}\right),
\]
so that of the five free constants which define a general totally symmetric spinor of rank four only two are available on the right hand side. 
It follows that equation (\ref{spin-B-comp-cond}), considered as an algebraic equation for the covector $\zeta_{AB}$, can be solved 
at points where $t_{ABCD}$ has the form (\ref{double-ev-t}) if and only if
\begin{equation}
\label{spec-fin-requ}
D_{A}\,^H\,t_{BCDH}  = i\,\{b\,(\xi_{AB}\,\lambda_{CD} + \lambda_{AB}\,\xi_{CD})
+ c\,(\xi_{AB}\,\eta_{CD} + \eta_{AB}\,\xi_{CD})\},
\end{equation}
at these points with some real coefficients $b$, $c$.

\vspace{.2cm}

Assume now that $t_{ABCD}$  satisfies (\ref{simple-eigenvalues})
so that 
\begin{equation}
\label{t-exp}
t_{ABCD} = 
- \frac{1}{2}\left( \alpha\,\xi_{AB}\,\xi_{CD} + \beta\,\lambda_{AB}\,\lambda_{CD}
+ \gamma\,\eta_{AB}\,\eta_{CD}\right).
\end{equation}
Using  again the expansion (\ref{zeta-expr})
and observing  that
\[
\xi_{(A}\,^H\,t_{BCD)H} = - \frac{i}{2}\,(\beta - \gamma)\,\lambda_{(AB}\,\eta_{CD)}
\]
\[
\lambda_{(A}\,^H\,t_{BCD)H} = - \frac{i}{2}\,(\gamma - \alpha)\,\eta_{(AB}\,\xi_{CD)}
\]
\[
\eta_{(A}\,^H\,t_{BCD)H} = - \frac{i}{2}\,(\alpha - \beta)\,\xi_{(AB}\,\lambda_{CD)},
\]
we get
\[
- 2\,\zeta_{(A}\,^H\,\,t_{BCD)H}  = 
 i\,\{x\,(\beta - \gamma)\,\lambda_{(AB}\,\eta_{CD)}
+ y\,(\gamma - \alpha)\,\eta_{(AB}\,\xi_{CD)}
+ z\,(\alpha - \beta)\,\xi_{(AB}\,\lambda_{CD)}\},
\]
so that we must have
\begin{equation}
\label{gen-fin-requ}
D_{A}\,^H\,t_{BCDH} = 
 i\,\{x\,(\beta - \gamma)\,\lambda_{(AB}\,\eta_{CD)}
+ y\,(\gamma - \alpha)\,\eta_{(AB}\,\xi_{CD)}
+ z\,(\alpha - \beta)\,\xi_{(AB}\,\lambda_{CD)}\}.
\end{equation}
Assuming a general  expansion
\[
D_{A}\,^H\,t_{BCDH}  = i\,\{a\,\xi_{AB}\,\xi_{CD} + d\,\lambda_{AB}\,\lambda_{CD}
+ f\,\eta_{AB}\,\eta_{CD}
\]
\[
+ b\,(\xi_{AB}\,\lambda_{CD} +   \lambda_{AB}\,\xi_{CD})
+ c\,(\xi_{AB}\,\eta_{CD} +   \eta_{AB}\,\xi_{CD})
+ e\,(\eta_{AB}\,\lambda_{CD} +   \lambda_{AB}\,\eta_{CD})\},
\]
with real coefficients such that $a + d + f = 0$ (to make the expression on the right hand side  symmetric) and  
observing the relations
\[
\xi^{AB}\,\lambda_{(AB}\,\eta_{CD)} = 0, \quad 
\xi^{AB}\,\eta_{(AB}\,\xi_{CD)} = - \eta_{CD}, \quad 
\xi^{AB}\,\xi_{(AB}\,\lambda_{CD)} = - \lambda_{CD},
\]
\[
\lambda^{AB}\,\lambda_{(AB}\,\eta_{CD)} = - \eta_{CD}, \quad 
\lambda^{AB}\,\eta_{(AB}\,\xi_{CD)} = 0, \quad 
\lambda^{AB}\,\xi_{(AB}\,\lambda_{CD)} = - \xi_{CD},
\]
\[
\eta^{AB}\,\lambda_{(AB}\,\eta_{CD)} = - \lambda_{CD}, \quad 
\eta^{AB}\,\eta_{(AB}\,\xi_{CD)} = - \xi_{CD}, \quad 
\eta^{AB}\,\xi_{(AB}\,\lambda_{CD)} = 0,
\]
we conclude by  contracting both expressions above with the frame vectors that  
(\ref{gen-fin-requ}) implies  the conditions
\[
a = d = f = 0, \quad \quad 
b = \frac{1}{2}\,z\,(\alpha - \beta), \quad 
c = \frac{1}{2}\,y\,(\gamma - \alpha), \quad 
e = \frac{1}{2}\,x\,(\beta - \gamma).
\]
While the last three conditions can be satisfied by suitable choices of $x$, $y$, $z$, the first three conditions  imply obstructions if not satisfied. Translating the previous result into the present notation we can state the results as follows.

\begin{lemma}
\label{algebraic-constraints}
At points of $W$, where $t_{ABCD}$ has simple eigenvalues 
$\alpha$, $\beta$, $\gamma$
with corresponding eigenvectors
$\xi_{AB}$, $\lambda_{AB}$, $\eta_{AB}$,
equation (\ref{spin-B-comp-cond}), considered as an algebraic equation for the covector $\zeta_{AB}$, is solvable if and only if 
\begin{equation}
\label{direct-Dt-restriction}
\xi^{AB}\,\xi^{CD} \,D_{A}\,^H\,t_{BCDH} = 0, \quad 
\lambda^{AB}\,\lambda^{CD}\,D_{A}\,^H\,t_{BCDH} = 0, \quad
\eta^{AB}\,\eta^{CD}\,D_{A}\,^H\,t_{BCDH} = 0,
\end{equation}
so that $D_{A}\,^H\,t_{BCDH}$ has an expansion 
\[
D_{A}\,^H\,t_{BCDH}  =
\]
\begin{equation}
\label{Dt-restriction}
i\,\{b\,(\xi_{AB}\,\lambda_{CD} +   \lambda_{AB}\,\xi_{CD})
+ c\,(\xi_{AB}\,\eta_{CD} +   \eta_{AB}\,\xi_{CD})
+ e\,(\eta_{AB}\,\lambda_{CD} +   \lambda_{AB}\,\eta_{CD})\}.
\end{equation}
Necessary for this to be true is that $t^{ABCD}\,D_{A}\,^H\,t_{BCDH} = 0$.

At points at which $t_{ABCD}$ has simple eigenvalue $\alpha = - 2\,f \neq 0$, 
and eigenvalues $\gamma = \beta = f $ with respective (orthogonal) eigenvectors
$\xi_{AB}$, $\lambda_{AB}$, $\eta_{AB}$ 
equation  (\ref{spin-B-comp-cond}) is solvable if and only if the conditions above hold and in addition the equation
\[
\lambda^{AB}\,\eta^{CD}\,D_{A}\,^H\,t_{BCDH} = 0,
\]
holds, so that $D_{A}\,^H\,t_{BCDH}$ has an expansion 
\[
D_{A}\,^H\,t_{BCDH}  = i\,\{b\,(\xi_{AB}\,\lambda_{CD} + \lambda_{AB}\,\xi_{CD})
+ c\,(\xi_{AB}\,\eta_{CD} + \rho_{AB}\,\eta_{CD})\}.
\]
\end{lemma}

The condition above can be understood as differential relations relating $t_{ab}$ and $D_at_{bc}$. While these allow us with the assumption (\ref{simple-eigenvalues}) to obtain pointwise expressions for a covector $\zeta_{AB}$ they do not tell us whether this field of covectors is in fact a differential of some function.

\vspace{.1cm}

The conditions above only depend on the conformal class of $h$. While  the eigenvalues and the (normalized) eigenvectors transform
under a rescaling $h_{ab} \rightarrow \hat{h}_{ab} = \vartheta^2\,h_{ab}$,
$\rho \rightarrow \hat{\rho} = \vartheta\,\rho$ as conformal densities, 
 it holds 
\[
\hat{D}_a\hat{t}_{bc} - \hat{D}_b\hat{t}_{ac} = 
\vartheta^{-1}(D_a t_{bc} - D_b t_{ac})
\]
\[
- \vartheta^{-2}\left(D^d\vartheta\,(t_{da}\,h_{bc} - t_{db}\,h_{ac})
+ 2\,(D_a\vartheta\,t_{bc} - D_b\vartheta_{ac})\right).
\]
In terms of space spinors this implies that $t_{ABCD}$ changes up to multiplicative  factors by an additive term of the form
\[
D_{(A}\,^H \vartheta\,t_{BCD)H}.
\]
It holds, however,   
\[
\xi^{AB}\,\xi^{CD}\,D_{(A}\,^H \vartheta\,t_{BCD)H} =
\frac{1}{3}\,(\xi^{AB}\,\xi^{CD} + \xi^{AD}\,\xi^{BC} + \xi^{AC}\,\xi^{DB})
D_{A}\,^H \vartheta\,t_{BCDH}
\]
\[
= \xi^{AB}\,\xi^{CD}D_{A}\,^H \vartheta\,t_{BCDH}
= \alpha\,\xi^{AB}\,\xi_{BH}  D_{A}\,^H \vartheta = 0,
\]
and similarly
$\lambda^{AB}\,\lambda^{CD}\,D_{(A}\,^H \vartheta\,t_{BCD)H} = 0$, 
$\eta^{AB}\,\eta^{CD}\,D_{(A}\,^H \vartheta\,t_{BCD)H} = 0 $. 
Moreover,
\[
\lambda^{AB}\,\eta^{CD}\,D_{(A}\,^H \vartheta\,t_{BCD)H} 
= \frac{1}{2}\,(
\lambda^{AB}\,\eta^{CD} 
+ \lambda^{CD}\,\eta^{AB})\,D_{A}\,^H\vartheta\, t_{BCDH}
\]
\[
= \frac{i}{2}\,(\gamma - \beta)\,\xi^{AB}\,D_{AB}\vartheta = 0
\quad
\mbox{if} \quad  \gamma = \beta.
\]

\vspace{.1cm}

 In the following we will need to take derivatives of the  eigenvalue functions and
eigenvector fields. {\it It will then always be assumed that the eigenvalues are simple}, because in  that case we have the following result.

\begin{lemma}
\label{eigenfields-smooth}
If no two eigenvalues coincide near $i$,  the eigenvalues and the eigenvector fields (as normalized above) are smooth
(real analytic) near the point $i$  if $h$ is.
\end{lemma}

\noindent
{\bf Proof}: We only discuss the analytic case here, the smooth case is similar.
Let $x^a$ denote real analytic coordinates on $W$. Because the eigenvalues are simple we can assume that they define functions  
$\alpha = \alpha(x^a)$, $\beta = \beta(x^a)$, $\gamma = \gamma(x^a)$ such that $\alpha(x^a) >  \beta(x^a) > \gamma(x^a)$   for $x^a$ near $x^a_* = x^a(i)$. These functions are then real analytic. In fact,
for given value of  $x^a$ the eigenvalues are the zeros of the polynomial 
$P(x^a, z) = \det(t^c\,_b(x^a) - z\,h^c\,_b(x^a))$ in $z \in \mathbb{R}$. The function
$P$ of $x^a$ and $z$ is real analytic and because the eigenvalues are simple the derivatives $P_{,z}(x^a, z)$ do not vanish for $x^a$ near $x^a_*$ if $z$ coincides with one of the eigenvalues. The assertion thus follows from the implicit function theorem.

Let $W \times \mathbb{R}^2 \ni (x^c, u^{A}) \rightarrow \xi^a(x^c, u^{A}) \in TW$ be a real analytical embedding into the tangent bundle such that $\xi_a\,\xi^a = - 2$ for 
$(x^c, u^{A}) \in W \times \mathbb{R}^2$ and consider the analytic function 
$f^a(x^c, u^A) = t^a\,_b(x^c)\,\xi^b(x^c, u^A) - \alpha(x^c)\,\xi^a(x^c, u^A)$. If for some 
$u_*^A$ the vector $\xi^b(x^c_*, u^A_*)$ is an eigenvector of $t^a\,_b$ so that 
$t^a\,_b(x^c_*)\,\xi^b(x^c_*, u^A_*) - \alpha(x^c_*)\,\xi^a(x^c_*, u^A_*) = 0$, the matrix
$f^a\,_{,B}(x^c_*, u^A_*)$ has rank $2$. Otherwise there would arise a contradiction to the assumption that the eigenvalues are simple. In fact,
there would exist $\nu^B \neq 0$ such that $0 = f^a\,_{,B}(x^c_*, u^A_*)\,\nu^B 
= t^a\,_b(x^c_*)\,k^b - \alpha(x^c_*)\,k^a$
with $k^a = \xi^a\,_{,B}(x^c_*, u^A_*)\,\nu^B$. But $k^a \neq 0$ because $\xi$ is an embedding so that  $k^a$ would be an eigenvector with eigenvalue  
$\alpha(x^c_*)$, orthogonal to to $\xi^a(x^c_*, u^A_*)$ by  the normalization of $\xi$.  
By the implicit function theorem there exist then an analytic function 
$u^A = u^A(x^c)$ near $x^a = x^a_*$ such that the normalized analytic vector field $\xi^a(x^c) \equiv \xi^a(x^c, u^A(x^c))$ satisfies  $t^a\,_b(x^c)\,\xi^b(x^c) = \alpha(x^c)\,\xi^a(x^c)$. \\
$\Box$

\vspace{.1cm}

Conditions (\ref{direct-Dt-restriction}), which read more explicitly 
\[
\alpha\,\xi^{AB}\,D^C\,_B\xi_{AC} = \frac{i}{2}\,
(\beta\,\eta^{AB}\,\lambda^{CD} - \gamma\,\lambda^{AB}\,\eta^{CD})\,D_{AB}\,\xi_{CD},
\]
\[
\beta\,\lambda^{AB}\,D^C\,_B\lambda_{AC} = \frac{i}{2}\,
(\gamma\,\xi^{AB}\,\eta^{CD} - \alpha\,\eta^{AB}\,\xi^{CD})\,D_{AB}\,\lambda_{CD},
\]
\[
\gamma\,\eta^{AB}\,D^C\,_B\eta_{AC} = \frac{i}{2}\,
(\alpha\,\lambda^{AB}\,\xi^{CD} - \beta\,\xi^{AB}\,\lambda^{CD})\,D_{AB}\,\eta_{CD},
\]
do not impose conditions on the derivatives of the eigenvalues of $t_{ABCD}$
and represent only two independent conditions because $t_{ab}$ is trace free.

\vspace{.2cm}

\noindent
{\it With (\ref{spin-Bab-tab-relation}) conditions (\ref{direct-Dt-restriction}) imply near $i$
the equivalent conformally invariant restrictions
\[
\xi^{AB}\,\xi^{CD} \,b_{ABCD} = 0, \quad 
\lambda^{AB}\,\lambda^{CD}\,b_{ABCD} = 0, \quad
\eta^{AB}\,\eta^{CD}\,b_{ABCD} = 0,
\]
and thus in particular} $t^{ABCD}\,b_{ABCD} = 0$.
Furthermore one gets the relations
\[
- i\,\xi^{AB}\,\lambda^{CD}\,b_{ABCD}  = 4\,b\,\rho + (\alpha - \beta)\,\eta^{AB}\,D_{AB}\rho,
\]
\[
- i\,\lambda^{AB}\,\eta^{CD}\,b_{ABCD}  = 4\,e\,\rho + (\beta - \gamma)\,\xi^{AB}\,D_{AB}\rho,
\]
\[
- i\,\eta^{AB}\,\xi^{CD}\,b_{ABCD}  = 4\,c\,\rho + (\gamma - \alpha)\,\lambda^{AB}\,D_{AB}\rho,
\]

\vspace{.32cm}

The expansions (\ref{t-exp}) and (\ref{Dt-restriction}) imply
\begin{equation}
\label{b-explicit}
b = \frac{1}{4}\,\eta^{AB}\,D_{AB}\,\alpha - \frac{i\,\alpha}{4}\,\lambda^{AB}\,D^C\,_A\,\xi_{BC}
- \frac{\gamma}{8}\,\xi^{AB}\,\eta^{CD}\,D_{AB}\,\xi_{CD},
\end{equation}
\begin{equation}
\label{c-explicit}
c = \frac{1}{4}\,\lambda^{AB}\,D_{AB}\,\gamma - \frac{i\,\gamma}{4}\,\xi^{AB}\,D^C\,_A\,\eta_{BC}
- \frac{\beta}{8}\,\eta^{AB}\,\lambda^{CD}\,D_{AB}\,\eta_{CD},
\end{equation}
\begin{equation}
\label{e-explicit}
e = \frac{1}{4}\,\xi^{AB}\,D_{AB}\,\beta - \frac{i\,\beta}{4}\,\eta^{AB}\,D^C\,_A\,\lambda_{BC}
- \frac{\alpha}{8}\,\lambda^{AB}\,\xi^{CD}\,D_{AB}\,\lambda_{CD}.
\end{equation}
Because of the equation $D^at_{ab} = 0$, which reads more explicitly 
\[
D^{AB}(\alpha\,\xi_{AB}) = t^{ABCD}\,D_{AB}\,\xi_{CD}, \quad
D^{AB}(\beta\,\lambda_{AB}) = t^{ABCD}\,D_{AB}\,\lambda_{CD}, 
\]
\[
D^{AB}(\gamma\,\eta_{AB}) = t^{ABCD}\,D_{AB}\,\eta_{CD},
\]
these equations can assume different forms. They can be used in the previous equations to express the non-vanishing components of the Cotton tensor, which is of third order in the metric, in terms of the function $\rho$, its differential, the eigenvalues and their differentials, and the eigenframe fields and their derivatives.

\vspace{.2cm}

 In the case (\ref{simple-eigenvalues}) suitable contractions of  (\ref{gen-fin-requ}) with the frame vectors give the solution formula

\begin{equation}
\label{zeta-sol-form}
\zeta_{AB} = 2\left(
\frac{e}{\beta - \gamma}\,\xi_{AB}
+ \frac{c}{\gamma - \alpha}\,\lambda_{AB}
+ \frac{b}{\alpha - \beta}\,\eta_{AB}
\right),
\end{equation}
and it follows from Lemma \ref{eigenfields-smooth} that the right hand side of this equation and thus $\zeta_{AB}$ is real analytic.
If (\ref{Dt-restriction}) holds, 
a direct calculation shows that the right hand side of (\ref{zeta-sol-form}) coincides with the right hand side of (\ref{t-contr-B-comp-cond}).

\noindent
Similarly  we obtain in the case  (\ref{double-eigenvalue})
\[
\zeta_{AB} = x\,\xi_{AB} 
 + 2\left(
\frac{c}{\gamma - \alpha}\,\lambda_{AB}
+ \frac{b}{\alpha - \beta}\,\eta_{AB}
\right),
\]
where the coefficient $x$ remains undetermined.

\vspace{.2cm}

We can now require that the 1-form $\zeta_{AB}$ given by equation (\ref{zeta-sol-form}) be closed. This condition is equivalent to $D^C\,_{(A}\,\zeta_{B)C} = 0$.
After contraction with the three independent eigenvector fields it assumes with (\ref{zeta-sol-form}) the form
\[
D^{AB}\left( \frac{b}{\alpha - \beta}\,\lambda_{AB}
-  \frac{c}{\gamma - \alpha}\,\eta_{AB}\right) = 
i \left(\frac{e}{\beta - \gamma}\,\xi^{AB}
+ \frac{c}{\gamma - \alpha}\,\lambda^{AB}
+ \frac{b}{\alpha - \beta}\,\eta^{AB}\right)D^C\,_A\,\xi_{BC},
\]
\[
D^{AB}\left(\frac{e}{\beta - \gamma}\,\eta_{AB}
-   \frac{b}{\alpha - \beta}\,\xi_{AB}\right) = 
i \left(\frac{e}{\beta - \gamma}\,\xi^{AB}
+ \frac{c}{\gamma - \alpha}\,\lambda^{AB}
+ \frac{b}{\alpha - \beta}\,\eta^{AB}\right)D^C\,_A\,\lambda_{BC},
\]
\[
D^{AB}\left( \frac{c}{\gamma - \alpha}\,\xi_{AB}
-  \frac{e}{\beta - \gamma}\,\lambda_{AB}\right) = 
i \left(\frac{e}{\beta - \gamma}\,\xi^{AB}
+ \frac{c}{\gamma - \alpha}\,\lambda^{AB}
+ \frac{b}{\alpha - \beta}\,\eta^{AB}\right)D^C\,_A\,\eta_{BC}.
\]
If these equations are written out more explicitly some of the derivatives may be reexpressed
by using the relation $D^{AB}\,t_{ABCD} = 0$. With  (\ref{b-explicit}), (\ref{c-explicit}), (\ref{e-explicit}),
these equations can be written as explicit differential conditions of second order on the eigenvalue functions  and the eigenframe fields.

\section{Concluding remarks}
\label{k-datevol}

In the case of general time reflection symmetric vacuum data the data $u$ for the conformal field equations are derived from the metric $h$ and the conformal factor $\Omega$ 
determined by  (\ref{Omega-and-derivs-at-i}), (\ref{Omega-equ}).
These data include in particular  the rescaled conformal Weyl tensor which represents,  in the conformal gauge in which space-like infinity is represented by the point $i$,  the most singular field comprised by $u$.
In spinor notation it is given by   
\[
\phi_{ABCD} = \frac{1}{\Omega^2}(D_{(AB}D_{CD)}\Omega + \,\Omega\,s_{ABCD}).
\]
Near $i$ it can be written in the form  $\phi_{ABCD} = \left(\phi'_{ABCD} +\phi^W_{ABCD}\right)$ where 
the part $\phi'_{ABCD}$, which carries only  local information on the metric $h$,
and the part $\phi^W_{ABCD}$, which contains global information on $(S, h)$,
have been referred to
 in \cite{friedrich:i-null} as the `massless'  and the `massive' part of the rescaled Weyl spinor respectively. It holds
\[
\phi^W_{ABCD} =
\Gamma^{-2}\left\{
\left(- \frac{3}{2}\,\Gamma^{-1/2}\,D_{(AB}\Gamma\,D_{CD)}\Gamma
+ \Gamma^{1/2}\,D_{(AB}D_{CD)}\Gamma\right)U\,W
\right.
\]
\[
+ 2\,\Gamma^{1/2}\,\left(
W\,D_{(AB}\Gamma\,D_{CD)}U - 3\,U\,D_{(AB}\Gamma\,D_{CD)}W
\right)
\]

\[
+ 2\,\Gamma^{3/2}\,\left(
- U\,D_{(AB}D_{CD)}W 
- W\,D_{(AB}D_{CD)}U
+ 6\,D_{(AB}U\,D_{CD)}W + U\,W\,s_{ABCD}
\right)
\]
\[
\left. + 2\,\Gamma^2\left(- W\,D_{(AB}D_{CD)}W
+ 3\,D_{(AB}W\,D_{CD)}W
+ \frac{1}{2}\,W^2\,s_{ABCD}
\right) \right\},
\]
so that $\phi^W_{ABCD} = O(\Gamma^{- 3/2})$ as $\Gamma \rightarrow \Gamma(i) = 0$.
If $h$ satisfies condition $(*)$ the massless part is given in the notation of section
\ref{j-subconds}
by 
\[
\phi'_{ABCD}  = \mu\,t_{ABCD}.
\]
It is thus regular also in the gauge in which the massive part is strongly singular
(this field is in fact  independent of $\mu$, the factor $\mu$ only  turns up on the right hand side because of our definition of $\rho$ in section \ref{c-problem}).

In the conformal gauge used in the picture with the cylinder $I$ and the boundary 
$I^0 = \bar{S} \cap I$ of the initial hypersurface these fields pick up a factor so that 
\[
\phi_{ABCD} = \frac{\varpi^3}{\Omega^2}(D_{(AB}D_{CD)}\Omega + \Omega\,s_{ABCD}),
\]
with a  function $\varpi$ that behaves as
\[
\varpi = O(\Gamma^{1/2})  \quad \mbox{as} \quad \Gamma \rightarrow \Gamma|_{I^0} = 0.
\]
With respect to the differential structure  underlying  this picture the field $\phi_{ABCD}$ then extends smoothly to $I^0$. The unknown $u$
comprises further fields (cf \cite{friedrich:i-null}) but the expression above may suffice to illustrate the kind of operations  needed in the transition  from the initial data $(h, \Omega)$ to the initial data $u$ on $S$.

If $h$ is conformally static so that, in the notation of sections \ref{c-problem} and  \ref{d-critrel}, the field $h' = \omega^{-2}\,h$ satisfies the conformal static field equations
$\Sigma_{ab}[h'\,\rho'] = 0$, $R[h'] = 0$, the corresponding vacuum data are given by
\[
\tilde{h}_* = \Omega_*^{-2}\,h \quad \mbox{with} \quad 
\Omega_*= \frac{\rho}{\mu\,(1 + \sqrt{\frac{\rho}{\mu}} \, W)^2}
\quad
\mbox{where} \quad  
W =  \sqrt{\frac{\mu}{\omega}},
\]
and  $\omega = \frac{\mu^2}{W^2}$ and $h$  satisfy the equation
$\Upsilon_{ab}[\omega, h] = 0$.

\vspace{.3cm}

Let $(S, h)$ be conformal vacuum data as specified  in the beginning of section \ref{c-problem}. We can then calculate from these the initial data $u$ for the conformal field equations and determine (in principle) the solution-jets $J^p_I(u)$ in $I$. It is known that  for low orders $p$ these will extend smoothly to the critical sets. Let $\hat{p}$ be the lowest order at which this is not the case any longer. There can be several reasons for this. It may happen that $h$ satisfies the conditions of the criterion so that 
$\tilde{h}$ is conformally static up to the relevant order  but it has a `wrong' scaling.
In that case the equation, $\Upsilon_{ab}[ \omega, h] = 0$, which may not be solvable in a neighbourhood of $i$ should at least be solvable up to the relevant order
and the situation can essentially be dealt with as
in \cite{j.a.v.kroon:2011}. 

A new and most interesting situation occurs if the data turn out to be such that the order $\hat{p}$  corresponds to  the first order at which one of the conditions of the criterion is violated (the order $\hat{p}$ of the solution-jet $J^{\hat{p}}_I(u)$ and  the order at which the data $h$ and $\Omega$ enter the calculation of $J^{\hat{p}}_I(u)$ on $I^0$
must be carefully distinguished here). In this situation one will have to 
inspect  the expression of $J^{\hat{p}}_I(u)$ at $I^{\pm}$ for the precise requirements 
which need to be imposed on the data at $I^{0}$ for the logarithmic terms to drop out.
{\it The critical question then is whether these requirements comprise the conditions of our criterion and the rescaling of the metric by the solution to  the equation 
$\Upsilon_{ab}[ \omega, h] = 0$ up to the relevant order}. 
(The need to satisfy the  conditions of the criterion 
may  explain the observation in \cite{j.a.v.kroon:2011} that several steps, involving the inspection of several $J^{p'}_I(u)$ with $p'$ close to $\hat{p}$, were needed to determine the correct scaling.)
A positive answer would give strong evidence that asymptotic staticity is the relevant regularity condition. Though it is hard to conceive of a reasonable condition somehow positioned between $(*)$ and asymptotic staticity, a priori we cannot exclude the possibility that such an unknown condition may be required for regularity at higher orders.

\section{Appendix}
\label{m-unique}

In the following we show that the conformal extensions of asymptotically flat Riemannian manifolds considered in this article are unique up to conformal diffeomorphisms and that a conformal diffeomorphism which maps one such manifold  
onto another one with a conformal factor that is bounded
induces a smooth conformal diffeomorphism
of the extended manifolds. The last statement is used to discuss data which satisfy requirements 
(i') - (iv') of Definition \ref{as-stat} in terms of conformal extensions. Moreover, it
strengthens the results of
\cite{friedrich:confstatic}, \cite{friedrich:exactstat}, where the existence of an extended conformal diffeomorphism has been assumed. 
We note that the following arguments  profited from  \cite{lelong-ferrand:1976} and the discussion of the Myers Steenrod Theorem in \cite{kobayashi:nomizu:I:1996}.

Let $(\tilde{S}, \tilde{h})$ be a smooth (negative definite) Riemannian manifold of dimension $n \ge 3$ with one (for convenience) asymptotically flat end. It 
admits a conformal extension if there exists a quintuple $(S, i, h, \Omega, \kappa)$ with the following properties. $S$ is a smooth, compact n-dimensional manifold and $i$ a point in $S$, $h$ is a smooth Riemannian metric and $\Omega \in C^2(S) \cap 
C^{\infty}(S')$, where $S' = S \setminus \{i\}$, so that 
\[
\Omega(i) = 0, \quad \quad d\Omega(i) = 0, \quad 
Hess_h\Omega(i) = - \gamma\,h(i), \quad \Omega > 0 \,\,\,\mbox{on} \,\,\,
S',
\]
with some constant $\gamma > 0$ 
and $\kappa$ is a diffeomorphism $S'   \rightarrow \tilde{S}$ which maps $i$-punctured neighbourhoods of $i$ in $S$ onto neighbourhoods of the asymptotically flat end and satisfies
\[
\kappa_* (\tilde{h}) = \Omega^{-2}\,h
\quad \mbox{on} \quad S'.
\]
We note that in the statements above `smooth' could be replaced by `real analytic' and  the results below would then be obtained in the category of real analytic manifolds and maps.

\begin{proposition}
\label{unique conformal extension}
Assume that $(\hat{S}, \hat{i}, \hat{h}, \hat{\Omega}, \hat{\kappa})$ satisfies with respect to $(\tilde{S}, \tilde{h})$ the same properties as $(S, i, h, \Omega, \kappa)$. Then there exist a smooth diffeomorphism $\phi:  S \rightarrow \hat{S}$ and a positive function 
$\Theta \in C^{\infty}(S)$ such that $\phi_* \hat{h} = \Theta^2\,h$.
\end{proposition}

\noindent
{\bf Proof}:
By our assumptions the map 
$\phi = \hat{\kappa}^{-1} \circ \kappa: S' \rightarrow \hat{S'} 
= \hat{S} \setminus \hat{i}$
is a $C^{\infty}$ diffeomorphism which extends to a homeomorphism
$\phi: S \rightarrow \hat{S}$ by setting $\phi(i) = \hat{i}$ and it holds
\[
\phi_*(\hat{h}) = \Theta^2\,h
\quad \mbox{with} \quad 
\Theta = \frac{\hat{\Omega} \circ \phi}{\Omega}
\quad \mbox{on}\quad S'.
\]
We show that $\Theta$ extends as a positive $C^{\infty}$-function to $S$. Because the definition above leaves the freedom to perform rescalings 
$(h, \Omega) \rightarrow (\vartheta^2 h, \vartheta\,\Omega)$ and similarly for 
$(\hat{h}, \hat{\Omega})$ the scalings of $h$ and $\hat{h}$ can be assumed  such that
$R[h] = 0$ and $R[{\hat{h}}] = 0$ on some open neighbourhoods $W$ and $\hat{W}$ of $i$ and $\hat{i}$ respectively with $\phi(W) = \hat{W}$.
Otherwise we choose a conformal factor $\vartheta > 0$ on $S$ which in normal coordinates $x^a$ near $i$ has the form $\vartheta = 1 + \beta\,|x|^2$ with a constant 
$\beta < - \frac{a_n\,R[h](i)}{2\,n}$ where $a_n = \frac{n - 2}{4\,(n - 1)}$. The transformation law
\[
\Delta_h \vartheta - a_n R[h]\,\vartheta = - a_n\,\vartheta^{\frac{n + 2}{n - 2}}\,
R[h'] \quad \mbox{with} \quad h' = \vartheta^\frac{4}{n - 2}\,h,
\]
then shows that $R[h']  < 0$ near $i$. We can thus assume that $R[h] < 0$
on some geodesics ball $B_{\alpha} (i)$ with center $i$ and sufficiently small radius  
$\alpha > 0$. 
The Dirichlet problem for the equation
\[
\Delta_h \vartheta - a_n R[h]\,\vartheta = 0,
\]
on $B_{\alpha}(i)$ with positive data on $\partial B_{\alpha}(i)$ has then a unique solution, which is positive  by the strong maximum principle. Extending the function $\vartheta$ as a smooth positive function to $S$
we see by the transformation law above that the Ricci scalar 
$R[h']$ of the metric $h' = \vartheta^\frac{4}{n - 2}\,h$ vanishes on $W \equiv B_{\alpha}(i)$. 
An analogous rescaling can be performed of $\hat{h}$ if necessary. 
If  $\alpha$ is small enough it can be arranged  that $\phi(W) = \hat{W}$.

Denoting  the inverse of $\psi$ by $\phi$,   we get for arbitrary $f \in C^2(\hat{S'})$ the relation 
\[
\Delta_h(f \circ \phi) =  (\Delta_{\psi_*h} \,f) \circ \phi
=  (\Delta_{\Sigma^2 \hat{h}} \,f) \circ \phi 
\quad \mbox{on} \quad S'
\quad \mbox{with} 
\quad 
\Sigma = (\Theta \circ \psi)^{-1} .
\]
Observing 
that
\[
R[\Sigma^2 \hat{h}] \circ \phi = R[\psi_* h] \circ \phi = R[h],
\]
and using  the general  transformation law of the conformally covariant Laplacian 
we get 
\[
(\Delta_h - a_n\,R[h])( \Theta^{\frac{- 2 + n}{2}}\,(f \circ \phi) )
= (\Delta_h - a_n\,R[h])( (\Sigma^{\frac{2 - n}{2}}\,f) \circ \phi) 
\]
\[
= \left((\Delta_{\Sigma^2 \hat{h}}  - a_n \,R[\Sigma^2 \hat{h}])\,(\Sigma^{\frac{2 - n}{2}}\,f)\right) \circ \phi 
+ a_n\left(R[\Sigma^2 \hat{h}] \circ \phi  - R[h]\right)\,(\Sigma^{\frac{2 - n}{2}}\,f) \circ \phi 
\]
\[
= \left(\Sigma^{- \frac{2 + n}{2}}\,(\Delta_{\hat{h}} - a_n\,R[\hat{h}])\, f\right) \circ \phi
= \Theta^{\frac{2 + n}{2}}\,\left((\Delta_{\hat{h}} - a_n\,R[\hat{h}])\, f\right) \circ \phi.
\]
With the choice $f = 1$ this gives 
\[
\Delta_h( \Theta^{\frac{n - 2}{2}}) = 0 \quad \mbox{on} \quad B_{\alpha}(i) \setminus \{i\},
\]
where $B_{\alpha}(i) \subset W$ is a suitable $h$-geodesic ball with center $i$. Because 
$\Theta > 0$ on $B_{\alpha}(i) \setminus \{i\}$ a result of  \cite{gilbarg:serrin:} applies which says
that 
\[
 \Theta^{\frac{n - 2}{2}} = c\,G + w
 \quad \mbox{on} \quad B_{\alpha}(i) \setminus \{i\},
 \]
where $c$ is a constant, $G$ is a solution to
$\Delta_h G = \delta_i$ on $B_{\alpha}(i)$, $\delta_i$ the Dirac measure with weight $1$ at $i$, and
and $w$ solves $\Delta_h w = 0$ on $B_{\alpha}(i)$. The function $G$ has a singularity at $i$ which is such that 
\[
(\Omega^{\frac{n - 2}{2}} \,G)(p) \rightarrow k = const. \neq 0
\quad \mbox{as} \quad p \rightarrow i.
\]
Writing the equation above in the form
\[
(\hat{\Omega} \circ \phi)^{\frac{n - 2}{2}} =   \Omega^{\frac{n - 2}{2}} \,( c\,G + w),
 \]
and observing that $\hat{\Omega}(\phi(p))  \rightarrow 0$
as $p \rightarrow i$ we conclude that $c = 0$. 
Since $\Theta$ is positive on $\partial B_{\alpha}(i)$ it follows that $w > 0$ on 
$B_{\alpha}(i)$, 
$\Theta$ extends to a smooth positive function on  $B_{\alpha}(i)$, and 
$ \bar{h} \equiv  \Theta^2\,h$ defines a smooth Riemannian metric on $B_{\alpha}(i)$.

The homeomorphism $\phi$ of $B_{\alpha}(i)$ onto $\phi(B_{\alpha}(i))$  thus induces an isometry $B_{\alpha}(i) \setminus \{i\}$ onto $\phi(B_{\alpha}(i)) \setminus \{\hat{i}\}$. We show that it extends smoothly to $B_{\alpha}(i)$. This is essentially a consequence of the Myers Steenrod Theorem. Because the assumptions made there are somewhat different from the situation given here, we include a proof.

Being an isometry, $\phi$ maps an $\bar{h}$-geodesic $\gamma(\tau)$, $\tau >0$, 
 in $B_{\alpha}(i) \setminus \{i\}$ onto an  $\hat{h}$-geodesic $\phi(\gamma(\tau))$
 in $\phi(B_{\alpha}(i)) \setminus \{\hat{i}\}$. 
 If $\lim_{\tau \rightarrow 0}\gamma(\tau) = i$, continuity of $\phi$ implies that 
$\lim_{\tau \rightarrow 0}\phi(\gamma(\tau)) = \hat{i}$. If 
$\lim_{\tau \rightarrow 0}\frac{d}{d\tau}\gamma = X$, we set
 $F(X) = \lim_{\tau \rightarrow 0}\frac{d}{d\tau}\phi(\gamma)$. The properties of geodesics under linear rescalings of affine parameters then imply that
 \begin{equation}
 \label{F--is-R-linear}
 F(c\,X) = c\,F(X), \quad c \in \mathbb{R}, \quad \quad 
 ||F(X)||_{\hat{h}} = ||X||_{\bar{h}}.
 \end{equation}
In particular, $\phi$ maps the $\bar{h}$-geodesic ball $B_{\alpha}(i)$
onto  the $\hat{h}$-geodesic ball $B_{\alpha}(\hat{i})$.

$F$ defines an isometry of $T_i S$ onto $T_{\hat{i}} \hat{S}$. To show this
we consider a certain general property of metrics. For suitably small  $\alpha_* > 0$
denote by  $B_{\alpha}(i)$ the $h$-geodesically convex  geodesic balls with center $i$ and radius $0 < \alpha \le \alpha_*$ and let $x^a$ be normal coordinates with origin at $i$ which cover these balls. In these coordinates we have
$\partial B_{\alpha}(i) =   \{y^a = \alpha\,x^a|\,x^a \in S^2\}$
where $S^2 \equiv \{x^a \in \mathbb{R}^3|\, x^a\,x^b\,\delta_{ab} = 1\}$.
Because $h_{ab}(x^c)  + \delta_{ab}  = O(|x|^2)$ we can write for  $x^a \in S^2$ and
$\alpha \le \alpha_*$
 \[
 h_{ab}(\alpha\,x^c) = - \delta_{ab} + \alpha\,
 \int _0^1  x^d\,h_{ab, d}(s\,\alpha\,x^c)\,ds.
 \]
Any $C^1$ curve $y^a(\tau)$ on  $\partial B_{\alpha}(i)$ can be written in the form 
$y^a(\tau) = \alpha\,z^a(\tau)$ with $z^a(\tau) \in S^2$, which gives 
 \[
-  h_{ab}(\alpha\,x^c)\,\dot{y}^a\,\dot{y}^b = 
 \alpha^2\left(\delta_{ab}\,\dot{z}^a\,\dot{z}^b 
  - \alpha\,
\dot{z}^a\,\dot{z}^b \, \int _0^1 x^d\,h_{ab, d}(s\,\alpha\,x^c)\,ds\right).
 \]
Because $h_{ab, d}(0) = 0$ we can choose $\alpha_* $ so  small  that 
\[
\left| \int _0^1 x^d\,h_{ab, d}(s\,\alpha\,x^c)\,ds \right| \le \frac{1}{3}
\quad \mbox{for} \quad 0 \le \alpha \le \alpha_*, \,\,\,x^a \in S^2,
 \]
which implies 
\[
\left|\dot{z}^a\,\dot{z}^b \,  \int _0^1 x^d\,h_{ab, d}(s\,\alpha\,x^c)\,ds \right| \le 
\delta_{ab}\,\dot{z}^a\,\dot{z}^b.
 \]
For  $0 < \alpha \le \alpha_* < 1$  this gives
\[
\sqrt{(1 - \alpha)}\,\sqrt{\delta_{ab}\,\dot{z}^a\,\dot{z}^b}
\le \frac{1}{\alpha}\,\sqrt{|h_{ab}(\alpha\,x^c)\,\dot{y}^a\,\dot{y}^b|} \le
\sqrt{(1 + \alpha)}\,\sqrt{\delta_{ab}\,\dot{z}^a\,\dot{z}^b}.
\]

Let $d$ denote the distance function on $S^2$ associated with the pull back to $S^2$ of the euclidean metric $\delta_{ab}$ on $T_iS$ and $d_{\alpha}$ the distance function on 
$\partial B_{\alpha}(i)$ associated with the pull-back to $\partial B_{\alpha}(i)$ of the metric 
$h_{ab}$ on $B_{\alpha}(i)$.

Consider the geodesics
$\gamma_k(\tau) = \tau\,x_k$ through $i$ where $x_k \in S^2$, $ k = 1, 2$,
$|\tau| \le \alpha_*$ and write $\gamma_k(1) = x_k$.
The estimates above  imply then
 \[
\sqrt{(1 - \alpha)}\,d(\gamma_1(1), \gamma_2(1))
 \le \frac{1}{\alpha}\,d_{\alpha}(\gamma_1(\alpha), \gamma_2(\alpha)) 
\le \sqrt{(1 + \alpha)}\,d(\gamma_1(1), \gamma_2(1)),
\]
and thus, with $X_k = \dot{\gamma}_k(0)$,

\[
h(X_1, X_2) = \cos(d(\gamma_1(1), \gamma_2(1)))
= 
\cos\left(\lim _{\alpha \rightarrow 0} \frac{1}{\alpha}\,d_{\alpha}(\gamma_1(\alpha),
\gamma_2(\alpha)) \right),
\]
and analogous relations hold with the metrics $\bar{h}$ and $\hat{h}$. Since $\phi$
maps $\partial B_{\alpha}(i)$ isometrically onto  $\partial B_{\alpha}(\hat{i})$ it follows 
that 
$\hat{d}_{\alpha}(\phi(\gamma_1)(\alpha), \phi(\gamma_2)(\alpha)) 
= \bar{d}_{\alpha}(\gamma_1(\alpha), \gamma_2(\alpha)) $
for $\alpha > 0$  which implies in the limit above with the first of properties (\ref{F--is-R-linear})
 that 
\begin{equation}
\label{F-is-isometry}
\hat{h}(F(X), F(Y)) = h(X, Y), \quad X, Y \in T_iS.
\end{equation}
A direct calculation using this relation then shows that 
\[
\hat{h}(F(X+ Y) - F(X) - F(Y), F(Z)) = 0, \quad X, Y, Z \in T_iS.
\]
Since $F$ maps by (\ref{F-is-isometry}) an $\bar{h}$-orthonormal basis 
$X_1$, \ldots $X_n$ onto an $\hat{h}$-orthonormal basis, letting $Z$ take the values $X_i$ allows us to conclude that $F$ is a linear map.

The properties obtained so far imply that 
$\phi \circ exp_{\,i} = exp_{\,\hat{i}} \circ F$
on a neighbourhood of the origin of $T_iS$. It follows that $\phi$  extends to a smooth 
$\bar{h}$ - $\hat{h} $ - isometry resp. to a smooth $h$ - $\hat{h}$ - conformal map 
at $i$.\\ $\Box$

\begin{corollary}
\label{smooth-conf-map-ext}
Let $(\tilde{S}, \tilde{h})$ and $(\hat{\tilde{S}}, \hat{\tilde{h}})$ be smooth, asymptotically flat Riemannian spaces of dimension $n \ge 3$ which admit smooth conformal extensions
$(S, i, h, \Omega, \kappa)$ and $(\hat{S}, \hat{i}, \hat{h}, \hat{\Omega}, \hat{\kappa})$ 
respectively in the sense described above and suppose that there exists a smooth diffeomorphism  $\Phi: \tilde{S} \rightarrow \hat{\tilde{S}}$,  
which maps open neighbourhoods of space-like infinity in $\tilde{S}$ onto such neighbourhoods in
$\hat{\tilde{S}}$ and which satisfies 
\[
\Phi_* \hat{\tilde{h}} = \lambda^2\, \tilde{h},
\]
with a conformal factor $\lambda$ which satisfies
\[
0 < \lambda \le K,
\]
with some constant $K > 0$.
Then there exists a smooth diffeomorphism 
$\phi: S \rightarrow \hat{S}$ which satisfies 
$\phi_*(\hat{h}) = \Theta^2\,h$, $\Theta > 0$, 
and
$\Phi = \hat{\kappa} \circ \phi \circ \kappa^{-1}$.
\end{corollary}

\noindent
{\bf Proof}: The map 
$\phi = \hat{\kappa}^{-1} \circ \Phi \circ \kappa: S' \rightarrow \hat{S}'$
satisfies
\[
\phi_*(\hat{h}) = \Theta^2\,h
\quad \mbox{with} \quad 
\Theta = \frac{(\hat{\Omega} \circ \phi)\,(\lambda \circ \kappa)}{\Omega} > 0.
\]
Again we can assume that the conformal extensions have been chosen such that $R[h]$ and $R[\hat{h}]$
vanish on some neighbourhoods of $B_{\alpha}(i)$ of $i$ and $B_{\hat{\alpha}}(\hat{i})$ of $\hat{i}$ respectively which satisfy 
$\phi(B_{\alpha}(i)) \subset B_{\hat{\alpha}}(\hat{i})$ and conclude that 
\[
 \Theta^{\frac{n - 2}{2}} = c\,G + w
 \quad \mbox{on} \quad B_{\alpha}(i) \setminus \{i\},
 \]
where $c$, $G$, and $w$ satisfy the same conditions as above. 
Writing the equation above in the form
\[
(\hat{\Omega} \circ \phi)^{\frac{n - 2}{2}} \, (\lambda \circ \kappa)^{\frac{n - 2}{2}}  
=   \Omega^{\frac{n - 2}{2}} \,( c\,G + w),
 \]
and observing that 
\[
\hat{\Omega}(\phi(p))  \rightarrow 0 \quad \mbox{and} \quad 
(\Omega^{\frac{n - 2}{2}} \,G)(p) \rightarrow k = const. \neq 0
\quad \mbox{as} \quad p \rightarrow i,
\]
we conclude that $c = 0$ because   
$\hat{\Omega}(\phi(p)) \cdot \lambda(\kappa(p))  \rightarrow 0$
as $p \rightarrow i$
by our assumptions on $\lambda$. As above it follows that $\Theta$ admits a smooth positive extension to $i$ and $\phi$ admits a smooth extension to $i$ so that  $\phi_*(\hat{h}) = \Theta^2\,h$.\\
$\Box$

}

\end{document}